\documentclass[aps,pra,twocolumn,superscriptaddress,letterpaper]{revtex4-1}
\usepackage{graphicx}
\usepackage{amsmath}
\usepackage{amssymb}
\usepackage{esint}
\usepackage{verbatim}
\usepackage{color}
\usepackage{hyperref}
\usepackage{tabulary}
\usepackage{tabularx}
\usepackage{booktabs}
\usepackage{tensor}

\newcommand{\vecb}[1]{\mathbf{#1}}
\newcommand{\be}{\begin{equation}}
\newcommand{\ee}{\end{equation}}
\newcommand{\ba}{\begin{array}}
\newcommand{\ea}{\end{array}}
\newcommand{\bqa}{\begin{eqnarray}}
\newcommand{\eqa}{\end{eqnarray}}

\newcommand{\rttensortwo}[1]{\bar{\bar{#1}}}
\newcommand{\tensoravg}[1]{\left\langle #1 \right\rangle_{\text{t}}}
\newcommand{\modeavg}[1]{\left\langle #1 \right\rangle_{\text{m}}}

\newcommand{\qavg}[1]{\left\langle #1 \right\rangle}

\renewcommand{\Re}{\text{Re}}
\renewcommand{\Im}{\text{Im}}
\newcommand{\nbar}{n_{\text{m}}}
\newcommand{\nm}{\nbar}
\newcommand{\dotnm}{\dot{n}_{\text{m}}}
\newcommand{\nbath}{n_{\text{b}}}
\newcommand{\nmnot}{n_{\text{0}}}
\newcommand{\nmoffset}{\tilde{n}_{\text{0}}}
\newcommand{\Tbath}{T_{\text{b}}}
\newcommand{\Tbathmin}{T_{\text{b,min}}}
\newcommand{\Tf}{T_{\text{f}}}

\newcommand{\Tm}{T_{\text{m}}}

\newcommand{\deltab}{\delta_{\text{b}}}

\newcommand{\tauth}{\tau_{\text{th}}}
\newcommand{\tauph}{\tau_{\text{ph},0}}
\newcommand{\taucoh}{\tau_{\text{coh},0}}
\newcommand{\Nshield}{N_{\text{shield}}}
\newcommand{\meff}{m_{\text{eff}}}

\newcommand{\nBose}{n_{\text{B}}}
\newcommand{\omegacutoff}{\omega_{\text{co}}}

\newcommand{\xm}{x_{\text{m}}}
\newcommand{\nbathp}{n_{\text{p}}}
\newcommand{\Tbathp}{T_{\text{p}}}
\newcommand{\gammap}{\gamma_{\text{p}}}
\newcommand{\nmi}{n_{\text{m}}^{\text{i}}}
\newcommand{\nmf}{n_{\text{m}}^{\text{f}}}
\newcommand{\nmfp}{\tilde{n}_{\text{m}}^{\text{f}}}
\newcommand{\nmzero}{\nmi}
\newcommand{\nmstar}{n_{\text{m}}^*}
\newcommand{\ncavRO}{n_{\text{c,RO}}}

\newcommand{\Gammanpon}{\theta_{\nbathp}}
\newcommand{\Gammanpoff}{\zeta_{\nbathp}}
\newcommand{\Gammagammapon}{\theta_{\gammap}}
\newcommand{\Gammagammapoff}{\zeta_{\gammap}}

\newcommand{\sigmahole}{\sigma_{\text{size}}}
\newcommand{\sigmapos}{\sigma_{\text{pos.}}}

\newcommand{\nbathgen}{n_{\text{b}}}
\newcommand{\Gth}{G_\text{th}}
\newcommand{\kB}{k_\text{B}}
\newcommand{\lambdac}{\lambda_{\text{c}}}
\newcommand{\ncav}{n_{\text{c}}}
\newcommand{\ncavCW}{n_{\text{c,CW}}}
\newcommand{\gzero}{g_{\text{0}}}

\newcommand{\kappae}{\kappa_{\text{e}}}
\newcommand{\gammaOM}{\gamma_{\text{OM}}}

\newcommand{\Qm}{Q_{\text{m}}}

\newcommand{\omegal}{\omega_{\text{l}}}
\newcommand{\ahat}{\hat{a}}
\newcommand{\adag}{\hat{a}^{\dagger}}

\newcommand{\omegac}{\omega_{\text{c}}}

\newcommand{\gammanotO}{\gamma_{\text{0}}}

\newcommand{\nNEP}{n_\text{NEP}}

\newcommand{\gammaSB}{\Gamma_\text{SB,0}}

\newcommand{\etaSPD}{\eta_\text{SPD}}

\newcommand{\gammai}{\gamma_{\text{i}}}
\newcommand{\gammam}{\gamma_{\text{m}}}
\newcommand{\omegam}{\omega_{\text{m}}}
\newcommand{\deltaf}{\delta f}
\newcommand{\Deltaf}{\Delta\omegam}
\newcommand{\Deltafj}{\Delta_{1/2}}

\newcommand{\Toff}{\tau}
\newcommand{\Tpulse}{T_\text{pulse}}

\newcommand{\ton}{t}
\newcommand{\Tper}{T_\text{per}}

\newcommand{\Vm}{V_{m}}
\newcommand{\esqm}{e^2_{\text{vac},m}}
\newcommand{\ndTLS}{n_{\text{0}}}
\newcommand{\ndTLSm}{n_{\text{0},m}}
\newcommand{\NTLS}{N_{\text{TLS},m}}
\newcommand{\LambdaTLS}{\Lambda_{\text{TLS}}}
\newcommand{\gts}{\bar{g}_{\text{t},s}}
\newcommand{\gls}{\bar{g}_{\text{l},s}}
\newcommand{\gtm}{\bar{g}_{\text{t},m}}
\newcommand{\glm}{\bar{g}_{\text{l},m}}
\newcommand{\dfmmax}{\delta f_{m,\text{max}}}

\newcommand{\omegaTLS}{\omega_{\text{TLS}}}
\newcommand{\tildeomegaTLS}{\tilde{\omega}_{\text{TLS}}}
\newcommand{\tildeomegaTLSdressed}{\tilde{\omega}^{\prime}_{\text{TLS}}}
\newcommand{\omegas}{\omega_{\mone}}
\newcommand{\tildeomegas}{\tilde{\omega}_{\mone}}
\newcommand{\tildeomegam}{\tilde{\omega}_{m}}
\newcommand{\tildeomegasdressed}{\tilde{\omega}^{\prime}_{\mone}}
\newcommand{\tildeDeltaTLSs}{\tilde{\Delta}_{\text{TLS},s}}

\newcommand{\gammas}{\gamma_{\mone}}
\newcommand{\Gammas}{\Gamma_{\mone}}
\newcommand{\Gammasphi}{\Gamma_{s,\phi}}
\newcommand{\GammaTLStwo}{\Gamma_{2,\text{TLS}}}

\newcommand{\GammaTLSphi}{\Gamma_{\phi,\text{TLS}}}
\newcommand{\GammaTLSone}{\Gamma_{1,\text{TLS}}}

\newcommand{\chirel}{\rttensortwo{\chi}_{\text{rel}}}

\newcommand{\GammathreephI}{\mathcal{F}^{\mthr}_{m\, \mtwo }}

\newcommand{\grun}{\gamma_{\text{G}}}
\newcommand{\vSit}{v_\text{t}}
\newcommand{\vSil}{v_\text{l}}
\newcommand{\vSibar}{v_\text{Si}}
\newcommand{\rhoSi}{\rho_\text{Si}}
\newcommand{\rhobarm}{\bar{\rho}_{\text{m}}}
\newcommand{\vbar}{\bar{v}}
\newcommand{\mone}{s}
\newcommand{\mtwo}{s^{\prime}}
\newcommand{\mthr}{s^{\prime\prime}}
\newcommand{\pone}{p}
\newcommand{\ptwo}{p^{\prime}}
\newcommand{\pthr}{p^{\prime\prime}}
\newcommand{\xone}{x}
\newcommand{\xtwo}{x^{\prime}}
\newcommand{\xthr}{x^{\prime\prime}}

\DeclareMathOperator{\sech}{sech}

\begin{document}

\title{Phononic bandgap nano-acoustic cavity with ultralong phonon lifetime}

%\title{Ultra-long-lived phonons in a microwave-frequency nano-acoustic cavity}
%\title{Ultra-long lifetime phonons in a microwave-frequency nano-acoustic cavity}
%\title{Ultra-long lifetime microwave phonons in a nano-acoustic cavity with full bandgap shielding}
%\title{Phononic bandgap microwave nano-acoustic cavity with ultralong phonon lifetime}

%\title{Phononic bandgap nano-acoustic cavity with ultralong hypersonic phonon lifetime}
%\title{Phononic bandgap nano-acoustic cavity with ultralong microwave-frequency phonon lifetime}
%\title{Phononic bandgap nano-acoustic cavity with ultra-high-$Q$ microwave-frequency phonon modes}
%\title{Ultra-high-$Q$ phononic bandgap nano-acoustic cavity at microwave frequencies}

%\title{Nano-acoustic cavity at microwave frequencies with second-long lifetime}
%\title{Hypersonic nano-acoustic cavity with second-long phonon lifetime}
%\title{A nano-acoustic cavity at microwave frequencies with second-long phonon lifetime}
\author{Gregory S. MacCabe}
\thanks{These authors contributed equally to this work.}
\author{Hengjiang Ren}
\thanks{These authors contributed equally to this work.}
\author{Jie Luo}
\affiliation{Kavli Nanoscience Institute, California Institute of Technology, Pasadena, California 91125, USA}
\affiliation{Institute for Quantum Information and Matter and Thomas J. Watson, Sr., Laboratory of Applied Physics, California Institute of Technology, Pasadena, California 91125, USA}
\author{Justin D. Cohen}
\author{Hengyun Zhou}
\altaffiliation[Current Address: ]{Department of Physics, Harvard University, Cambridge, Massachusetts 02138, USA}
\author{Alp Sipahigil}
\author{Mohammad Mirhosseini}
\affiliation{Kavli Nanoscience Institute, California Institute of Technology, Pasadena, California 91125, USA}
\affiliation{Institute for Quantum Information and Matter and Thomas J. Watson, Sr., Laboratory of Applied Physics, California Institute of Technology, Pasadena, California 91125, USA}
\author{Oskar Painter}
\affiliation{Kavli Nanoscience Institute, California Institute of Technology, Pasadena, California 91125, USA}
\affiliation{Institute for Quantum Information and Matter and Thomas J. Watson, Sr., Laboratory of Applied Physics, California Institute of Technology, Pasadena, California 91125, USA}
\email{opainter@caltech.edu}

\date{\today}

\begin{abstract} %100-150 words, unreferenced

  We present measurements at millikelvin temperatures of the microwave-frequency acoustic properties of a crystalline silicon nanobeam cavity incorporating a phononic bandgap clamping structure for acoustic confinement.  Utilizing pulsed laser light to excite a co-localized  optical mode of the nanobeam cavity, we measure the dynamics of cavity acoustic modes with single-phonon sensitivity.  Energy ringdown measurements for the fundamental $5$~GHz acoustic mode of the cavity shows an exponential increase in phonon lifetime versus number of periods in the phononic bandgap shield, increasing up to $\tauph \approx 1.5$~seconds.  This ultralong lifetime, corresponding to an effective phonon propagation length of several kilometers, is found to be consistent with damping from non-resonant two-level system defects on the surface of the silicon device.  Potential applications of these ultra-coherent nanoscale mechanical resonators range from tests of various collapse models of quantum mechanics to miniature quantum memory elements in hybrid superconducting quantum circuits.
  
\end{abstract}
%\pacs{}

\maketitle

In optics, geometric structuring at the nanoscale has become a powerful method for modifying the electromagnetic properties of a bulk material, leading to metamaterials~\cite{Smith2000} capable of manipulating light in unprecedented ways~\cite{Joannopoulos_book}.  In the most extreme case, photonic bandgaps can emerge in which light is forbidden from propagating, dramatically altering the emission of light from within such materials~\cite{Yablonovitch1987,John1991,Fujita2005}.   More recently, a similar phononics revolution~\cite{Maldovan2013} in the engineering of acoustic waves has led to a variety of new devices, including acoustic cloaks that can shield objects from observation~\cite{SZhang2011}, thermal crystals for controlling the flow of heat~\cite{Narayana2012,Maldovan2013b}, optomechanical crystals that couple photons and phonons via radiation pressure~\cite{Eichenfield2009a}, and phononic topological insulators whose protected edge states can transport acoustic waves with minimal scattering~\cite{Brendel2018,Cha2018}.

Phononic bandgap structures, similar to their electromagnetic counterparts, can be used to modify the emission or scattering of phonons.  These ideas have recently been explored in quantum optomechanics~\cite{Alegre2011,Chan2012,Yu2014,Tsaturyan2017,Ghadimi2018} and electromechanics~\cite{Kalaee2018} experiments to greatly reduce the mechanical coupling to the thermal environment through acoustic radiation.  At ultrasonic frequencies and below, one can combine phononic bandgap clamping with a form of `dissipation dilution' in high stress films~\cite{Unterreithmeier2010} to realize quality ($Q$) factors in excess of $10^8$ in two-dimensional nanomembranes~\cite{Tsaturyan2017} and approaching $10^9$ in one-dimensional strain-engineered nanobeams~\cite{Ghadimi2018}.  At higher, microwave frequencies the benefit of stress-loading of the film fades as local strain energy dominates~\cite{Ghadimi2018} and one is left once again to deal with intrinsic material absorption.

%-- corresponding to the energy storage time in number of acoustic cycles --

To date, far less attention has been paid to the impact of geometry and phononic bandgaps on acoustic material absorption~\cite{Behunin2016,Hauer2018}.  Fundamental limits to sound absorption in solids are known to result from the anharmonicity of the host crystal lattice~\cite{Landau1937,Srivastava_book,Woodruff1961}.  At low temperatures $T$, in the Landau-Rumer regime ($\omega\tauth \gg 1$) where the thermal phonon relaxation rate ($\tauth^{-1}$) is much smaller than the acoustic frequency ($\omega$), a quantum model of three-phonon scattering can be used to describe phonon-phonon mixing that results in damping and thermalization of acoustic modes~\cite{Landau1937,Srivastava_book}.  Landau-Rumer damping scales approximately as $T^\alpha$, where $\alpha\approx4$ depends upon the phonon dispersion and density of states (DOS)~\cite{Srivastava_book}.  At the very lowest lattice temperatures ($\lesssim 10$~K), where Landau-Rumer damping has dropped off, a residual damping emerges due to material defects.  These two-level system (TLS) defects~\cite{Phillips1987}, typically found in amorphous materials, correspond to a pair of nearly degenerate local arrangements of atoms in the solid which can have both an electric and an acoustic transition dipole, and couple to both electric and strain fields.  Recent theoretical analysis shows that TLS interactions with acoustic waves can be dramatically altered in a structured material ~\cite{Behunin2016}.

\begin{figure*}[btp]
\begin{center}
\includegraphics[width=2\columnwidth]{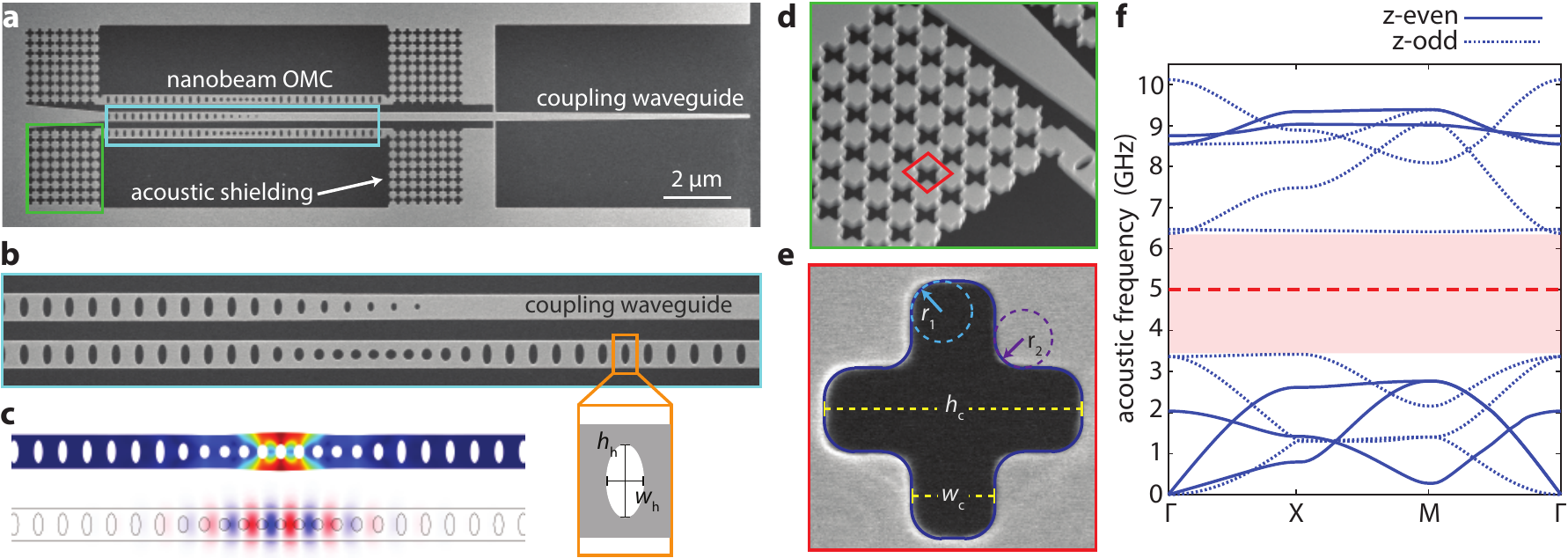}
\caption{\textbf{Nanobeam optomechanical crystal and phononic shield design.} \textbf{a}, Scanning electron microscope (SEM) image of a full nanobeam optomechanical crystal (OMC) device fabricated on SOI with $N=7$ periods of acoustic shielding. A central coupling waveguide allows for fibre-to-chip optical coupling as well as side-coupling to individual nanobeam OMC cavities. \textbf{b},  SEM image of an individual nanobeam OMC and the coupling waveguide, with enlarged illustration of an individual unit cell in the end-mirror portion of the nanobeam. \textbf{c}, FEM simulations of the mechanical (top; total displacement) and optical (bottom; transverse electric field) modes of interest in the nanobeam. Distortion of the mechanical displacement profile is exaggerated for clarity. \textbf{d}, SEM image showing the nanobeam clamping geometry. \textbf{e},  SEM image of an individual unit cell of the cross-crystal acoustic shield. The dashed lines show fitted geometric parameters used in simulation, including cross height ($h_{\text{c}} = 474$~nm), cross width ($w_{\text{c}} = 164$~nm), inner fillet radius ($r_1$), and outer fillet radius ($r_2$). \textbf{f}, Simulated acoustic band structure of the realized cross-crystal shield unit cell, with the full acoustic bandgap highlighted in pink. Solid (dotted) lines correspond to modes of even (odd) symmetry in the direction normal to the plane of the unit cell. The dashed red line indicates the mechanical breathing-mode frequency at $\omegam / 2\pi = 5.0$~GHz.}
\label{fig:fig1}
\end{center}
\end{figure*}

Here we explore the limits of acoustic damping and coherence of a microwave acoustic nanocavity with a phononic crystal shield that possesses a wide bandgap for all polarizations of acoustic waves.  Our nanocavity, formed from an optomechanical crystal (OMC) nanobeam resonator~\cite{Eichenfield2009a,Chan2012}, supports an acoustic breathing mode at $\omegam/2\pi \approx 5$~GHz and a co-localized optical resonant mode at $\omegac/2\pi \approx 195$~THz ($\lambdac \approx 1550$~nm) which allows us to excite and readout mechanical motion using radiation pressure from a pulsed laser source. This minimally invasive pulsed measurement technique avoids a slew of parasitic damping effects $-$ typically associated with electrode materials and mechanical contact~\cite{Galliou2013}, or probe fields for continuous readout $-$ and allows for the sensitive measurement of motion at the single phonon level~\cite{Meenehan2015}.  The results of acoustic ringdown measurements at millikelvin temperatures show that damping due to radiation is effectively suppressed by the phononic shield, with breathing mode quality factors reaching $Q=4.9 \times 10^{10}$, corresponding to an unprecedented frequency-$Q$ product of $f$-$Q = 2.6 \times 10^{20}$.  The temperature and amplitude dependence of the residual acoustic damping is consistent with relaxation damping of non-resonant TLS, modeling of which indicates that not only does the phononic bandgap directly eliminate the acoustic radiation of the breathing mode but it also reduces the phonon damping of TLS in the host material.  
%The anomalously high measured acoustic $Q$-factors are thus likely a result of the suppression of phonon emission by TLS, in analogy to the quantum electrodynamics of an atom in a photonic bandgap~\cite{John1991}.        

%% Cavity design and fabrication

% \begin{itemize}
%     \item \textbf{Cavity design and fab:}
%     \item start with high level description of optomechanical cavity
%     \item discuss acoustic bandgap shield design and issues with realization
% \end{itemize}

The devices studied in this work are fabricated from the $220$~nm device layer of a silicon-on-insulator (SOI) microchip.  Details of the fabrication process are provided in App.~\ref{App:A}. In Figs.~\ref{fig:fig1}(a-b) we show scanning electron microscope images of a single fabricated device, which consists of a coupling optical waveguide, the nanobeam OMC cavities that support both the microwave acoustic and optical resonant modes, and the acoustic shield that connects the cavity to the surrounding chip substrate.  Fig.~\ref{fig:fig1}(c) shows finite-element method (FEM) simulations of the microwave acoustic breathing mode and fundamental optical mode of the nanobeam cavity. We use the on-chip coupling waveguide to direct laser light to the nanobeam OMC cavities.  A pair of cavities with slightly different optical mode frequencies are evanescently coupled to each waveguide.  An integrated photonic crystal back mirror in the waveguide allows for optical measurement in a reflection geometry.  The design of the OMC cavities, detailed in Ref.~\cite{Chan2012}, uses a tapering of the etched hole size and shape in the nanobeam to provide strong localization and overlap of the breathing mode and the fundamental optical mode, resulting in a vacuum optomechanical coupling rate~\cite{Aspelmeyer2014book} between photons and phonons of $\gzero/2\pi \approx 1$~MHz. 

\begin{figure*}[btp]
\begin{center}
\includegraphics[width=2\columnwidth]{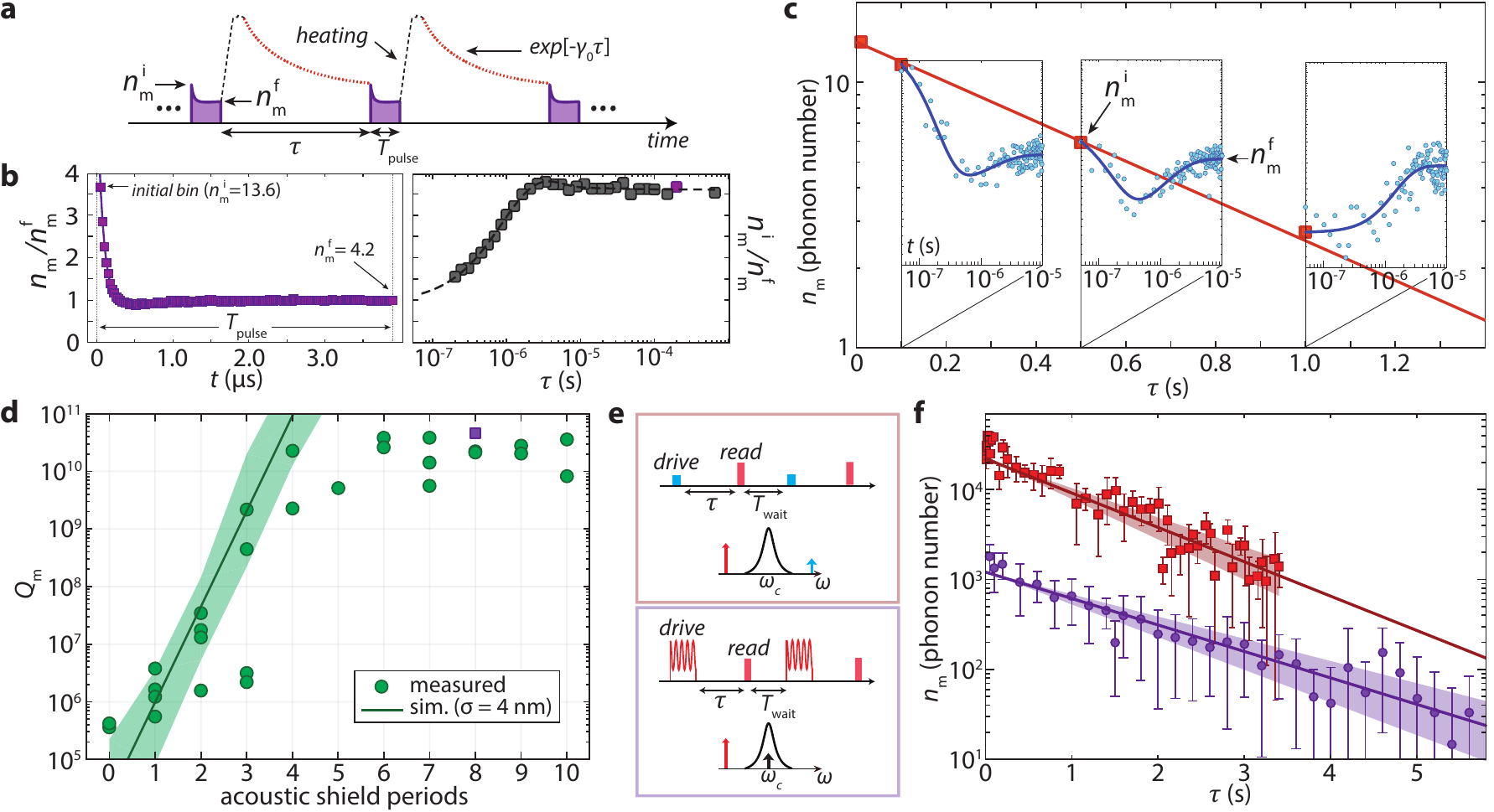}
\caption{\textbf{Ringdown measurements of the acoustic breathing mode. a}, Illustration of the ringdown measurement performed using a red-detuned ($\Delta = +\omegam$) pulsed laser for excitation and readout. \textbf{b}, Normalized phonon occupancy measured during (left) and after (right) the laser readout pulse ($\ncav=569$; optomechanical back-action rate $\gammaOM/2\pi = 1.07$~MHz) for a 6-shield device (device B). Squares are measured data points. Solid and dashed lines are a best fit to the dynamical model of the hot bath (see App.~\ref{App:D}). The displayed pulse-on-state plot (left) corresponds to a delay of $\Toff = 200$~$\mu$s, with $\nmi = 4.2$ and $\nmf=13.6$ phonons. \textbf{c}, Ringdown measurements of a 7-shield device (device C) for readout pulse amplitude of $\ncav = 320$.  The series of inset panels show the measured (and fit; solid blue curve) anti-Stokes signal during the optical pulse at a series of pulse delays. \textbf{d}, Plot of the measured breathing mode $Q$-factor versus number of acoustic shield periods $\Nshield$.  The solid green line is a fit to the corresponding simulated radiation-limited $Q$-factor (see App.~\ref{App:B}) for devices with standard deviation (SD) $\sigma=4$~nm disorder in hole position and size, similar to the value measured from device SEM image analysis.  The shaded green region corresponding to the range of simulated $Q$ values (ensemble size $10$) within one SD of the mean. The square purple data points represents the measured $Q$ in (\textbf{f}). \textbf{e}, Acoustic excitation is performed coherently by using either a blue-detuned pump (upper diagram) to drive the breathing mode into self-oscillation, or using an RF-modulated red-detuned pump~\cite{Safavi-Naeini2011} (lower diagram). See App.~\ref{App:E} for details of the coherent excitation and readout parameters. \textbf{f}, Ringdown measurements performed on an eight-shield device (device D) at large phonon amplitude. For blue-detuned driving (red squares) the fit decay rate is $\gammanotO/2\pi = (0.122 \pm 0.020)$~Hz. For modulated-pump driving (purple circles) the fit decay rate is $\gammanotO/2\pi = (0.108 \pm 0.006)$~Hz.  The error bars are $90\%$ confidence intervals of the measured values of $\nmi$.  The shaded regions are the $90\%$ confidence intervals for the exponential fit curves.}

%During the pulse, back-action cooling occurs at a timescale $\gammaOM^{-1} \approx 100$~ns. The optical-absorption-induced bath simultaneously heats the mode at a rate $\gammap \nbathp$, such that at long $\Tpulse$ a steady-state mode occupancy $\nmf$ is reached. In the pulse-off state (gray squares), the residual phonon bath heats the mode at a rate $\gammap(t) \nbathp(t)$, where the bath damping and effective occupancy are explicitly time-dependent. 

\label{fig:fig2}
\end{center}
\end{figure*}

In order to minimize mechanical clamping losses, the nanobeam is anchored to the Si bulk with a periodic cross structure which is designed to have a complete phononic bandgap at the breathing mode frequency~\cite{Chan2012}. Through tuning of the cross height $h_{\text{c}}$ and width $w_{\text{c}}$ (c.f., Figs.~\ref{fig:fig1}(d-e)), bandgaps as wide as $\sim 3$~GHz can be achieved as shown in Fig.~\ref{fig:fig1}. We analyze SEM images of realized structures to provide accurate structure dimensions for our FEM models, and in particular we include in our modeling a filleting of the inner and outer corners ($r_1$ and $r_2$ in Fig.~\ref{fig:fig1}e) of the crosses arising from technical limitations of the patterning of the structure.  To investigate the efficacy of the acoustic shielding we fabricate and characterize arrays of devices with a scaling of the cross period number from $\Nshield = 0$ to $10$, with all other design parameters held constant.  FEM modeling indicates (see App.~\ref{App:B}) that the addition of the cross shield provides significant protection against nanometer-scale disorder which is inherently introduced during device fabrication.

Optical measurements of the acoustic properties of the OMC cavity are performed at millikelvin temperatures in a dilution refrigerator.  The sample containing an array of different OMC devices is mounted directly on a copper mount attached to the mixing chamber stage of the fridge, and a single lensed optical-fiber is positioned with a 3-axis stage to couple light into and out of each device~\cite{Meenehan2015}. In a first set of measurements of acoustic energy damping, we employ a single pulsed laser scheme to perform both excitation and readout of the breathing mode.  In this scenario, depicted in Fig.~\ref{fig:fig2}(a), the laser frequency ($\omegal$) is tuned to the red motional sideband of the OMC cavity optical resonance, $\Delta \equiv \omegac-\omegal \approx +\omegam$, and is pulsed on for a duration $\Tpulse$ and then off for a variable time $\Toff$.  This produces a periodic train of photon pulses due to anti-Stokes scattering of the probe laser which are on-resonance with the optical cavity.  The anti-Stokes scattered photons are filtered from the probe laser and sent to a single photon detector producing a photon count rate proportional to the number of phonons in the acoustic resonator (see Apps.~\ref{App:C} for details of the measurement set-up and phonon number calibration methods).

We display in Fig.~\ref{fig:fig2}(b) a typical readout signal, showing the normalized phonon occupancy during and immediately after the application of a $4$~$\mu$s pulse.  The initial optomechanical back-action cooling of the acoustic breathing mode is followed by a slower turn-on of heating of the mode during the pulse.  After the pulse, with the back-action cooling turned off, a transient heating of the acoustic mode occurs over several microseconds.  The parasitic heating is attributable to very weak optical absorption of the probe pulse in the Si cavity which produces a hot bath coupled to the breathing mode~\cite{Meenehan2015}.  Here we use the transient heating of the acoustic mode to perform ringdown measurements of the stored phonon number.  A phenomenological model of the dynamics of the induced damping ($\gammap$) and effective occupancy ($\nbathp$) of the hot bath (see App.~\ref{App:D}) allows us to fit the anti-Stokes decay signal.  Plotting the initial mode occupancy at the beginning of the fit readout pulse ($\nmi$) versus delay time $\Toff$ between pulses (c.f., Fig.~\ref{fig:fig2}(a)), we plot the ringdown of the stored phonon number in the the breathing mode as displayed in Fig.~\ref{fig:fig2}(c) for a device with $\Nshield = 7$.  

Performing a series of ringdown measurements over a range of devices with varying $\Nshield$, and fitting an exponential decay curve to each ringdown we produce the $Q$-factor plot in Fig.~\ref{fig:fig2}(d).  We observe an initial trend in $Q$-factor versus shield number which rises on average exponentially with each additional shield period, and then saturates for $\Nshield \ge 5$ to $\Qm \gtrsim 10^{10}$.  As indicated in Fig.~\ref{fig:fig2}(c) these $Q$ values correspond to ringdown of small, near-single-phonon level amplitudes.  We also perform ringdown measurements at high phonon amplitude using two additional methods displayed schematically in Fig.~\ref{fig:fig2}(e) and described in detail in App.~\ref{App:E}.  These methods use two laser tones to selectively excite the acoustic breathing mode using a $\times 1000$ weaker excitation and readout optical pulse amplitude ($\ncav \lesssim 0.3$).  The measured ringdown curves, displayed in Fig.~\ref{fig:fig2}(f), show the decay from initial phonon occupancies of $10^{3}$-$10^4$ of an 8-shield device (device D; square purple data point in Fig.~\ref{fig:fig2}(d)).  The two methods yield similar breathing mode energy decay rates of $\gammanotO/2\pi = 0.108$~Hz and $0.122$~Hz, the smaller of which corresponds to a $Q$-value of $\Qm = 4.92^{+0.39}_{-0.26} \times 10^{10}$ and a phonon lifetime of $\tauph = 1.47^{+0.09}_{-0.08}$~s.  Comparing all three excitation methods with widely varying optical-absorption-heating and phonon amplitude, we consistently measure $\Qm \gtrsim 10^{10}$ for devices with $\Nshield \ge 5$.

%Applying laser light at the red motional sideband produces predominantly anti-Stokes scattering in which a phonon in the acoustic breathing mode is annihilated and a photon is created on resonance with the optical cavity.  This phonon scattering occurs at a rate (in the weak coupling limit) of $\gammaOM = 4\gzero^2\ncav/\kappa$, where $\ncav$ is the intra-cavity photon number due to the probe laser and $\kappa$ is the optical linewidth.  Ideally, the laser light damps and cools the breathing mode at the rate $\gammaOM$.  In practice, weak optical absorption of the probe laser, thought to occur through surface states of the patterned silicon device, results in parasitic heating of the acoustic mode.    

\begin{figure}[btp]
\begin{center}
\includegraphics[width=\columnwidth]{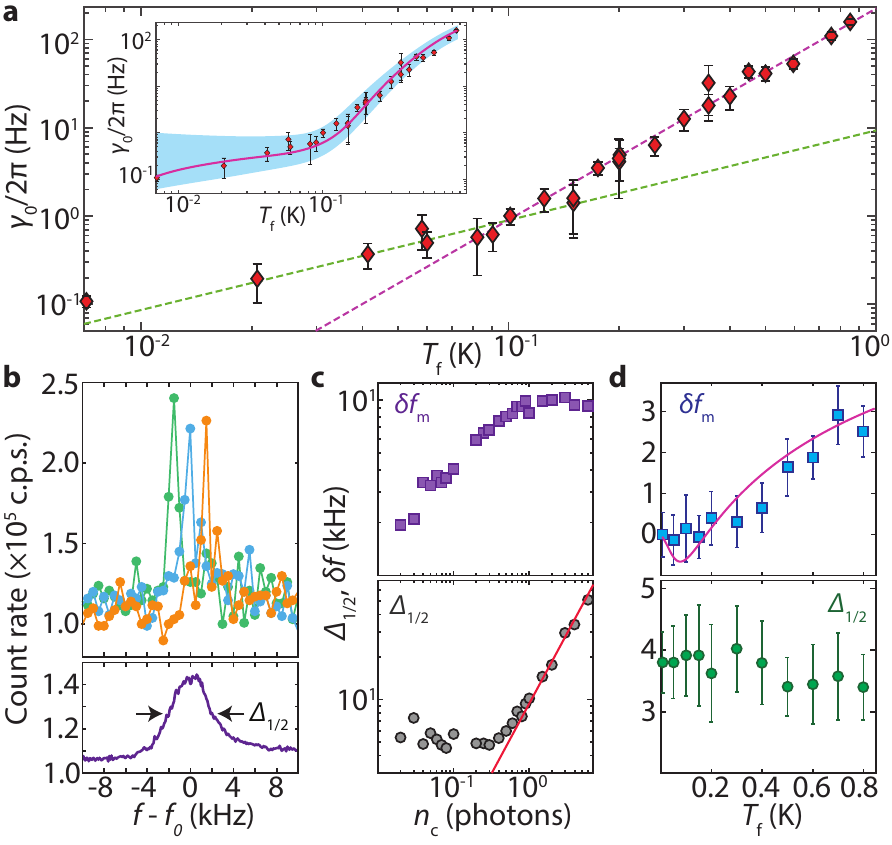}
\caption{\textbf{Temperature dependence of acoustic damping, frequency, and frequency jitter.} \text{a}, Plot of the measured breathing mode energy damping rate, $\gamma_0/2\pi$, as a function of fridge temperature ($\Tf$). Dashed green (magenta) curve is a fit with temperature dependence $\gammanotO \sim \Tf^{1.01}$ ($\gammanotO \sim \Tf^{2.39}$).  Error bars are $90\%$ confidence intervals of the exponential fit to measured ring down curves. Inset: Plot of measured damping data with estimated energy damping from a TLS model (see App.~\ref{App:G}).  The shaded blue region corresponds to the standard deviation of $\log{(\gammanotO/2\pi)}$ for $100$ different random TLS distributions.  \textbf{b}, Two-tone coherent spectroscopy signal. Upper plot: three individual spectrum of rapid frequency sweeps with a frequency step size of $500$~Hz and dwell time of $1$~ms (RBW $\approx 0.5$~kHz). Lower plot: average spectrum of rapid scan spectra taken over minutes, showing broadened acoustic response with FWHM linewidth of $\Deltafj/2\pi = 4.05$~kHz. The large on-resonance response corresponds to an estimated optomechanical cooperativity of $C \equiv \gammaOM/(\gammanotO+\gammap) \gtrsim 1.1$, consistent with the predicted magnitude of back-action damping $\gammaOM/2\pi \approx 817$~Hz and bath-induced damping $\gammap/2\pi \approx 120$~Hz at the measurement pump power level $\ncav = 0.1$. \textbf{c}, Breathing mode resonance frequency shift $\deltaf$ and ensemble average FWHM-linewidth $\Deltafj$ as a function of pump photon number $\ncav$. Solid red curve is fit to back-action limited linewidth, yielding $\gzero/2\pi = 1.15$~MHz. \textbf{d}, Measured $\delta f$ (upper plot) and $\Deltafj$ (lower plot) versus $\Tf$. Error bars are $90\%$ confidence intervals from Voigt fit to measured spectra.  Data presented in (\textbf{a}-\textbf{d}) are for device D.  The solid magenta curves in the inset of (\textbf{a}) and top panel of (\textbf{d}) corresponds to simulations of a single random TLS distribution.} 
\label{fig:fig3}
\end{center}
\end{figure}

% note that gzero for Device D in table is 1.15MHz, whereas originally we had 1.08MHz in Fig.4 caption.  Changed to make consistent.

%% Temperature dependence of damping and frequency shift

% \begin{itemize}
%     \item \textbf{Temperature dependence of damping and frequency shift:}
%     \item discuss measured temperature dependence in 4a
%     \item discuss coherent measurement of mechanical resonance line, and measured frequency jitter in 4b
%     \item discuss power dependence and temperature dependence of frequency shift and time-averaged linewidth (4c,d)
% \end{itemize}

In order to understand the origin of the residual damping for large $\Nshield$ we also measured the temperature dependence of the energy damping rate, breathing mode frequency, and full width at half maximum (FWHM) linewidth of the breathing mode for the highest $Q$ 8-shield device (device D).  In Fig.~\ref{fig:fig3}(a) we plot the energy damping rate which shows an approximately linear rise in temperature up to $\Tf\approx 100$~mK, and then a much faster $\sim (\Tf)^{2.4}$ rise in the damping.  Using the two-tone coherent excitation method~\cite{Safavi-Naeini2011}, we plot in Fig.~\ref{fig:fig3}(b) the measured breathing mode acoustic spectrum at $\Tf=7$~mK and pump power $\ncav=0.1$.  The top plot shows rapid spectral scans ($40$~ms per scan) in which the probe frequency is swept across the acoustic resonance.  These rapid scans show a jittering acoustic line with a roughly $\Deltaf/2\pi \approx 1$~kHz linewidth, consistent with the predicted magnitude of optical back-action ($\gammaOM/2\pi \approx 820$~Hz) and hot bath damping ($\gammap/2\pi \approx 120$~Hz) at the $\ncav=0.1$ measurement power.  An ensemble average of these scans, taken over several minutes, yields a broadened and reduced contrast acoustic line of FWHM $\Deltafj/2\pi = 4.05$~kHz.    

Note that in Fig.~\ref{fig:fig3}(b) we are measuring the acoustic line with the laser light on, as opposed to the ringdown measurements of Fig.~\ref{fig:fig2} in which the laser light is off.  Lowering the optical pump power to reduce back-action and absorption-induced damping limits further the already low signal-to-noise ratio, and scanning more slowly begins to introduce frequency jitter into the measured line.  As such, we can only bound the intrinsic low temperature coherence time of the breathing mode to $\taucoh \gtrsim 2/\Deltaf \approx 0.3$~ms.  Further information can, however, be gleaned by measuring the linewidth and center frequency of the ensemble averaged spectrum as a function of $\ncav$ (Fig.~\ref{fig:fig3}(c)) and $\Tf$ (Fig.~\ref{fig:fig3}(d)).  The width of the frequency jitter spectrum, averaged over minutes, is roughly independent of optical pump power and temperature down to the lowest measurable pump powers ($\ncav=0.02$) and up to $\Tf = 800$~mK.  The center frequency, on the other hand, shifts up in frequency with both temperature and optical power.  The frequency shift versus $\Tf$ is consistent with the frequency shift versus $\ncav$ if the hot bath temperature (see App.~\ref{App:D}) is used as a proxy for the fridge temperature.  

%% Discussion of reasons for ultra-low damping, and connection to non-resonant TLS damping

% \begin{itemize}
%     \item \textbf{Discuss modeling in SI and connection to TLS:}
%     \item discuss impacts of bandgap on TLS damping (res TLS suppression and non-res dominant)
%     \item discuss impact of small number of TLS due to nanoscale cavity volume
%     \item discuss how whole picture of frequency jitter, time-averaged linewidth, and damping level and lack of saturation point to TLS model
% \end{itemize}

Estimates of the magnitude of Landau-Rumer damping of the breathing mode (see App.~\ref{App:F}) indicate that 3-phonon scattering in Si is far too weak at $\Tf \lesssim 1$~K to explain the measured damping.  Analysis of the interactions of TLS with the localized acoustic modes of the confined geometry of the OMC cavity structure, however, show that TLS interactions can explain all of the observed breathing mode behavior.  In this analysis, detailed in App.~\ref{App:G}, FEM simulation is used to find the frequencies and radiation-limited damping rates of the acoustic quasi-normal modes of the OMC cavity structure.  An estimate of the spectral density of TLS within the breathing mode volume ($\Vm \approx 0.11$~$(\mu$m$)^3$) is ascertained from estimated surface oxide ($\sim 0.25$~nm~\cite{Yabumoto1981}) and etch-damage ($\sim 15$~nm~\cite{Oehrlein1989}) layer thicknesses in the Si device, and bulk TLS density found in amorphous materials~\cite{Kleiman1987,Phillips1987}.  Using the resulting effective spectral density of interacting TLS, $\ndTLSm \approx 20$~GHz$^{-1}$, and average TLS transverse and longitudinal deformation potentials of $\bar{M} \approx 0.04$~eV  and $\bar{D} \approx 3.2$~eV, respectively, yields breathing mode damping and frequency shifts which are in excellent agreement with the measured data (see Fig.~\ref{fig:fig3}). The estimated level of frequency jitter is also found in agreement with the measured value, assuming all TLS are being pumped via the same optical absorption that drives the hot bath.

Several key observations can be drawn from the TLS damping modeling.  The first is that the typical $T^{3}$ dependence of TLS relaxation damping of acoustic waves is dependent on the phonon bath DOS into which the TLS decay~\cite{Behunin2016,Hauer2018}.  In the OMC cavity the phonon DOS is strongly modified from a three-dimensional bulk material.  This directly results in the observed weak temperature dependence of the acoustic damping for $\Tf\lesssim 100$~mK, where the thermally acitvated TLS interact resonantly with an approximately one-dimensional phonon DOS.  A second point to note is that the TLS resonant damping is strongly suppressed due to the phononic bandgap surrounding the OMC cavity.  Estimates of the phonon-induced spontaneous decay rate of TLS in the bandgap is on the order of Hz; combined with the discrete number of TLS in the small mode volume of the breathing mode, acoustic energy from the breathing mode cannot escape via resonant coupling to TLS. The observed lack of saturation of the breathing mode energy damping with either temperature or phonon amplitude is further evidence that non-resonant relaxation damping $-$ due to dispersive coupling to TLS $-$ is dominant~\cite{Phillips1987}.  Finally, the small average number of estimated TLS in $\Vm$ which are thermally activated at the lowest temperatures ($\sim 2$), leads to significant variation in the simulated TLS relaxation damping at $\Tf \sim 10$~mK (see shaded blue region of the inset to Fig.~\ref{fig:fig3}(a)).  This is consistent with the observed fluctuations from device-to-device in the low-temperature $\Qm$ for devices with $\Nshield > 5$ (see Fig.~\ref{fig:fig2}(d)).  

Utilizing the advanced methods of nanofabrication and cavity optomechanics has provided a new toolkit to explore quantum acoustodynamics in solid-state materials.  Continued studies of the behavior of TLS in similar engineered nanostructures to the OMC cavity of this work may lead to, among other things, new approaches to modifying the behavior of quasi-particles in superconductors~\cite{Rostem2018}, mitigating decoherence in superconducting~\cite{GaoPhD,Martinis2014} and color center~\cite{Sohn2018,Astner2018} qubits, and even new coherent TLS-based qubit states in strong coupling with an acoustic cavity~\cite{Ramos2013}.  The extremely small motional mass ($\meff = 136$~fg~\cite{Chan2012}) and narrow linewidth of the OMC cavity also make it ideal for precision mass sensing~\cite{Hanay2015} and in exploring limits to alternative quantum collapse models~\cite{Nimmrichter2014}.  Perhaps most intriguing is the possibility of creating a hybrid quantum architecture consisting of acoustic and superconducting quantum circuits~\cite{Devoret2013,Pirkkalainen2013,Gustafsson2014,Chu2017,Manenti2017,Satzinger2018,Moores2018,Arrangoiz_Arriola2018}, where the small scale, reduced cross-talk, and ultralong coherence time of quantum acoustic devices may provide significant improvements in connectivity and performance of current quantum hardware.

%% Summary and Conclusions

% \begin{itemize}
%     \item \textbf{Summary and Conclusions:}
%     \item summarize results and how this points to methods for even further reducing the mechanical dissipation and improving coherence
%     \item discuss opportunities for using these new ultra-coherent acoustic oscillators for quantum metrology
%     \item discuss opportunities for integration with superconducting quantum circuitry
% \end{itemize}

\begin{acknowledgments} 
This work was supported by the ARO Quantum Opto-Mechanics with Atoms and Nanostructured Diamond MURI program (grant N00014-15-1-2761),  the ARO-LPS Cross-Quantum Systems Science \& Technology program (grant W911NF-18-1-0103), the Institute for Quantum Information and Matter, an NSF Physics Frontiers Center (grant PHY-1733907) with support of the Gordon and Betty Moore Foundation, and the Kavli Nanoscience Institute at Caltech.  H.R. gratefully acknowledges support from the National Science Scholarship from A*STAR, Singapore.
\end{acknowledgments}

%\bibliographystyle{ieeetr}
%\bibliography{references_v4}


\clearpage
\onecolumngrid

\appendix

\section{Device Fabrication}
\label{App:A}

The devices were fabricated using a silicon-on-insulator wafer with a silicon (Si) device layer thickness of $220$~nm and buried-oxide layer thickness of 3~$\mu$m. The device geometry was defined by electron-beam lithography followed by inductively coupled plasma reactive ion etching (ICP-RIE) to transfer the pattern through the $220$~nm Si device layer. Photoresist was then used to define a `trench' region of the chip to be etched and cleared for fiber access to device waveguides. In the unprotected trench region of the chip, the buried-oxide layer is etched using a highly anisotropic plasma etch, and the handle Si layer is cleared to a depth of 100 $\mu$m using an isotropic plasma etch. The devices were then undercut using a vapor-HF etch and cleaned in a piranha solution before a final vapor-HF etch to remove the chemically-grown oxide. In fabrication, devices were spatially grouped into arrays in which the number of acoustic radiation shield periods is scaled from zero to ten while all other geometric parameters are held nominally identical.

\section{Modeling of Disorder in the OMC Cavity}
\label{App:B}

%% Modeling of effects of imperfect fabrication

% \begin{itemize}
%     \item \textbf{Fabrication non-idealities:}
%     \item discuss various fabrication imperfections and how shield is impervious to them
%     \item discuss modeling results in Fig. 2
% \end{itemize}

%Imperfections in the realized structure due to randomness and disorder in the fabrication process will impact the energy loss rate of the mechanics, primarily by breaking symmetries of the structure and introducing coupling between the mode of interest and other lossy vibrational modes which radiate into the bulk. 

\begin{figure*}[bp]
\begin{center}
\includegraphics[width=\columnwidth]{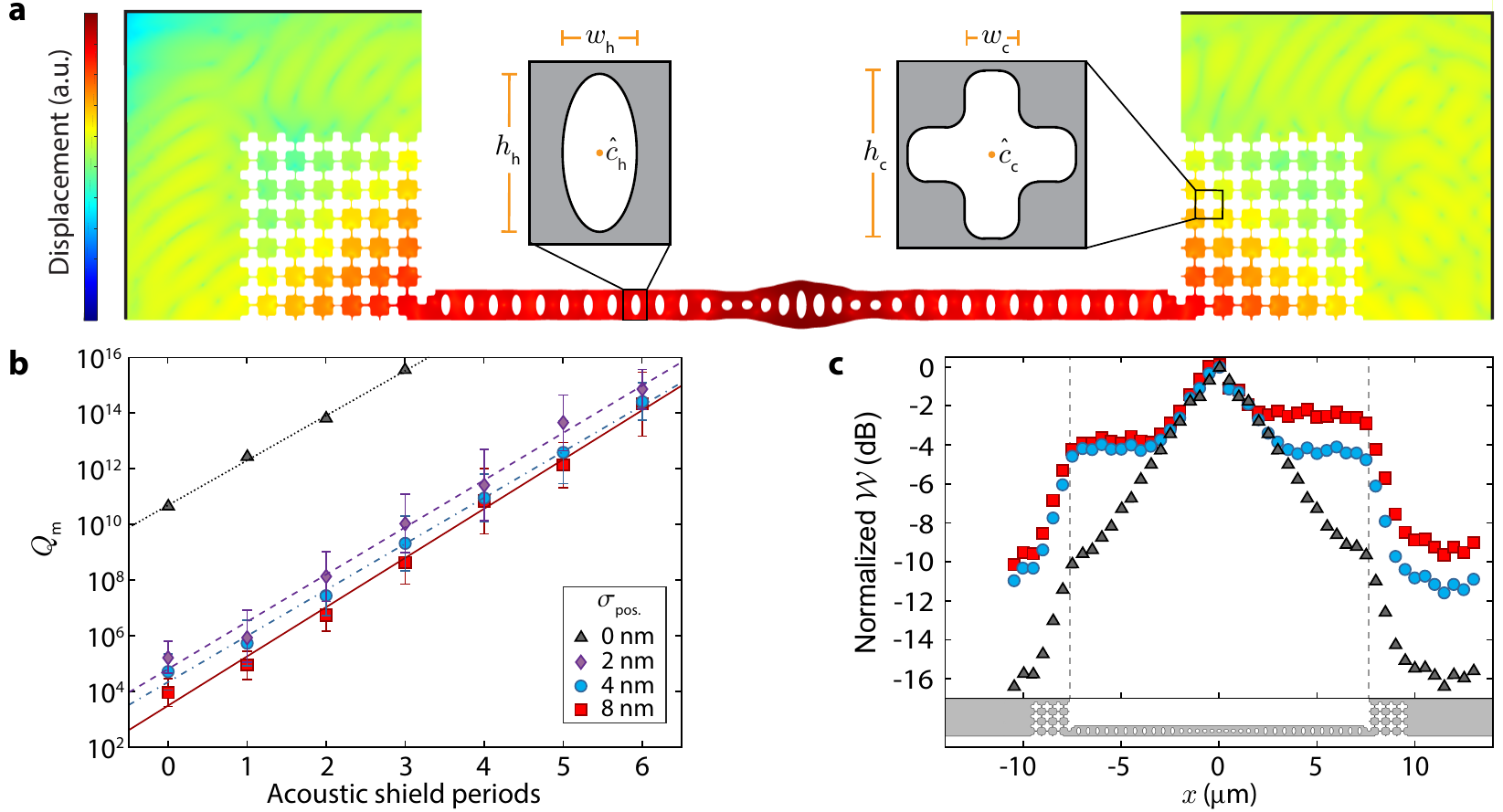}
\caption{\textbf{Impact of fabrication imperfections.} \textbf{a}, FEM simulation of the breathing-mode mechanical displacement field for a nanobeam OMC with $N=6$ periods of acoustic shielding, illustrating localization of the vibrational energy. The geometry in the simulation consists of the nanobeam OMC, acoustic shielding, and the surrounding Si substrate. The borders of the simulation geometry are modeled as an absorbing perfectly-matched layer (PML, outlined with solid black lines). The insets show critical parameters of the device geometry. To introduce disorder into the simulations, each of these geometric parameters is drawn from independent Gaussian distributions centered on the nominal design parameter value with standard deviation $\sigmapos$ for the center positions and $\sigmahole$ for the diameter or length of holes. \textbf{b}, Plot of the simulated mechanical $Q$-factor due to acoustic radiation from the cavity through the acoustic shielding. The straight lines are exponential fits to the mean data points of the simulation (the error bars indicate the standard deviation of ensemble of simulations for each shield number and disorder level). \textbf{c}, Plots of the normalized acoustic energy density $\mathcal{W}$ along a line cut through the center of the beam for $\sigmapos = 0$~nm (black triangles), $4$~nm (blue circles), and $8$~nm (red squares).}
\label{fig:disorder_sims}
\end{center}
\end{figure*}

The use of a phononic bandgap shield is necessitated by the lack of a full gap for the nanobeam cavity.  In Fig.~\ref{fig:disorder_sims} we present a numerical study of the effects of random fabrication imperfections on the radiative damping of the shielded OMC cavity mode.  We compare in Fig.~\ref{fig:disorder_sims}(b) the simulated acoustic $Q$-factor of the ideal, unperturbed cavity structure to that of cavity structures with a fixed level of disorder in the hole sizes (standard deviation, $\sigmahole = 4$~nm) and varying levels of disorder in the hole centers ($\sigmapos = 2,4,8$~nm).  An absence of perturbations to the nanobeam cavity, even without any shielding, yields large radiation-limited $Q$-factors in excess of $10^{10}$.  This is a result of the quasi-bandgap that exists in the nanobeam mirror section for modes of a specific symmetry about the center-line of the beam; however, any perturbation that breaks this symmetry results in a compromised quasi-bandgap in the nanobeam ($Q$ drops from $\gtrsim 10^{10}$ to $\lesssim 10^{5}$ for nanometer-scale perturbations).  Conversely, the exponential trend of the radiation-limited $Q$-factor with the number of shield periods is consistently a factor of $\times 5.5$ per additional period, independent of the disorder level.  This interpretation is further bolstered by the plots in Fig.~\ref{fig:disorder_sims}(c) comparing the linear acoustic energy density along the axis of the nanobeam, $\mathcal{W}$, for the mode of the unperturbed cavity and the modes of two different realizations of disordered cavities.

\section{Measurement Setup, Optical Characterization, and Optomechanical Calibration}
\label{App:C}

\subsection{Measurement Setup}

\begin{figure*}[btp]
\begin{center}
\includegraphics[width=0.75\columnwidth]{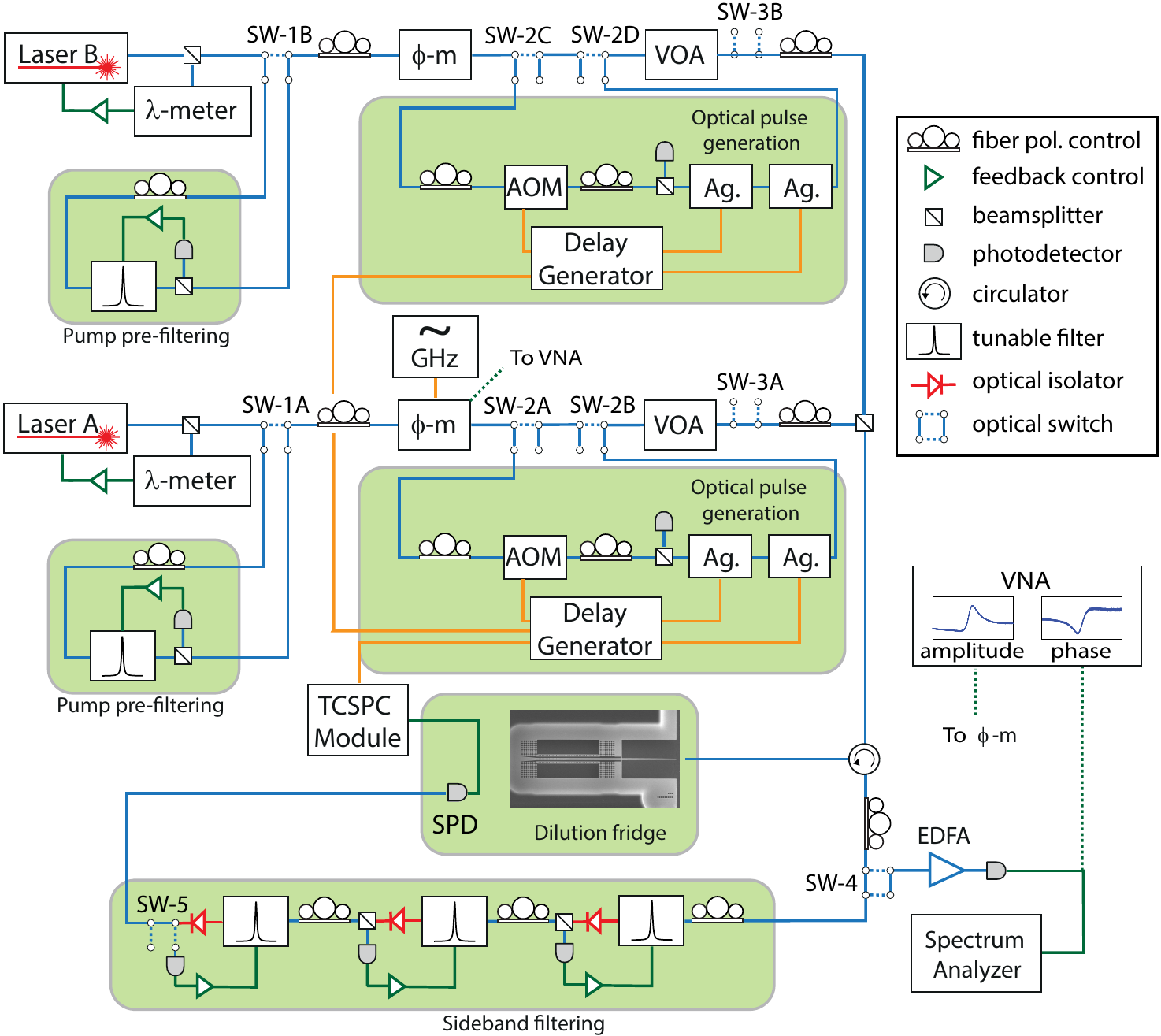}
\caption{\textbf{Pulsed-excitation phonon counting measurement setup.} Simplified diagram of the experimental setup used for low-temperature optomechanical device characterization and phonon-counting measurements. Lasers A and B are passed through $50$~MHz-bandwidth filters to suppress broadband spontaneous emission noise. Both lasers are equipped with modulation components (AOM, Ag.) for generating high-extinction optical pulses. The modulation components are triggered by a digital delay generator (Laser B components are triggered by the `master' Laser A generator). Upon reflection from the device under test, a circulator routes the outgoing light to either (1) an EDFA and spectrum analyzer, or (2) a sideband-filtering bank consisting of three cascaded fiber Fabry-Perot filters (Micron Optics FFP-TF2) and the SPD operated at $\sim 760$~mK. $\lambda$-meter: wavemeter, $\phi$-m: electro-optic phase modulator, EOM: electro-optic intensity modulator, AOM: acousto-optic modulator, Ag.: Agiltron 1x1 MEMS switch, SW: optical $2\times2$ switch, VOA: variable optical attenuator, EDFA: erbium-doped fiber amplifier, VNA: vector network analyzer, SPD: single photon detector, TCSPC: time-correlated single photon counting module (PicoQuant PicoHarp 300).} 
\label{SI_setupFig}
\end{center}
\end{figure*}

The full measurement setup used for device characterization is shown in Fig.~\ref{SI_setupFig}. The light source is a fiber-coupled tunable external-cavity diode laser, of which a small portion is sent to a wavemeter ($\lambda$-meter) for frequency stabilization. The light is then sent to high-finesse tunable fiber Fabry-Perot filter (Micron Optics FFP-TF2, bandwidth $50$~MHz, FSR $20$~GHz) to reject laser phase noise at the mechanical frequency, which can contribute to noise-photon counts on the SPDs. After this prefiltering, the light is routed to an electro-optic phase modulator ($\phi$-mod) which is driven by an RF signal generator at the mechanical frequency to generate optical sidebands used for locking the detection-path filters. The light is then directed via 2$\times$2 mechanical optical switches into a "high-extinction" path consisting of a series of modulator components which are driven by a digital pulse generator to generate high-extinction-ratio optical pulses. The digital pulse generator is used to synchronize the switching of the modulation components as well as to trigger the time-correlated single-photon-counting (TCSPC) module. Of these modulation components, two are electro-optic intensity modulators which together provide $\sim$60 dB of fast extinction ($\sim$20 ns rise and fall times), and two are Agiltron NS 1$\times$1 switches (rise time 100 ns, fall time $\sim$30 $\mu$s) which provide a total of 36 dB of additional extinction. The total optical extinction used to generate our optical pulses is approximately 96 dB, which is greater than the cross-talk specification of our mechanical optical switches. For this reason we use two 2$\times$2 switches in parallel to isolate the high-extinction path to ensure that our off-state optical power is limited by our high-extinction modulation components rather than by cross-talk through the mechanical switches. The light is then passed through a variable optical attenuator (VOA) to control the input pulse on-state power level to the cavity, and sent to a circulator which directs the light to a lensed-fiber tip for end-fire coupling to devices inside a dilution refrigerator. The reflected signal is then routed back to either one of two detection setups. The first includes an erbium-doped fiber amplifier (EDFA) and a high-speed photodetector (PD) connected to a spectrum analyzer (SA) and a vector network analyzer (VNA). The second detection path is used for the phonon counting measurements. Here the light passes through three cascaded high-finesse tunable fiber Fabry-Perot filters (Micron Optics FFP-TF2) inside an insulating housing and then to the SPD inside the dilution refrigerator.

The cascaded fiber Fabry-Perot (FP) filters are aligned to the optical cavity resonance frequency $\omegac$ during measurement such that the signal reaching the SPDs consists of sideband-scattered photons and a small contribution of laser-frequency pump-bleed-through. In total the filters suppress the pump by $>$100 dB. This bleed-through is calibrated by positioning the laser far off-resonance of the optical cavity, such that the device acts simply as a mirror, while fixing the relative detuning of the filters and the pump laser at the mechanical frequency $\omegam/2\pi$ and measuring the photon count rate on the SPDs as a function of laser power. 

Additionally, both the FP-filters and the EOMs will drift during measurement and must be periodically re-locked. We therefore regularly stop the measurement and perform a re-locking routine. First, we re-lock the EOMs by applying a sinusoidal dithering signal of $\sim$1 V to them while monitoring the optical transmission, then decrease the dithering amplitude gradually to lock to the minimum of transmission. Next we switch out of the high-extinction pulse path (SW-2A,2B) and out of the SPD path (SW-5), drive the phase modulator with a large RF power at $\omegam/2\pi$ to generate large optical sidebands at the cavity resonance frequency, and send this light into the FP-filter stack. The transmission through each filter is monitored while a dithering sinusoidal voltage is applied to each filter successively, and the amplitude and DC offset of the dithering signal are adjusted until the optical transmission signal at the desired sideband is maximized. The offset voltage is then held fixed during the subsequent measurement run. The filters will drift due to both thermal fluctuations and acoustic disturbances in their environment, so in order to further improve the filters' stability we have placed them inside a custom-built insulated housing.

The SPDs used in this work are amorphous WSi-based superconducting nanowire single-photon detectors developed in collaboration between the Jet Propulsion Laboratory and NIST. The SPDs are mounted on the still stage of the dilution refrigerator at $\sim700$~mK. Single-mode optical fibers are passed into the refrigerator through vacuum feedthroughs and coupled to the SPDs via a fiber sleeve attached to each SPD mount. The radio-frequency output of each SPD is amplified by a cold-amplifier mounted on the 50 K stage of the refrigerator as well as a room-temperature amplifier, then read out by a triggered PicoQuant PicoHarp 300 time-correlated single photon counting module. We have observed SPD dark count rates as low as $\sim 0.6$~c.p.s. and SPD quantum efficiency $\etaSPD \simeq 60\%$.

\subsection{Optical Characterization}

Each device we have measured in this work was characterized optically in order to determine its optical resonance frequency $\omegac$, total optical linewidth $\kappa$, waveguide-cavity coupling rate $\kappae$, waveguide-cavity coupling efficiency $\eta_{\kappa} = \kappae / \kappa$, and fiber-to waveguide coupling efficiency $\eta_{\text{cpl}}$. In particular, the waveguide-cavity coupling efficiency $\eta_{\kappa}$ is measured by placing the laser far off-resonance and using the VNA to drive an intensity modulator to sweep an optical sideband through the cavity frequency and measure the optical response on a high-speed PD (after amplification by the EDFA) connected to the VNA signal port. From this we obtain the amplitude and phase response of the cavity, which are fitted to determined $\eta_{\kappa}$. \par

\subsection{Vacuum Optomechanical Coupling Rate and Mode Occupancy Calibration}

\begin{figure}[btp]
	\begin{center}
		\includegraphics[width=0.4\columnwidth]{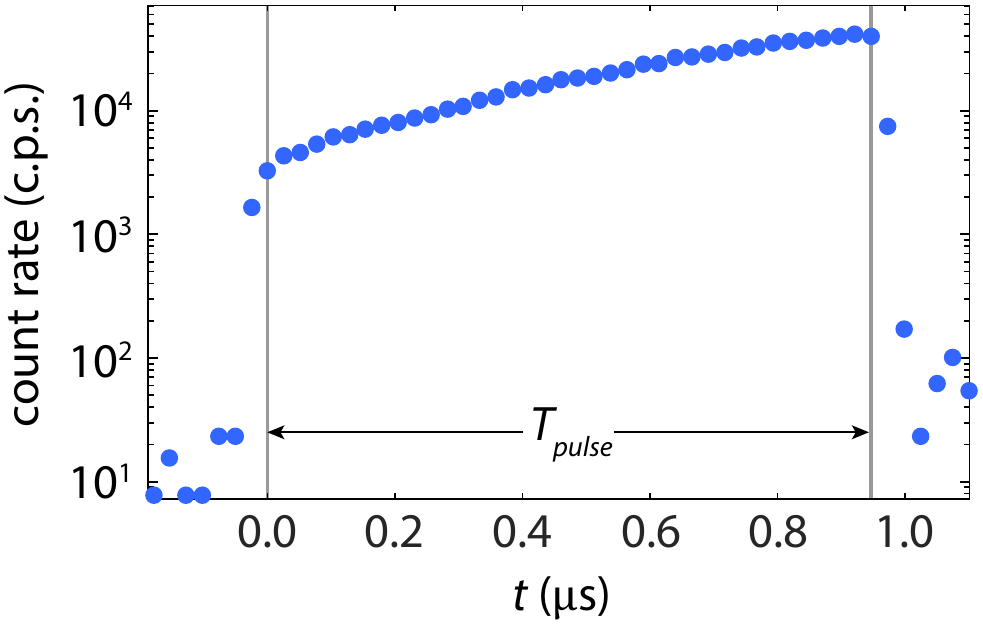}
		\caption{\textbf{Blue-detuned calibration of sideband photon scattering rate.} \textbf{a}, Plot of photon count rate during optical pulse for $\Delta = -\omegam$, $\ncav = 101$ and $T_{\text{per}} = 0.2$~ms. Measurements performed on device A with parameters ($\kappa$, $\kappae$, $\gzero$, $\omegam$, $\gammanotO$) $=$ $2\pi$($1.507$~GHz, $778$~MHz, $713$~kHz, $5.053$~GHz, $14.1$~kHz), and zero acoustic shielding periods.} 
		\label{fig:SI_blueCalib_fig}
	\end{center}
\end{figure}

The measurements presented in this work rely on calibration of the vacuum optomechanical coupling rate $\gzero$. After fitting the total optical linewidth $\kappa$ from an optical reflection spectrum and calibrating $\eta_{\kappa}$ as described above, the intracavity photon number $\ncav$ for a specified detuning is known and is proportional to the optical power input to the cavity:

\begin{equation}
\ncav = \langle \adag \ahat \rangle = \frac{P_\text{in}}{\hbar \omegal}\frac{\kappae}{\Delta^2 + (\kappa/2)^2},
\label{eqn:nc_expression}
\end{equation}

\noindent where $\omegal$ is the applied laser frequency.  In these phonon-counting measurements, a critical calibration parameter is the photon scattering rate per phonon in the mechanical mode. In particular, from Ref.~\cite{Cohen2015,Meenehan2014,Meenehan2015} the photon count rate at the SPD for a red- or blue-detuned pump is:

\begin{align}
\Gamma(\Delta = \pm \omegam) = \Gamma_{\text{dark}} + \Gamma_{\text{pump}} + \Gamma_{\text{SB,0}}(\nbar + \frac{1}{2}(1 \mp 1)),
\end{align}

\noindent where $\nbar$ is the average phonon occupancy of the breathing mode, $\gammaSB = \eta_{\text{det}} \eta_{\text{cpl}} \eta_{\kappa} \gammaOM$ is the \textit{detected} photon scattering rate per phonon (including experimental set-up efficiencies) and $\gammaOM = 4 \gzero^2 \ncav / \kappa$ is the optomechanical damping rate. In the absence of mechanical occupancy, pump photons may be spontaneously scattered by the mechanics owing to the nonzero mechanical susceptibility at the pump frequency. These real photons scattered by the mechanical vacuum noise in the presence of a pump laser are detected at a rate $\gammaSB$, providing a calibration of the per-phonon count rate directly to the vacuum noise. Here $\eta_{\text{det}}$ is the measured overall detection efficiency of the setup, including losses in the fiber runs and circulator, insertion losses in the filters, the fiber run inside the dilution refrigerator, and the detection efficiency of the SPD. To calibrate $\gammaSB$ (for a fixed intracavity photon number $\ncav$), a blue-detuned pump ($\Delta = - \omegam$) pulsed with a repetition time of $\Tper$ drives the mechanics. In the initial time bin during the pulse, $\nbar \ll 1$ (if $1/\Tper \ll \gammanotO$) and we can approximate the sideband photon count rate $\Gamma \approx \gammaSB$. Including detection non-idealities such as bleed-through of the pump laser to the SPD and dark counts on the SPD, the detected count rate is $\Gamma(\Tpulse  = 0) = \Gamma_\text{DCR} + \Gamma_\text{pump} + \gammaSB$. This measurement provides an absolute calibration of the detection photon count rate to the mechanical vacuum noise, where the count rate is proportional to intracavity photon number, allowing calibrated thermometry with a precision that is independent of knowledge of the losses in the optical path. Additional knowledge of the optical path losses enables the inference of $\gammaOM$ from a measurement of $\gammaSB$, which can be used to calibrate the vacuum optomechanical coupling rate. For one representative device as shown in Figure~\ref{fig:SI_blueCalib_fig} we measure $\Gamma_{\text{SB,0}} = 3.263 \times 10^3$~c.p.s. using a measurement photon number of $\ncav = 101$, and with $\kappa$ known from independent measurements we extract $\gzero/2\pi = 713$~kHz, consistent with previous measurements on similar devices with similar orientations relative to the Si crystal axes~\cite{Meenehan2015}.  \par

An important figure of merit for phonon-counting measurements is the sensitivity, expressed as a noise-equivalent phonon number $\nNEP$. This $\nNEP$ represents the equivalent occupancy inferred from noise counts only:

\begin{align}
\nNEP = \frac{\Gamma_{\text{dark}} + \Gamma_{\text{pump}}}{\Gamma_{\text{SB,0}}}.
\end{align}

\begin{figure*}[btp]
\begin{center}
\includegraphics[width=0.5\columnwidth]{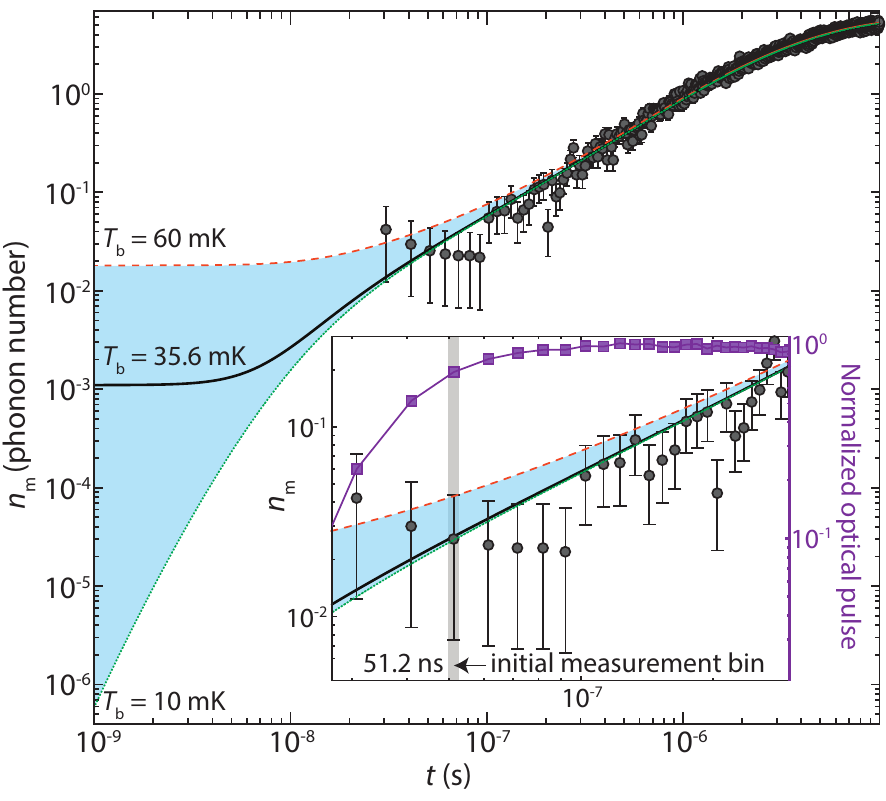}
\caption{\textbf{Base occupancy measurement and pulse turn-on dynamics.} Mode occupancy during the pulse on-state of the zero-shield device (device A) presented in Fig.~\ref{fig:SI_bath_properties}. The photon number $\ncav=10$ is chosen to be small to minimize parasitic heating during the initial time bins of the pulse (bin size is $10.24$~ns.). The model best-fit corresponds to $\Tbath = 35.6$~mK. Bounding curves to the fit are shown for $\Tbath = 60$~mK (orange dashed line) and $\Tbath = 10$~mK (green dotted line). Inset: Overlay plot of the initial time bins of the mode occupancy curve and the input optical pulse (purple squares). Time bins earlier than $51.2$~ns occur during the fast rise of the pulse, which occurs at a timescale set by the rise of the EOMs and optical switches. The first measurement bin is chosen at $\ton=51.2$~ns, where the input optical pulse has reached $>70\%$ of its nominal on-state value (here $\ncav=10$). For $10.24$~ns binning as shown here, the initial measurement bin is bin $\#5$. For $25.6$~ns binning as shown in the Main Text figures, the initial bin is bin $\#2$.} 
\label{SI_minoccFig}
\end{center}
\end{figure*}

The acoustic breathing mode thermalizes to a mode temperature $\Tm$ which is related to the applied fridge temperature $\Tf$ through the thermal conductance $G_{\text{th}}$ of the structure. In order to measure the minimum temperature $\Tbathmin$ to which the mode thermalizes at the lowest fridge temperature $\Tf = 10$~mK, we use a low-power ($\ncav=10$) red-detuned pulsed probe and a device having relatively low $Q_m = 3.57 \times 10^5$ (chosen so that the mode thermalizes to its base temperature rapidly between each incident optical pulse). The initial mode occupancy during the pulse then approximately corresponds to the `off-state' occupancy $\nmnot$. However, as the optical probe turns on during the first several time bins of the pulse, the mode is heated such that the initial observed occupancy exceeds $\nmnot$. We therefore extract $\nmnot$ from a fit to the pulsed heating model which is extrapolated back to $\ton = 0$.

Figure~\ref{SI_minoccFig} shows a fit to the pulse occupancy in the on-state, which yields $\Tm \approx 35.6$~mK. The lower bound of $10$~mK is set by the minimum applied fridge temperature, and the upper bound $60$~mK corresponds to the directly observed occupancy value in the initial measurement bin.

\subsection{Summary of Device Parameters}
\label{subsec:devparam}

Detailed measurements of several devices are presented in the Main Text and Appendices.  For reference, here we provide a look-up table for each of these devices and their measured optical and mechanical properties.

\begin{table}[ht!]
\begin{center}
\caption{\textbf{Measured optical and acoustic device parameters.} $\rho_{r}$ is the Si device layer resistivity of the SOI wafer from which the device was fabricated (as provided by the manufacturer), $\lambdac$ is the optical mode wavelength, $\kappa$ is the measured total optical linewidth, $\kappae$ is the measured coupling rate between the OMC cavity mode and the on-chip waveguide, $\gzero$ is the measured vacuum optomechanical rate, $\omegam$ is the measured breathing mode frequency, $\gammanotO/2\pi$ is measured breathing mode damping from ringdown measurements at $\Tf=7$~mK. \label{tab:devices}}
%\begin{ruledtabular}
\begin{tabular}{|l|l|l|l|l|l|l|l|l|}
%\toprule
\hline\hline
 \textbf{Device} & shield \# & $\rho_{r}$ [$\Omega$-cm] & $\lambdac$ [nm] & $\kappa/2\pi$ [GHz] & $\kappae/2\pi$ [MHz] & $\gzero/2\pi$ [MHz] & $\omegam/2\pi$ [GHz] & $\gammanotO/2\pi$ (Hz) \\
\hline
  A & 0 & $5$-$20$ & $1541.850$ &  $1.507$ & $778$ & $0.713$ & $5.053$ & $14.1\times 10^3$ \\ 
  B & 6 & $5$-$20$ & $1539.285$ & $1.13$ & $605$ & $\sim 0.713$ & $5.013$ & $0.21$ \\ 
  C & 7 & $5$-$20$ & $1538.716$ & $1.21$ & $362$ & $\sim 0.713$ & $5.014$ & $0.27$ \\
  D & 8 & $> 5 \times 10^3$ & $1538.971$ & $0.575$ & $131$ & $1.15$ & $5.31$ & $0.108$ \\
  E & 7 & $5$-$20$ & $\sim 1540$ & $1.244$ & $261$ & $0.833$ & $4.98$ & $0.33$ \\
 \hline\hline 
%\bottomrule
\end{tabular}
%\end{ruledtabular}
\end{center}
\end{table}

% note that I had changed damping of device C to gamma_0/2\pi = 0.13 (put back to original value of 0,27)

\section{Optical-Absorption-Induced Bath}
\label{App:D}

\subsection{Theoretical Model of the Bath}
\label{subsec:properties_of_heating_bath}

Optical absorption is found to induce additional parasitic heating and damping of the high-$Q$ acoustic breathing mode of the Si OMC devices at millikelvin temperatures.  This absorption heating is thought to proceed through excitation of sub-bandgap electronic defect states at the Si surfaces which undergo phonon-assisted decay, generating a local bath of thermal phonons coupled to the high-$Q$ breathing mode~\cite{Meenehan2014}. We may gain some understanding of the optically-induced bath by considering a simple model of phonon-phonon interactions which can couple the optically-induced hot phonon bath to the breathing mode. As we are concerned in this work with the phonon dynamics at low bath temperature ($\Tbath \lesssim 10$~K), and the acoustic mode of interest is at microwave frequencies, the phonon-phonon interactions leading to heating and damping of the breathing mode can be understood in terms of a Landau-Rumer scattering process~\cite{Srivastava_book,Zyryanov1966} (see App.~\ref{App:F}). In this context, we may consider a simple model in which our mode of interest at frequency $\omegam$ is coupled to higher-frequency bath phonon modes at frequencies $\omega_1$ and $\omega_2$, with $\omega_2 - \omega_1 = \omega_m$. Then we may write the scattering rates into and out of the mode of interest to first order in perturbation theory~\cite{Srivastava_book,Meenehan2014} as $\Gamma_{+} = A (n_m+1)(n_2 + 1)n_1$ and $\Gamma_{-} = A n_m n_2 (n_1+1)$, respectively, where $n_1$, $n_2$, and $n_m$ are the number of phonons in each mode involved in the scattering and $A$ is a constant describing the $\text{Si}$ lattice anharmonicity. Then the overall rate of change in the occupancy of the mode of interest is, 

\begin{equation}
\label{eqn:simplemodel_np_gammap}
\dot{n}_m = \Gamma_{+} - \Gamma_{-} = - A(n_1 - n_2)n_m + A n_2 (n_1 + 1).
\end{equation}

\noindent This expression has exactly the form of a harmonic oscillator coupled to a thermal bath with rate $\gammap = A(n_1 - n_2)$ and effective occupancy $\nbathp = A n_2 (n_1 + 1)/\gammap$. Assuming thermal occupancies for each of the higher frequency phonon modes of the hot bath, $n_{1,2} = \nBose[\hbar\omega_{1,2}/\kB\Tbathp] \equiv 1/(\exp{[\hbar\omega_{1,2}/\kB T]}-1)$, and using the identity $\nBose[x+x^{\prime}](\nBose[x]+1) = (\nBose[x] - \nBose[x+x^{\prime}])\nBose[x^{\prime}]$~\cite{Srivastava_book}, one finds that the mode $m$ thermalizes with the hot bath via 3-phonon scattering to an effective occupancy which is $\nbathp = \nBose[\hbar\omegam/\kB\Tbathp]$.  This result holds when the hot bath thermalizes to some temperature independent of the interactions with mode $m$.  

\begin{figure}[btp!]
\centering
\includegraphics[width=0.7\textwidth]{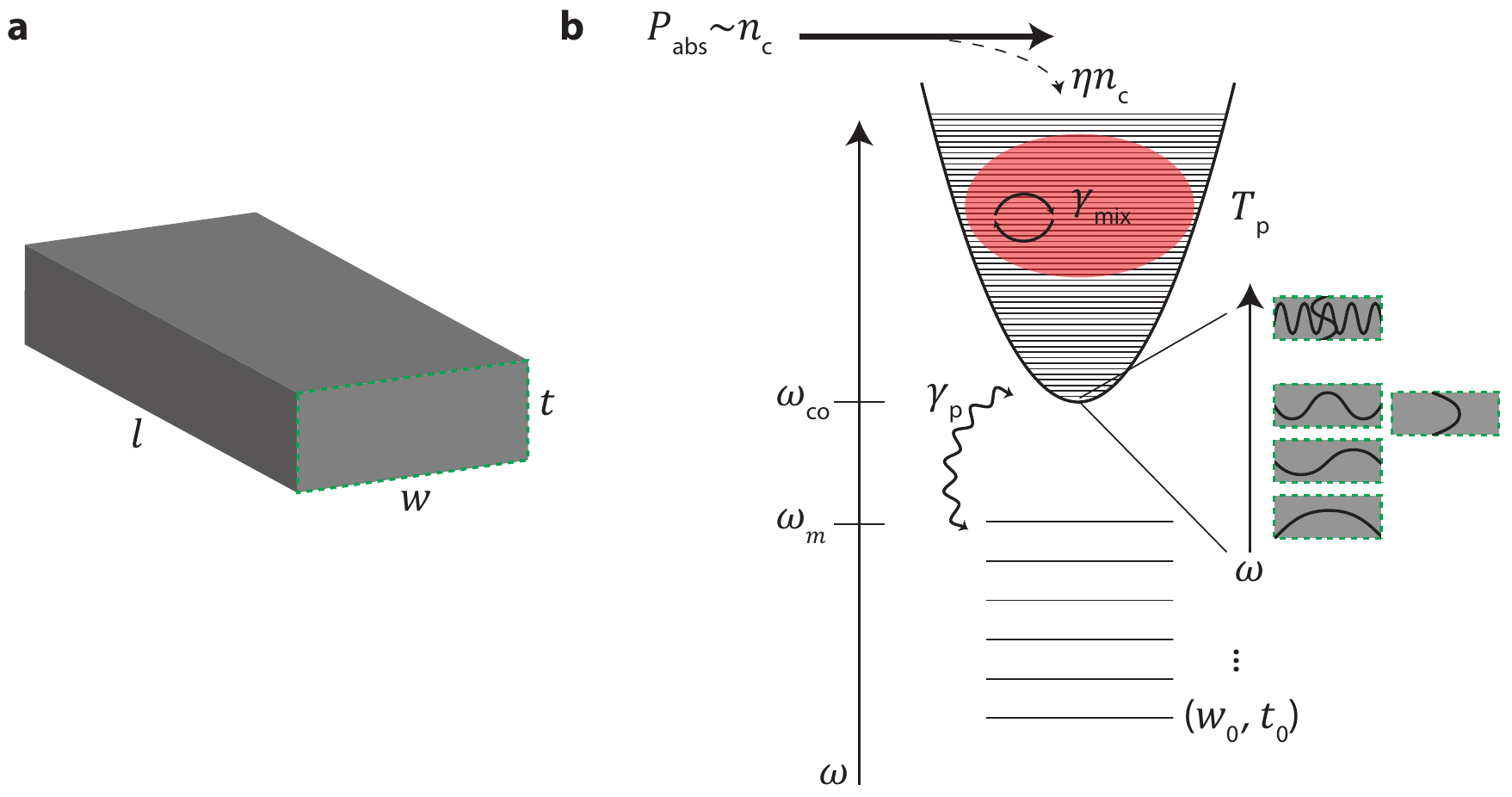}
\caption[Impact of the phonon bottleneck on the optical-absorption bath]{\textbf{Impact of the phonon-bottleneck on the optical-absorption bath. a}, Cross-sectional dimensions of the thin-film nanobeam. \textbf{b}, Absorption of sub-$\text{Si}$-bandgap photons gives rise to phonon-assisted decay of THz phonons into a local bath of GHz phonons in the nanobeam. This bath is expected to experience a bottleneck at a cutoff frequency corresponding to the cross-sectional dimensions of the nanobeam, such that a high-frequency phonon bath accumulates and thermalizes among itself to a local temperature $\Tbath$ at rate $\gamma_\text{mix}$. In the vicinity of the bottleneck frequency the relevant normal modes of the beam are those shown in the inset (black lines are schematics of the local strain in the beam). The lowest-lying discrete mode $(w_0,t_0)$ is a fundamental bowstring mode of the nanobeam at $\sim20$~MHz.}
\label{fig:phonon_bottleneck}
\end{figure}

In the real material system of the nanobeam, the local hot phonon bath at elevated temperature $\Tbath$ is expected to be generated as electronic states at $\sim$ eV energy undergo phonon-assisted relaxation processes, emitting a shower of high-frequency phonons which subsequently decay by a cascade of nonlinear multi-phonon interactions into a bath of GHz phonons. Due to the geometric aspect ratio of the thin-film nanobeam, the local density of phonon states becomes restricted at lower frequency, decreasing the rates of phonon-phonon scattering at low frequency relative to those of a bulk crystal with a 3D Debye density of states. The beam thickness ($t = 220$~nm, width $w \approx 560$~nm, length $l \approx 15$~$\mu$m) corresponds to a relatively high cutoff frequency in the vicinity of $\omegacutoff/2\pi \approx v_l / (2 t) \approx 20$~GHz, where $v_l = 8.433$~km/s is the longitudinal-phonon velocity in $\text{Si}$. This cutoff frequency imposes an effective phonon \textit{bottleneck} preventing further rapid thermalization to lower-lying modes and a resulting buildup in the bath phonon population above the bottleneck. For phonon frequencies below the cutoff, where the wavelength is large enough to approach the lattice constant of the acoustic bandgap clamping region, the reflectivity of the clamping region increases as ballistic radiation out of the nanobeam is suppressed. The result is a reduced density of phonon states near and below the cutoff, where the nanobeam supports quasi-discrete (and long-lived, especially in the vicinity of the mirror bandgap and acoustic shield bandgap) phonon modes at lower frequency as outlined in Fig.~\ref{fig:phonon_bottleneck}. The phenomenological coupling rate $\gammap$ describes the rate at which the lower-lying modes---in particular the breathing mode at $5$~GHz---are coupled to the elevated-temperature bath of higher-frequency phonons above the bottleneck.

In the context of this proposed phonon-bottleneck model, we now consider instead of a discrete pair of modes $n_1$ and $n_2$ a quasi-continuum of high-frequency bath modes coupled to the mode of interest via some anharmonicity matrix element $A(\omega;\omega_m)$. We will assume that the thermal phonons populating the bath have sufficient time to thermalize amongst each other before decaying, or in other words, that they couple to each other at a mixing rate $\gamma_\text{mix}$ much greater than their coupling rates to the external environment or to the lower-lying phonon modes. Under this assumption, we may define an effective local temperature $\Tbathp$ such that the occupancy of a bath phonon at frequency $\omega$ is given by the Bose-Einstein occupation factor

\begin{equation}
\nbathgen[\omega;\Tbathp] \equiv \nBose[\hbar(\omega-\omegacutoff)/\kB \Tbathp] =  \frac{1}{e^{\hbar (\omega - \omegacutoff) / \kB \Tbathp} - 1},
\end{equation}

\noindent where $\omegacutoff$ represents the new effective ground-state frequency due to the phonon bottleneck effect, and $\nBose[x]=1/(\exp[x] -1)$ is the Bose distribution.

The temperature of the optically-induced hot phonon bath, $\Tbathp$, can then be related to the absorbed optical power $P_\text{abs}$ using a model of the lattice thermal conductivity in the low temperature limit. Assuming the optical absorption process is linear, we can write the absorbed optical power as a fraction $\eta$ of the optical pump power: $P_\text{abs} = \eta P_\text{in} = \eta' \ncav$. In steady state, the power output into the phonon bath is equal to its input, $P_\text{out} = P_\text{abs} \sim \ncav$. The lattice thermal conductivity at low temperatures, where phonon transport is ballistic, scales as a power law of the phonon bath temperature~\citep{Holland1963,Callaway1959}, $\Gth \sim (\Tbathp)^\alpha$.  The power law exponent $\alpha$ is equal to the effective number of spatial dimensions $d$ of the material/structure under consideration.  Effectively, the hot phonon bath radiates energy as a black body, with radiated power scaling as $(\Tbathp)^{\alpha+1}$ via Planck's law.  In the case of a structure with 2D phonon density of states, such as the OMC cavity in the frequency range from $10$-$100$~GHz (c.f., Fig.~\ref{fig:sim_ph_TLS_prop}), $\alpha=d=2$ and the hot phonon bath temperature scales as $\Tbathp \sim P_\text{out}^{1/3} \sim \ncav^{1/3}$. This approximate scaling is expected to be valid so long as phonons in the hot phonon bath approximately thermalize each other upon creation from optical absorption events, and then radiate freely (balistically) into the effective zero temperature substrate.  The picture one has then is that the hot bath phonons make multiple passes within the OMC cavity region, scattering with other phonons leading to thermalization, and then eventually radiating into the substrate, i.e., the OMC cavity is still a good cavity for many phonons in the acoustic frequency region above the phononic bandgap.

In analogy with Equation~\ref{eqn:simplemodel_np_gammap}, for a phonon bath density of states $\rho(\omega)$ we can calculate the effective coupling rate $\gammap$ between the hot phonon bath and the mode of interest due to 3-phonon scattering:

\begin{equation}
\gammap = \int_{\omegacutoff}^{\infty} \mathrm{d}\omega~ A[\omega;\omegam] \rho[\omega] \rho[\omega + \omegam] \left(\nbathgen[\omega] - \nbathgen[\omega + \omegam]\right),
\end{equation}

\noindent In a simple continuum elastic model~\citep{Srivastava_book,Meenehan2014}, the product of the anharmonicity matrix element $A[\omega;\omega_m]$ and the density of states is taken to obey a polynomial scaling $A[\omega;\omega_m] \rho[\omega] \rho[\omega+\omegam] = A' (\omega-\omegacutoff)^a$ for some constants $A'$ and $a$, where we have introduced the cut-off frequency below which we assume the density of states is zero. With this assumption,

\begin{align}
\gammap &\cong A' \int_{\omegacutoff}^{\infty} \mathrm{d}\omega~(\omega-\omegacutoff)^a \left(\nbathgen[\omega] -\nbathgen[\omega + \omegam]\right) \\
&= A' \int_{\omegacutoff}^{\infty} \mathrm{d}\omega~(\omega-\omegacutoff)^a \left(\frac{\nbathgen[\omega + \omegam](\nbathgen[\omega]+1)}{\nBose[\hbar\omegam/\kB \Tbathp]}\right) \\
&= \frac{A'}{\nBose[\hbar\omegam/\kB \Tbathp]+1} \int_{\omegacutoff}^{\infty} \mathrm{d}\omega~(\omega-\omegacutoff)^a \left(\nbathgen[\omega](\nbathgen[\omega + \omegam]+1)\right) \label{eq:gammapint}, 
\end{align}

\noindent where in the last line we used the identity $\nBose[x+x^{\prime}](\nBose[x]+1)/\nBose[x^{\prime}] = (\nBose[x+x^{\prime}]+1)\nBose[x]/(\nBose[x^{\prime}]+1)$.  Making a change of variables to $x \equiv \hbar (\omega-\omegacutoff) / \kB \Tbathp$ in the integral in Eq.~(\ref{eq:gammapint}), we have 

\begin{equation}
\gammap \cong \left(\frac{A'}{\nBose[\xm]+1}\right)\left(\frac{\kB \Tbathp}{\hbar} \right)^{a+1} \int_{0}^{\infty} \mathrm{d}x~x^a \left(\nBose[x](\nBose[x + \xm]+1)\right)
\label{eqn:gammapDerivGenExp}
\end{equation}

\noindent where $\xm = \hbar\omegam / \kB \Tbathp$.  The integral in Equation~\ref{eqn:gammapDerivGenExp} depends on temperature only through $\xm$, and in the small and large $\xm$ limit (corresponding to low and high temperature), is relatively independent of $\xm$. If we assume that the anharmonicity element $A[\omega;\omega_m]$ is approximately frequency independent, and the only frequency dependence in $A'(\omega-\omegacutoff)^{a}$ comes from the phonon density of states, then $a \approx 2(d-1)$ for a phonon bath of dimension $d$.   We can thus make a general observation about the scaling of the bath-induced damping rate $\gammap$ in the low ($\xm \gg 1$) and high ($\xm \ll 1$) temperature regimes: 

\begin{align}
\label{eqn:gammap_general_scaling}
\gammap \propto \begin{cases} \big( \frac{\kB \Tbathp}{\hbar} \big)^a \sim \ncav^{2(d-1)/(d+1)} & \mbox{for } \Tbathp \gg \frac{\hbar \omega_m}{\kB}, \\ 
 \big( \frac{\kB \Tbathp}{\hbar} \big)^{a+1} \sim \ncav^{(2d-1)/(d+1)} & \mbox{for } \Tbathp \ll \frac{\hbar \omegam}{\kB},
\end{cases}
\end{align}

\noindent for a generic hot phonon bath of dimension $d$.  In a structure such as the OMC nanobeam cavity we expect the dimensionality of the effective bath density of states to be reduced relative to the Debye 3D density of states for a bulk crystal. Here we will assume - consistent with numerical simulations of the OMC structure in Section~\ref{App:F} - that the phonon bath has a two-dimensional density of states corresponding  to $a = 2$. In this case, we have the following scaling of the damping factor with intra-cavity photon number,

\begin{align}
\label{eqn:gammap_twoDbathmodel_scaling}
\gammap \propto \begin{cases} \big( \frac{\kB\Tbathp}{\hbar} \big)^2 \sim \ncav^{2/3} & \mbox{for } \Tbathp \gg \frac{\hbar \omegam}{\kB}, \\ 
 \big( \frac{\kB \Tbathp}{\hbar} \big)^3 \sim \ncav & \mbox{for } \Tbathp \ll \frac{\hbar \omegam}{\kB}.
\end{cases}
\end{align}

Upon thermalizing with the hot phonon bath, the effective thermal occupancy $\nbathp$ of the high-$Q$ breathing mode of the acoustic cavity  can be found from a similar rate equation analysis as considered for the 3-mode scattering in Eq.~(\ref{eqn:simplemodel_np_gammap}).  Integrating over all the possible 3-phonon scattering events involving the mode of interest at frequency $\omegam$ yields,

\begin{align}
\label{eqn:nP_scaling_derivation}
\nbathp &= \frac{1}{\gammap} \int_{\omegacutoff}^{\infty} \mathrm{d}\omega~ A[\omega;\omegam] \rho[\omega] \rho[\omega + \omegam] \nbathgen[\omega + \omegam]\left(\nbathgen[\omega]+1]\right) \\
&\cong \frac{\nbath[\omegacutoff+\omegam] A'}{\gammap}  \int_{\omegacutoff}^{\infty} \mathrm{d}\omega~ \omega^a\left(\nbathgen[\omega] - \nbathgen[\omega+\omegam]\right) \\
&= \nBose[\hbar\omegam/\kB\Tbathp].
\end{align}

\noindent We therefore have a characteristic scaling behavior for the effective phonon occupancy $\nbathp$ coupled to the cavity mode of interest that is,

\begin{align}
\label{eqn:np_twoDbathmodel_scaling}
\nbathp \propto \begin{cases} \big( \frac{\kB \Tbathp}{\hbar \omegam} \big) \sim \ncav^{1/(d+1)} \overset{\makebox[0pt]{\mbox{\normalfont\tiny\sffamily d=2}}}{=} \ncav^{1/3}  & \mbox{for } \Tbathp \gg \frac{\hbar \omegam}{\kB}, \\ 
 \exp{[-\hbar\omegam/\kB\Tbathp]} & \mbox{for } \Tbathp \ll \frac{\hbar \omegam}{\kB}.
\end{cases}
\end{align}

\subsection{Measurement of optical-absorption-induced damping, $\gammap$}
\label{subsec:meas_gammap}

\begin{figure}[btp!]
\centering
\includegraphics[width=0.75\textwidth]{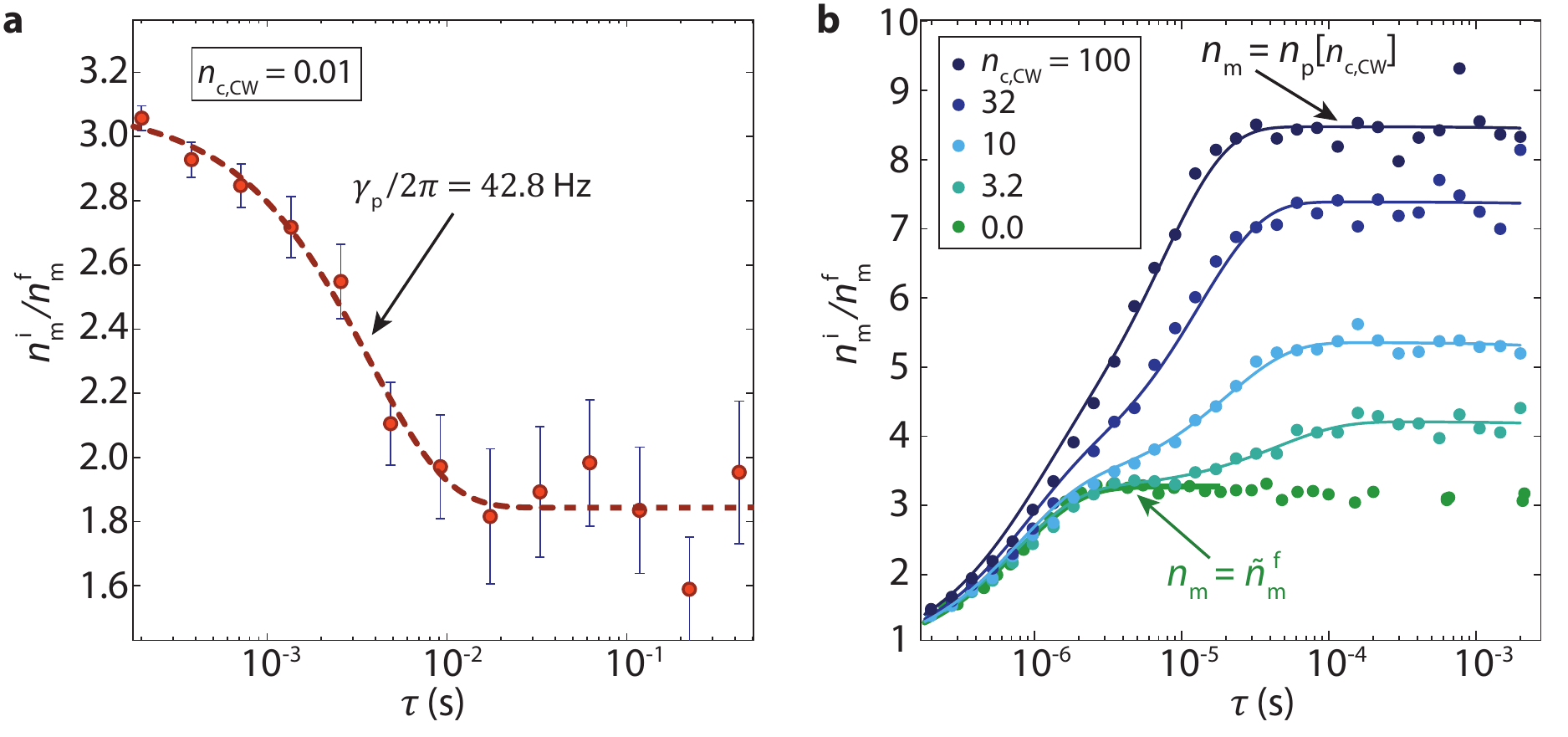}
\caption[Techniques for extracting the optical-bath-induced damping rate $\gammap$.]{\textbf{Measurement techniques for extracting the optical-bath-induced damping rate $\gammap$.} \textbf{a}, Ringdown measurement in the presence of a continuous-wave pump laser with an average intracavity photon number of $\ncavCW = 10^{-2}$. The total decay rate is $\gamma = \gammap + \gammanotO$, and with $\gammanotO/2\pi=0.21$~Hz known from separate measurements, $\gammap/2\pi=42.8$~Hz is extracted directly from the fitted decay rate. \textbf{b}, At larger $\ncav$, the bath-heating induced by the pump laser causes net heating in the pulse-off state. Here, the heating is fitted to the phenomenological model of Eq.~(\ref{eqn:nbar_bathdecaymodel}) to extract $\gammap$ due to the CW laser pump. Measurements were performed on a six-acoustic-shield device (device B) with parameters ($\kappa$, $\kappae$, $\gzero$, $\omegam$, $\gammanotO$) $=$ $2\pi$($1.13$~GHz, $605$~MHz, $713$~kHz, $5.013$~GHz, $0.21$~Hz) and with a readout photon number $\ncavRO = 569$.}
\label{fig:nanobeamgammaP_meas}
\index{figures}
\end{figure}

\begin{figure}[btp!]
	\begin{center}
		\includegraphics[width=0.7\columnwidth]{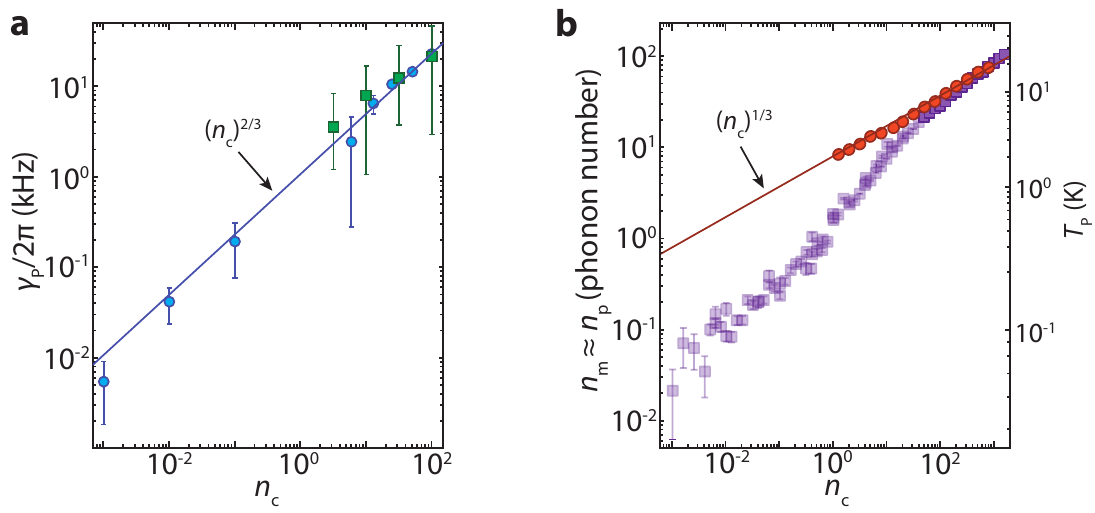}
		\caption{\textbf{Measured steady-state properties of the optical-absorption-induced bath.} \textbf{a}, Plot of $\gammap$ versus $\ncav$ for six-shield (blue circles) and zero-shield (green squares) devices. The solid line is a power-law fit to the six-shield device data: $\gammap/2\pi = (1.07~\text{kHz}) \times \ncav^{2/3}$. The zero-shield device (device A) has parameters ($\kappa$, $\kappae$, $\gzero$, $\omegam$, $\gammanotO$) $=$ $2\pi$($1.507$~GHz, $778$~MHz, $713$~kHz, $5.053$~GHz, $14.1$~kHz). The six shield device (device B) has parameters ($\kappa$, $\kappae$, $\gzero$, $\omegam$, $\gammanotO$) $=$ $2\pi$($1.13$~GHz, $605$~MHz, $713$~kHz, $5.013$~GHz, $0.21$~Hz). \textbf{b}, Plot of $\nbathp$ versus $\ncav$ for zero-shield (purple symbols) and six-shield (orange circles) devices. Purple squares represent the measured mode occupancy corrected for heating induced by the readout laser tone. The right-hand axis gives the effective bath temperature $\Tbathp$ which corresponds to the measured bath occupancy. Translucent squares show data taken in the regime where the intrinsic decay rate $\gammanotO$ is comparable to the bath-induced damping $\gammap$, indicating that the raw measured occupancy begins to deviate substantially from the inferred occupancy given in the plot. The solid line is a fit to the six-shield data giving $\nbathp = (7.94) \times \ncav^{1/3}$. \textbf{c}, Normalized phonon occupancy during and after the optical pulse. Squares are data points and the solid line is a best fit to the dynamical model. During the pulse, back-action cooling occurs at a timescale $\gammaOM^{-1} \approx 100$~ns. The optical-absorption-induced bath simultaneously heats the mode at a rate $\gammap \nbathp$, such that at long $\text{T}_\text{pulse}$ a steady-state mode occupancy $n_\text{f}$ is reached. In the pulse-off state (gray squares), the residual phonon bath heats the mode at a rate $\gammap(t) \nbathp(t)$, where the bath damping and effective occupancy are explicitly time-dependent. A full dynamical model of the bath heating is used to generate the fit (dotted line). The purple data point in the off-state plot ($\Toff = 200$~$\mu$s) corresponds to off-state delay for the measured intra-pulse data shown in the on-state plot.} 
		\label{fig:SI_bath_properties}
	\end{center}
\end{figure}

In order to measure the additional bath-induced damping rate $\gammap$, we use a pump-probe technique employing two laser sources. The pump laser is tuned to optical resonance ($\Delta = 0$) to eliminate dynamical back-action effects ($\gammaOM = 0$), and impinges on the cavity in continuous-wave (CW) operation. The pump laser generates a steady-state intracavity photon population $\ncavCW$ and an absorption-induced bath at elevated temperature in the steady state. A second pulsed laser, the probe laser, is tuned to the red motional sideband of the cavity ($\Delta = +\omegam$) and is used to periodically read out the phonon occupancy, where the scattering rate of the probe laser at the beginning of the probe pulse provides an estimate of $\nbathp$ due to the CW laser alone. Application of the probe laser not only allows readout of the breathing mode occupancy, but also produces an excess absorption-induced bath above and beyond that of the background CW laser alone.  When the readout probe pulse is turned off, the breathing mode initially heats due to the excess hot bath created by the probe pulse (over several microseconds; see Fig.~\ref{fig:SI_bath_properties}(c)), and then after this excess hot bath evaporates away leaving a breathing mode occupancy of $n_\text{f}^{\prime}$, relaxes back to its steady-state occupancy set by the CW laser, $\nbar[\ncavCW] = (\gammap[\ncavCW] \nbathp[\ncavCW] + \gammanotO \nmnot)/(\gammap[\ncavCW] + \gammanotO)$.  The rate of relaxation is set by the modified total damping rate of $\gammanotO + \gammap[\ncavCW]$. By observing this modified exponential decay rate we directly extract $\gammap[\ncavCW]$, with $\gammanotO$ known from independent ringdown measurements in the absence of the CW background laser. For example, in Fig.~\ref{fig:nanobeamgammaP_meas}(a) we show the measured ringdown of a high-$Q$ six-shield device (device B; $\gammanotO/2\pi=0.21$~Hz) for a CW pump laser photon number of $\ncavCW = 10^{-2}$, from which we extract $\gammap/2\pi = 42.8$~Hz. 

For large $\ncavCW$ ($\gtrsim 1$) the steady-state occupancy of $\nbar[\ncavCW]$ becomes larger than the occupancy $\nmfp$ at the end the readout pulse.  The readout pulse \emph{should} cool the breathing mode, after all, and it is only the absorption-induced heating caused by the readout pulse itself that manifests as a ring down in absence of heating from the CW laser.  For large $\ncavCW$ then, $\gammap$ is estimated by observing a ring-\textit{up} in the pulse-off state from the final pulse occupancy $\nmfp$ to the elevated $\nbar[\ncavCW]$.  Figure~\ref{fig:nanobeamgammaP_meas}(b) shows a representative data set for extracting $\gammap$ at $\ncavCW > 1$, where an initial fast rise is observed in the mode occupancy in the pulse-off state from $\nmf$ to $\nmfp$ due to the aforementioned excess bath created by the readout pulse, followed by a slower second heating stage from $\nmfp$ to $\ncavCW$.  As discussed in more detail in sub-Section~\ref{subsec:meas_npgammap_dynamics}, we can fit the ring up curve after the readout pulse is turned off by considering a phenomenological model including decay of the readout-induced hot bath, 

\begin{multline}
\dot{\nbar} = - \left\{\gammanotO + \gammap[\ncavRO] e^{-\Gammagammapoff \Toff} + \gammap[\ncavCW]\right\} \nbar \\
+ \left\{\gammap[\ncavRO] e^{-\Gammagammapoff \Toff} + \gammap[\ncavCW]\right\}(\nbathp[\ncavRO] e^{-\Gammanpoff \Toff} + \nbathp[\ncavCW]).
\label{eqn:nbar_bathdecaymodel}
\end{multline}

\noindent We first measure the transient readout-induced bath in the absence of the CW laser (dark green curve in Fig.~\ref{fig:nanobeamgammaP_meas}b), from which a fit to Eq.~(\ref{eqn:nbar_bathdecaymodel}) yields $\nbathp[\ncavRO] = 40$ phonons, $\gammap[\ncavRO]/2\pi = 9.55$~kHz, $\Gammagammapoff/2\pi = 143$~kHz, and $\Gammanpoff/2\pi = 15.9$~kHz. With these readout-induced bath values known, Eq.~(\ref{eqn:nbar_bathdecaymodel}) is numerically integrated to fit the entire heating curve in the pulse-off state to extract the additional CW-pump-induced damping $\gammap[\ncavCW]$. 

%The data are fitted to a full heating model in which the decay of the bath induced by the readout laser at a total rate $\gammaS$ is modeled by an effective decay of $\gammap$ at rate $\gammaR$ and of $\nbathp$ at rate $\gammaT$. The readout-induced bath parameters, with a subscript $1$, satisfy in the model $\gamma_{p,1}(\Toff) = \gamma_{p,1}(0)e^{-\gammaR \Toff}$ and $n_{p,1}(\Toff) = n_{p,1}(0)e^{-\gammaT \Toff}$. The bath generated by the pump laser, with subscript $2$, is maintained in steady-state. The resulting rate equation for the mode occupancy is:

%However, as we have previously discussed, the probe laser itself introduces a short-lived optical heating bath which results in net heating of the mechanics in the pulse-off state, and this probe heating must be distinguished from the heating caused by the pump laser in order to accurately extract the pump-induced damping $\gammap(\ncav)$.

The results of the measured optical-absorption-induced damping $\gammap$ versus $\ncav$ are summarized in Fig.~\ref{fig:SI_bath_properties}(a) for measurements on both a six-shield (device B) and a zero-shield (device A) nanobeam device. The observed power law scaling fits well to $\gammap/2\pi = (1.07$~kHz$) \times \ncav^{2/3}$, in agreement with the scaling predicted in Eq.~(\ref{eqn:gammap_twoDbathmodel_scaling}) for a 2D density of states for the bath phonon population.  Note that the much lower $\gammanotO$ of the six-shield device allows a much wider range of $\gammap$ (and thus $\ncav$) to be explored.

\subsection{Measurement of optical-absorption-induced bath occupancy, $\nbathp$}
\label{subsec:meas_np}

In order to measure the bath occupancy $\nbathp$, again two different methods are used to probe the high- and low-photon-number dependencies of the bath. To measure the bath occupancy at photon numbers $\ncav \gtrsim 1$, a simple readout technique may be used in which a single readout laser is sent to the cavity in continuous-wave operation. The laser is tuned to cavity resonance ($\Delta = 0$) and the resulting sideband scattered photon count rate appearing at either the lower or upper frequency mechanical sideband ($\Delta = \pm \omega_m$) will be 

\begin{equation}
\Gamma = \Gamma_\text{noise} + \bigg(\frac{\kappa}{2 \omega_m}\bigg)^2\Gamma_\text{SB,0}\nbar.
\label{eq:npcountsCW}
\end{equation}

\noindent With the sideband filters aligned to either of the mechanical sidebands of the cavity, the observed count rate is used to extract an equivalent occupancy $\nbar = \nbathp$ at various pump powers $\ncav$. The results are shown in Fig.~\ref{fig:SI_bath_properties}b (orange circles) for a six-shield device (device B), exhibiting a power-law scaling of $\nbathp  = (7.94) \times \ncav^{1/3}$ in agreement with the model in the limit of high bath temperature $\Tbathp \gg \hbar \omega_m /\kB \approx 200$~mK. The right-hand axis of Fig.~\ref{fig:SI_bath_properties} gives the effective bath temperature $\Tbathp$ corresponding to the measured occupancy, indicating that the measurement regime is indeed well in the high temperature limit.

\begin{figure}[btb!]
	\begin{center}
		\includegraphics[width=0.63\textwidth]{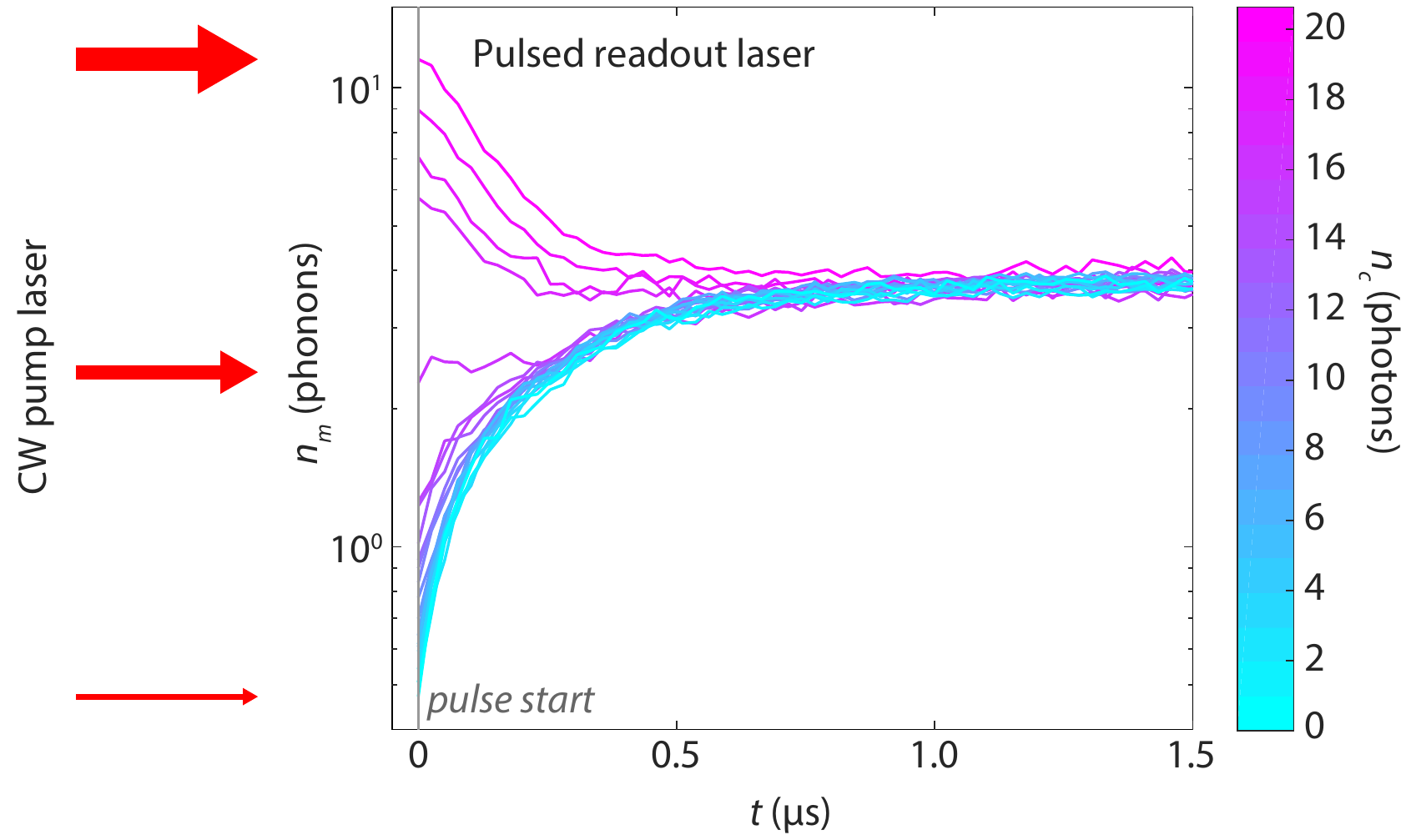}
		\caption[Pulsed measurements of the bath occupancy in a low-$Q$ nanobeam cavity]{\textbf{Pulsed measurements of the bath occupancy in a low-$Q$ nanobeam.} A continuous-wave background laser (red arrows, detuning $\Delta/2\pi \approx 1$~GHz) is used to generate a constant stead-state absorption bath, while a pulsed readout laser (readout $\ncavRO = 50.6$) is used to probe the resulting bath occupancy for various background laser powers $\ncavCW$. The initial measured occupancy during the pulse is given by $\nmzero \approx (\nbathp \gammap + \nmnot \gammanotO)/(\gammap + \gammanotO) + \nmoffset$, where $\nmoffset$ is residual occupancy due to the finite heating occurring before the first readout time bin. Measurements were performed on the zero-shield device (device A) with parameters ($\kappa$, $\kappae$, $\gzero$, $\omegam$, $\gammanotO$) $=$ $2\pi$($1.507$~GHz, $778$~MHz, $713$~kHz, $5.053$~GHz, $14.1$~kHz).}
		\label{fig:nanobeam_pulsenmnPmethods}
	\end{center}
\end{figure}

At lower photon numbers $\ncav \lesssim 1$, and corresponding lower $\nbathp \approx \nbar$, the SNR of the counting of photons scattered from cavity resonance into either mechanical sideband begins to drop below $1$ due to the large sideband resolution factor $(2\omegam/\kappa)^2$ of the OMC cavity (c.f., Eq.~(\ref{eq:npcountsCW})). In this regime, an alternative measurement method is employed in which a CW pump laser generates a steady-state optical-absorption bath while a second pulsed readout laser is used to probe the breathing mode occupancy (see Fig.~\ref{fig:nanobeam_pulsenmnPmethods}). The background pump laser is detuned to $\Delta/2\pi = 1$~GHz from the cavity resonance to minimize back-action as well as bleed-through counts through the sideband filters aligned at $\Delta = 0$. The initial measured occupancy $\nmzero$ during the pulse is a measure of the pump-induced bath occupancy; however, it includes a small residual occupancy $\tilde{n}_\text{0} \approx 0.04$ due to heating caused by the readout laser prior to the first measurement time bin of the pulse-on state. We define a corrected occupancy $\nmstar \equiv \nmzero - \nmoffset$ which denotes the measured mode occupancy which is coupled to the fridge bath as well as the absorption-bath induced by the pump laser:

\begin{equation}
\nmstar = \frac{\nbathp \gammap + \nmnot \gammanotO}{\gammap + \gammanotO}.
\end{equation}

\noindent With $\nmnot$, $\gammanotO$, and the power-dependence of $\gammap$ known from independent measurements, we can estimate the equivalent bath occupancy 

\begin{equation}
\nbathp[\ncav] = \frac{\nmstar\gammap[\ncav] + (\nmstar - \nmnot)\gammanotO}{\gammap[\ncav]}.
\end{equation}

Using this second method, over a much larger span of $\ncav$, the behavior of the effective bath occupancy $\nbathp$ for a zero-shield device with intrinsic damping rate $\gammanotO/2\pi = 14.1$~kHz is shown in Fig.~\ref{fig:SI_bath_properties}b as purple squares. Note that measurement of the very high-$Q$ six-shield device (device B) using the pulsed readout scheme is not practical due to the extremely long relaxation times required between readout pulses (we did, however, verify for a few values of $\ncav$ that the two schemes give consistent results).  For $\ncav \gtrsim 1$, again we find good agreement for the zero-shield device with a power-law scaling $\nbathp \propto \ncav^{1/3}$ for $\Tbath \gg \hbar \omega_m/\kB$. Not only is the scaling of $\nbathp$ versus $\ncav$ the same for both zero-shield and six-shield devices, but so is the absolute value of $\nbathp$.   For $\ncav \lesssim 1$, $\gammap(\ncav) \approx \gammanotO$ for the zero-shield device and the measured occupancy $\nmstar$ deviates substantially from $\nbathp$ as the breathing mode thermalizes more strongly with the external substrate temperature set by the fridge ($\Tf \approx 10$~mK). In this range we have plotted $\nmstar$ in translucent purple squares to distinguish it from the region of parameter space where $\nmstar$ is expected to faithfully represent $\nbathp$.  

\subsection{Measurement of optical-absorption-induced bath dynamics}
\label{subsec:meas_npgammap_dynamics}

The hot bath created by the application of laser light resonant with the optical mode of the OMC cavity does not instantaneously appear when the laser light is turned on, nor does it instantaneously vanish once the laser is turned off.  Rather, there is a somewhat complicated bath dynamics that can be inferred from careful study of the temporal variation of the scattered photon signal from the readout pulse due to excitation of the mechanical breathing mode by the hot phonon bath.  Using the measured breathing mode occupancy as a proxy one can infer many subtle features of the bath dynamics.

\begin{figure}[btp!]
	\begin{center}
		\includegraphics[width=0.55\columnwidth]{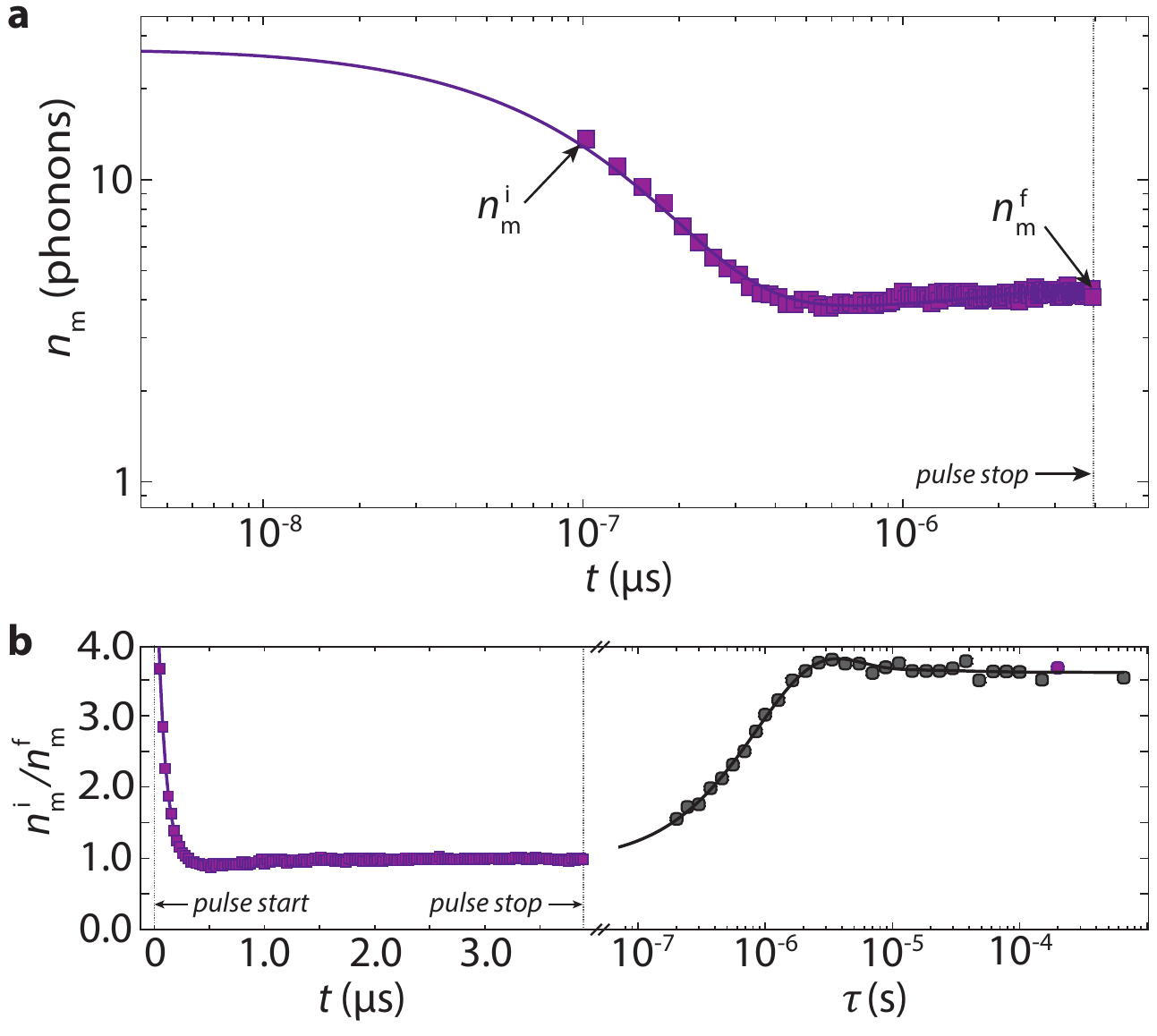}
		\caption{\textbf{Measured pulse dynamics of the breathing mode occupancy and the optical-absorption-induced bath.} \textbf{a}, Phonon occupancy of the breathing mode as a function of time $\ton$ during the red-detuned ($\Delta = +\omega_m$) optical excitation pulse.  Here, the time delay between successive pulses is $\Toff = $~$654\mu$s. Squares are data points and the solid line is a best fit to the dynamical model. The nanobeam device is device B with six periods of acoustic shielding, and device parameters ($\kappa$, $\kappae$, $\gzero$, $\omegam$, $\gamma_\text{0}$) = $2\pi$ ($1.13$~GHz, $605$~MHz, $713$~kHz, $5.013$~GHz, $0.21$~Hz). During the pulse, back-action cooling occurs at a timescale $\gammaOM^{-1} \approx 100$~ns. Note that the initial mode occupancy $\nbar[0] = 27$~phonons is determined by extrapolating the model fit back to $\ton = 0$, while the earliest measurement bin has an occupancy of $\nmi = 13.6$~phonons. The optical-absorption-induced bath heats the mode at a rate $\gammap(\ton) \nbathp(\ton)$, such that for long enough $\Tpulse$ a steady-state mode occupancy $\nmf$ is reached. Here $\nmf = 4.2$~phonons. The measurement time resolution bin size is $10.24$~ns. \textbf{b}, Normalized breathing mode phonon occupancy during (left) and after (right) the optical pulse. In the pulse-off state (gray squares), the residual phonon bath heats the mode at a rate $\gammap(\Toff) \nbathp(\Toff)$, where the bath damping and effective occupancy are explicitly time-dependent. The purple data point in the off-state plot at $\Toff = 200$~$\mu$s indicates the pulse shown in the on-state plot.  For all panels, the measurements were performed using an on-state readout intra-cavity photon number of $\ncav = 569$, and the solid curves correspond to the phenomenological model including the dynamics of the optical-absorption-induced bath, fit to the data using the parameters shown in Table~\ref{tab:bath}.} 
		\label{fig:SI_bath_dynamics}
	\end{center}
\end{figure}

Figure~\ref{fig:SI_bath_dynamics}(a) shows the measured scattered photon signal due to a pulsed readout tone applied on the lower motional sideband of the optical cavity ($\Delta=+\omegam$) of a high mechanical $Q$-factor six-shield OMC device (device B).  Here the readout pulses are $\Tpulse=4$~$\mu$s long and a variable delay $\Toff$ is applied between each successive optical readout pulse.  The scattered photons from the readout pulse are filtered by the filter bank resonantly aligned with the optical cavity resonance ($\Delta=0$), thus yielding a photon count rate throughout the readout pulse which is proportional to the average occupancy of the mechanical breathing mode $\nbar$.  This is shown in Fig.~\ref{fig:SI_bath_dynamics}(a) for a time bin resolution of $10.24$~ns, with the first measurement bin occurring at $\ton = 100$~ns after the pulse-on signal is applied in order to ensure that the optical pulse amplitude has settled and reached its maximum value.  In the left panel of Fig.~\ref{fig:SI_bath_dynamics}(b) we plot time-varying normalized breathing mode occupancy, corresponding to the ratio of the measured signal during the pulse to that at the very end of the pulse.  This curve is not a single-shot measurement, but rather averaged over thousands of pulses, for which the normalized signal avoids small, slow drifts in the efficiency of the measurement apparatus.  In the right panel we plot the normalized initial measurement bin occupancy (still taken to be at $100$~ns after the readout pulse is turned on) as a function of the off-state delay time $\Toff$ between successive pulses.

Several things are quickly evident from these plots of the measured breathing mode occupancy during and after the applied optical pulse.  During the pulse we expect the optomechanical back-action to induce damping and cooling of the breathing mode at a rate $\gammaOM[\ncav]$.  Without any parasitic heating effects from the applied optical pulse, the breathing mode should cool down to its equilibrium occupancy, ideally very close to zero at the fridge temperature ($\Tf = 10$~mK).  This does not occur, but rather the breathing mode occupancy is seen to initially cool to a few phonons over $\sim 300$~ns, and then slowly heat to a steady-state phonon occupancy at the end of the pulse of $\nmf = 4.2$~phonons (c.f., Fig.~\ref{fig:SI_bath_dynamics}(a)).  Similarly, once the optical pulse is turned off and light has left the optical cavity, the breathing mode occupancy starts to heat again, levelling off after a few microseconds following a slight overshoot to a modified post-pulse value of $\nm[0] = 27$~phonons ($\rightarrow \nmi = 13.6$~phonons in the first masurement bin; c.f., Fig.~\ref{fig:SI_bath_properties}(b)).  This strange dynamics is a result of the coupling of the breathing mode to the optical-absorption-induced hot bath.  The slight undershoot of the cooling and slow heating in the pulse-on state is a result of a slow turn on of the hot bath.  Similarly, the transient post-optical-pulse heating results from the slow decay of the hot bath, now without the cooling from optomechanical back-action.  

Noticeably, the timescales for the turn on ($\sim 400$~ns) of the bath and the turn off ($\sim 3$~$\mu$s) of the bath are different.  Less evident from these plots, but nonetheless very clear when attempting to model the hot bath dynamics, is that there seems to be two components to the bath, one whose turn on and turn off transients are very rapid (effectively instantaneous with the optical field), and one with much slower relaxation times.  Even more subtle is that to get very good agreement with the measured initial transient in the breathing mode occupancy in the immediate aftermath of turning off the optical pulse, it seems that the hot-bath damping factor, $\gammap$, should be modeled with a more rapid relaxation rate than that of the hot bath occupancy, $\nbathp$.  It may be that this is also the case in the transient dynamics during the pulse-on state, however, in the pulse-off state the relaxation rate of the measured breathing mode occupancy is far more sensitive to the value of $\gammap$ as it dominates the total relaxation rate of the breathing mode in the absence of appreciable $\gammaOM$. 

The model used to fit the data in Fig.~\ref{fig:SI_bath_dynamics} consists of a set of coupled differential equations involving the breathing mode occupancy, the hot bath damping factor, and the effective hot bath occupancy.  The rate equation for the breathing mode occupancy is given by,

\begin{equation}
\dotnm = -(\gammap + \gammaOM + \gammanotO)\nm + \gammap\nbathp + \gammanotO\nmnot
\label{eq:nmrate}
\end{equation}

\noindent where in the pulse-on state $\gammaOM = \gammaOM[\ncav]$ will take on a large value on the order of $1$~MHz for a readout pulse amplitude of a few hundred intra-cavity photons, and in the pulse-off state $\gammaOM\approx 0$ due to the large extinction ($\gtrsim 80$~dB) and rapid timescale of the turn-off the optical pulse (~$20$~ns).  During the pulse-on state the rate equations for the fast (F) and slow (S) components of the hot bath damping factor and effective occupancy are,

\begin{equation}
\dot{(\gammap)}_{\text{F(S)}}(\ton) = -(\Gammagammapon)_{\text{F(S)}}\left\{ (\gammap)_{\text{F(S)}}(\ton) - (\deltab)_{\text{F(S)}}\gammap[\ncavRO]\right\},
\label{eq:gammappulseon}
\end{equation}

\noindent and

\begin{equation}
\dot{(\nbathp)}_{\text{F(S)}}(\ton) = -(\Gammanpon)_{\text{F(S)}}\left\{ (\nbathp)_{\text{F(S)}}(\ton) - (\deltab)_{\text{F(S)}}\nbathp[\ncavRO]\right\},
\label{eq:nppulseon}
\end{equation}

\noindent where $t=\{0,\Tpulse\}$ is the time from the start of the pulse to the end of the pulse, $(\Gammagammapon)_{\text{F(S)}}$ and $(\Gammagammapon)_{\text{F(S)}}$ are the pulse-on relaxation rate constants for the damping factor and occupancy of the two different bath components, respectively, and $(\deltab)_{\text{F(S)}}$ is the F(S) fraction of the hot bath.  $\gammap[\ncavRO]$ and $\nbathp[\ncavRO]$ are the steady-state bath values reached at the end of the optical readout pulse.   The corresponding rate equations for the hot bath in the pulse-off state are,  

\begin{equation}
\dot{(\gammap)}_{\text{F(S)}}(\Toff) = -(\Gammagammapoff)_{\text{F(S)}}\left\{ (\gammap)_{\text{F(S)}}(\Toff) - (\deltab)_{\text{F(S)}}\gammap[\ncavRO]\right\},
\label{eq:gammappulseoff}
\end{equation}

\noindent and

\begin{equation}
\dot{(\nbathp)}_{\text{F(S)}}(\Toff) = -(\Gammanpoff)_{\text{F(S)}}\left\{ (\nbathp)_{\text{F(S)}}(\Toff) - (\deltab)_{\text{F(S)}}\nbathp[\ncavRO]\right\}.
\label{eq:nppulseoff}
\end{equation}

\noindent where $\Toff$ is the time from the end of the optical pulse, and $(\Gammagammapoff)_{\text{F(S)}}$ and $(\Gammagammapoff)_{\text{F(S)}}$ are the pulse-off relaxation rate constants for the damping factor and occupancy of the two different bath components, respectively.

The model parameters used to fit the specific measured data for the six-shield device (device B) presented in Fig.~\ref{fig:SI_bath_dynamics} are listed in Table~\ref{tab:bath}.  Similar bath dynamical parameters are found for all of the measured devices we have studied.  Independent of the optical readout pulse power, the fraction of the bath which reacts quickly seems to be consistently close to a value of $(\deltab)_{F}=0.65$.  The fast component of the bath turns on faster than we can resolve ($\gtrsim 50$~MHz), while slow component of the bath turns on with a rate constant of approximately $\theta_{S}/2\pi = 600$~kHz (for both damping factor and occupancy).  The fast component of the bath turns off with an exponential rate constant of $(\Gammagammapoff)_{F}/2\pi = 150$~kHz for $\gammap$ and $(\Gammanpoff)_{F}/2\pi =70$~kHz for $\nbathp$.  Even more slowly, the slow component of the bath turns off with a rate constant of $(\Gammagammapoff)_{S}/2\pi = 90$~kHz and $(\Gammanpoff)_{S}/2\pi =24$~kHz for the two different bath factors.  

Our ability to measure the bare damping rate of the acoustic breathing mode relies on the fact that the hot bath evaporates prior to the actual measurement of the free decay of the breathing mode.  This means that the first $\sim 10$~$\mu$s of the pulse-off state is dead time in which the dynamics of the breathing mode occupancy is still coupled to that of the hot bath.  Crucial to the measurement of a ringdown curve using the red-detuned optical pulse as both a readout signal and an excitation source, is that after this dead time there remain a residual, elevated phonon occupancy of the breathing mode from which  the mode can decay.  This is clearly the case for the data measured in Fig.~\ref{fig:SI_bath_dynamics}, and is a result of the fact that at this readout power the peak magnitude of $\gammap$ ($2\pi(85$~kHz$)$) is still smaller than the fastest decay of the hot bath ($(\Gammagammapoff)_{F}/2\pi = 150$~kHz), so that the breathing mode occupancy cannot follow that of the fast dynamics of the hot bath.  This non-adiabatic quenching leaves the breathing mode with an elevated occupancy after the dead time.  At readout powers beyond $\ncav=1000$ this stops being the case, hence our choice of readout pulse powers $\ncav \lesssim 600$ in the ringdown measurements.  

A few observational comments are warranted.  The fact that the hot bath should have faster pulse-on rate constants than pulse-off rate constants might be explained by the fact that there are likely a wide spectrum of phonons which are created by absorption of the optical pulse.  This may lead to a hierarchy of phonon baths.  Consider for instance a two bath scenario, consisting of a high and a low frequency phonon bath.  The high frequency bath is assumed to be directly populated from optical absorption events, while the low frequency bath is predominantly responsible for coupling to the breathing mode of interest.  In the high frequency phonon bath, phonons rapidly mix with each other due the large density of states and mode occupancy.  The high frequency phonon bath is also well thermalized to the external substrate through acoustic radiation.  Phonons in the low frequency bath are fed from the phonon-phonon scattering processes in the high frequency phonon bath, and are less connected via radiation to the substrate.  When the optical pulse is on, the high frequency bath is rapidly populated.  The high frequency bath not only acts as a source of phonons for the low frequency bath, but through nonlinear phonon mixing also helps bring it into some quasi-equilibrium temperature.  When the optical pulse is turned off, the high frequency bath rapidly decays away, leaving the low frequency bath of phonons to more slowly decay away due to the absence of the phonons in the high frequency bath to mix with.  This scenario would also explain the difference in the decay of the low frequency bath $\gammap$ damping rate, which depends on the phonon number density, to that of the effective occupancy $\nbathp$, which is set by the quasi-equilibrium temperature of the bath.  The absence of the high frequency bath could greatly slow down the low frequency bath equilibriation rate, and thus the rate of change of the effective bath temperature, while the low frequency bath acoustic coupling to the external substrate will provide a constant decay channel for bath phonons and thus $\gammap$. 

We should further note that the dynamical bath parameters reported in Table~\ref{tab:bath} are consistent for devices fabricated from low resistivity SOI.  In the case of our high resistivity SOI samples, we have measured devices with much slower post-read-pulse decay of the hot bath.  Hot bath decay times as long as tens of milliseconds have been observed.  Although requiring further study, we believe that this very slow decay dynamics of the optical-absorption-induced bath indicate that phonons are not the only parties involved in the optically-induced hot bath, but that the hot bath is likely also composed of much longer lifetime two-level system (TLS) defects.  The high resistivity SOI seems to harbor much longer-lived TLS states, possibly due to the reduction of electronic relaxation pathways.  Our two-phonon-bath scenario above may in fact be a two-bath scenario consisting of one phonon bath coupled to a longer lived TLS bath.  Other evidence for this interpretation is the high values of measured $\gammap$ which is more consistent with estimated TLS damping rates (damping due to three-phonon scattering is shown to be too slow, at least for bath temperatures below $1$~K, in sub-Section~\ref{subsec:3phononnumerical}).     

\begin{table}[btp]
\begin{center}
\caption{\textbf{Dynamical model parameters of the optical-absorption-induced bath.} These parameters were used to fit the measured phonon occupancy shown in Fig.~\ref{fig:SI_bath_dynamics} during the entire pulse on period and during the transient initial millisecond after the pulse is turned off.  The specific device measured was a six-shield device, device B in Table~\ref{tab:devices}, fabricated from low resistivity SOI ($\rho = 5$-$20$~$\Omega$-cm). \label{tab:bath}}
%\begin{ruledtabular}
\begin{tabular}{|l|l|l|l|}
\hline\hline
%\toprule
 Parameter & Description & Value & Refs./Notes \\
\hline
$(\deltab)_{\text{F(S)}}$ & fast (slow) fractional component of the optically-induced bath & 0.65 (0.35) &  from model fit \\
$(\Gammanpon)_{\text{F(S)}}/2\pi$ & $\nbathp$ exponential relaxation rate with optical \emph{pulse on} & $\gtrsim 50$~MHz ($600$~kHz) &  from model fit \\
$(\Gammagammapon)_{\text{F(S)}}/2\pi$ & $\gammap$ exponential relaxation rate with optical \emph{pulse on} & $\gtrsim 50$~MHz ($600$~kHz) &  from model fit \\
$(\Gammanpoff)_{\text{F(S)}}/2\pi$ & $\nbathp$ exponential relaxation rate with optical \emph{pulse off} & $70$~kHz ($24$~kHz) &  from model fit \\
$(\Gammagammapoff)_{\text{F(S)}}/2\pi$ & $\gammap$ exponential relaxation rate with optical \emph{pulse off} & $150$~kHz ($90$~kHz) & from model fit \\
$\ncavRO$ & readout pulse amplitude & 569 photons & \\
$\gammaOM/2\pi$ & optomechanical back-action rate during \emph{pulse on} & $1.4$~MHz & measured ind. \\
$\nbathp$ & steady-state hot bath occupancy for $\ncavRO$ intra-cavity photons & $73$ phonons & measured ind. \\
$\gammap$ & steady-state hot bath damping rate for $\ncavRO$ intra-cavity photons & $85$~kHz & measured ind. \\
$\Toff$ & delay time between successive readout pulses & $200$~$\mu$s & \\
$\Tpulse$ & readout pulse length & $4$~$\mu$s & \\
$\nmf$ & breathing mode occupancy at end ($t=4$~$\mu$s) of optical readout pulse & 4.2 phonons & measured \\
$\nmi$ & breathing mode occupancy in first measurement bin of readout pulse & 13.6 phonons & measured \\
$\nbar[0]$ & breathing mode occupancy referred back to $t=0$ of optical readout pulse & 27 phonons & from model fit \\
\hline\hline
%\bottomrule
\end{tabular}
%\end{ruledtabular}
\end{center}
\end{table}

\section{Coherent Excitation Methods: Blue-Detuned Pumping and RF-Modulation Amplification}
\label{App:E}

\subsection{Low-Threshold Acoustic Self-Oscillation}

Owing to the extremely slow intrinsic damping rate $\gammanotO$ observed in the ultra-high-$Q$ nanobeam devices at low temperature, it is possible to drive the mechanics into the regime of self-sustained oscillations with a blue-detuned pumping laser at very low input optical powers, or equivalently, a very low rate of measurement back-action. The total effective damping rate experienced by the mechanics in the presence of a blue-detuned drive laser is $\gamma = \gammai - \gammaOM$, where the intrinsic damping rate $\gammai = \gammanotO + \gammap$ includes damping $\gammanotO$ from both the cold fridge bath (with occupancy $\nmnot \approx 10^{-3}$) and from the optical absorption-induced phonon bath at rate $\gammap$. The usual condition for self-oscillation is that the damping rate is matched by the back-action amplification rate $\gammaOM$:

\begin{equation}
\gammaOM = \gammanotO + \gammap.
\end{equation}

\begin{figure}[hbt]
	\begin{center}
	\includegraphics[width=0.666\textwidth]{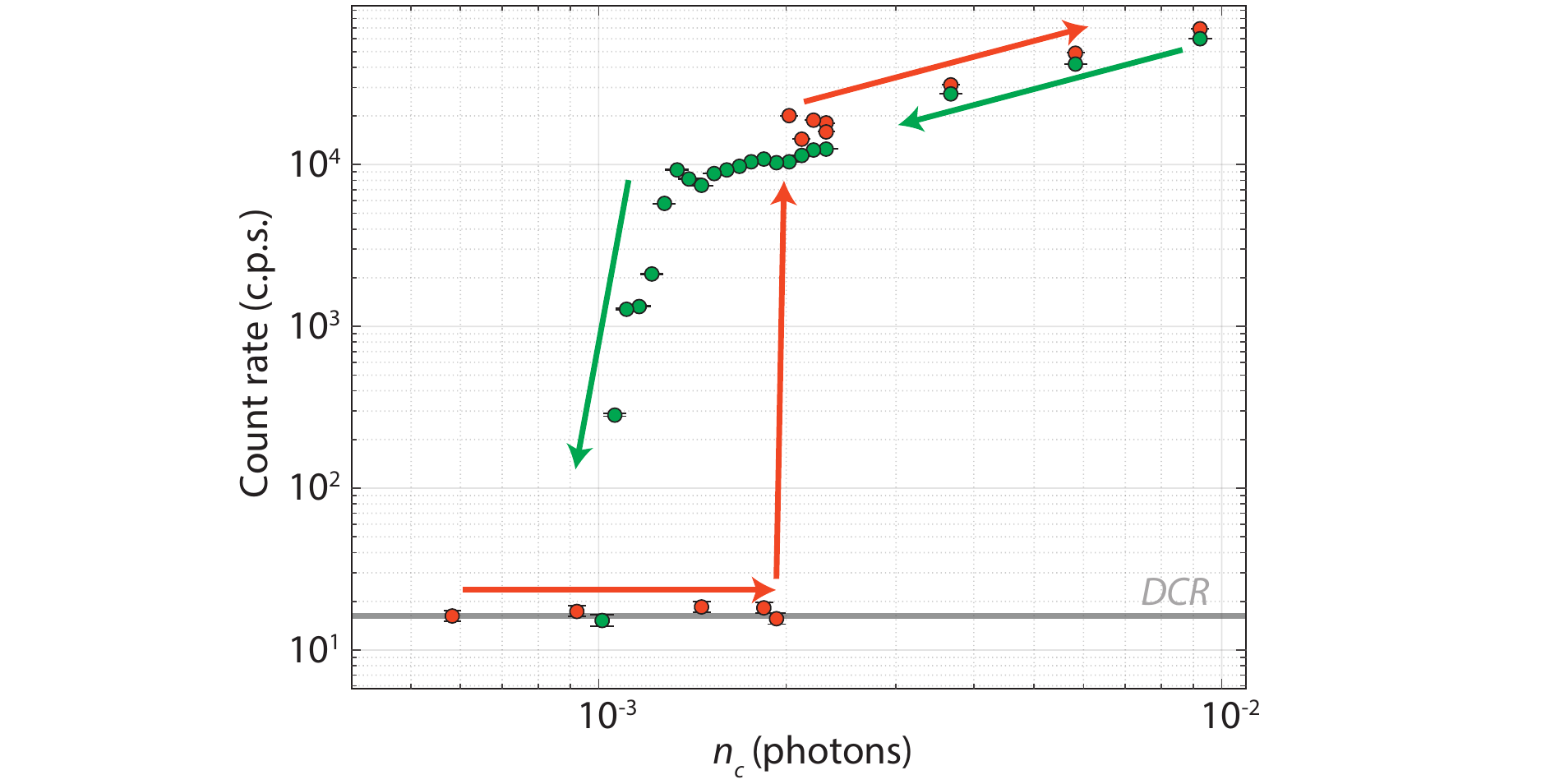}
	\caption[Low-temperature measurement of the self-oscillation threshold in a high-$Q$ nanobeam]{\textbf{Low-temperature measurement of the self-oscillation threshold in a high-$Q$ nanobeam at $\Tf = 10$~mK.} Under blue-detuned ($\Delta = - \omega_m$) driving the self-oscillation threshold is reached when $\gammaOM = \gammai$, here for $n_\text{c,thresh} = 2 \times 10^{-3}$, or $\gammaOM/2\pi = 8$~Hz, for increasing optical power (orange points). Measurements were performed on device D.}
	\label{fig:nb_highQ_selfoscillation}
	\index{figures}
	\end{center}
\end{figure}

We observe the onset of mechanical self-oscillation at $\Tf = 10$~mK, in which a CW blue-detuned pump laser drives the cavity and the sideband filters are aligned to the cavity resonance ($\Delta = 0$). The scattered photon count rate $\gammaSB$ is measured in steady-state. In the setup configuration used for these measurements, an additional VOA is placed in the optical path, elevating the measured SPD dark count rate to $10.8$~c.p.s. Sweeping the input power (photon number) $\ncav$ results in a sharp increase in detected count rate at the self-oscillation threshold $n_\text{c,thresh} = 2 \times 10^{-3}$ as shown in Figure~\ref{fig:nb_highQ_selfoscillation}, where we estimate the resulting steady-state phonon occupancy to be of order $\nbar \sim 5 \times 10^4$. At the threshold $n_\text{c,thresh}$ we can estimate the back-action amplification rate $\gammaOM/2\pi = 4 \gzero^2 n_\text{c,thresh}/\kappa \approx 8$~Hz from the known optical device parameters, indicating that the intrinsic damping $\gammai$ is dominated by the bath damping rate: $\gammap(n_\text{c,thresh}) = \gammaOM(n_\text{c,thresh}) - \gammanotO \approx 2\pi (7.9$~Hz$)$, in good quantitative agreement with the trend measured on a similar device in Figure~\ref{fig:SI_bath_properties}a. Upon decreasing the driving power (green data in Figure~\ref{fig:nb_highQ_selfoscillation}), self-oscillation appears to relax at a decreased threshold of $\ncav = 1.4 \times 10^{-3}$, indicating a hysteresis in the measured count rates as a function of input power. This apparent hysteresis likely arises from a change in the \textit{true} intracavity photon number as a function of driving power $P_\text{in}$. We have so far adhered to Equation~\ref{eqn:nc_expression} in determining the $\ncav$ as a function of $P_\text{in}$; this expression is used to generate the horizontal axis of Figure~\ref{fig:nb_highQ_selfoscillation}, and so does not represent the true intracavity photon number. However, in order to unambiguously calculate $\ncav$ in the presence of large phonon amplitude $\nbar$, a more thorough calculation is needed which accounts for the effective optical reflection profile in the presence of strong modulation by the mechanical motion.

\subsection{Electromagnetically Induced Transparency Mechanical Spectroscopy}

\begin{figure}[btp]
	\begin{center}
		\includegraphics[width=\columnwidth]{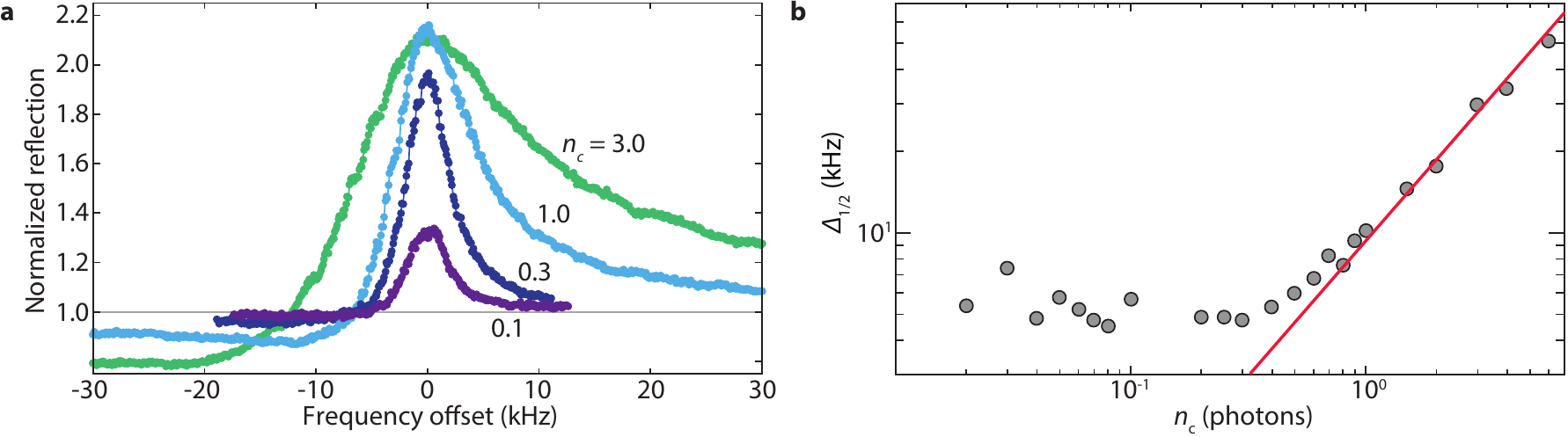}
		\caption{\textbf{Mechanical mode time-averaged linewidth versus probe power. a}, Time-averaged mechanical mode linewidth as a function of optical pump power. Broadening of the linewidth is due to optical back-action. Solid line is a fit to the back-action damping rate $\gammaOM = 4 \gzero^2 \ncav / \kappa$, from which we extract an estimated $\gzero/2\pi = 1.15$~MHz. Measurements are performed on Device D.} 
		\label{fig:SI_EITfig}
	\end{center}
\end{figure}

Electromagnetically induced transparency (EIT) in optomechanical systems allows for a spectral measurement of the mechanical response via observation of a transparency window in the optical reflection spectrum. A pump laser tone at $\omegac$ is amplitude modulated to generate a weak probe tone at ${\omegas}_{,\pm} = \omegac \pm \Delta_\text{p}$. If the pump-cavity detuning is fixed on either the red- or blue-side of the optical cavity ($\Delta = \pm \omegam$), the optical susceptibility of the cavity strongly suppresses one of the probe sidebands (at ${\omegas}_{,\mp}$) and only the other probe sideband will have an appreciable intracavity population. For a red-detuned pump, the interaction of the pump tone and mechanics with the probe sideband yields a reflection coefficienct $r(\Delta,\delta)$ for the probe which contains a transparency window having a width on the scale of the mechanical mode linewidth:

\begin{align}
r(\Delta,\delta) = 1 - \frac{\kappae}{\kappa/2 + i (\Delta - (\delta + \omegam)) + \frac{\lvert G \rvert^2}{ -i \delta + \gammai/2}},
\end{align}

\noindent where we have defined $\delta \equiv \Delta_\text{p} - \omegam$ and $G \equiv \gzero \sqrt{\ncav}$. We measure the reflection amplitude $R = \lvert r \rvert^2$ by driving an EOM weakly to generate a probe tone and observing the count rates of sideband-scattered probe photons. The pump is locked at $\Delta = +\omegam$ and the cascaded filter stack is locked to the cavity frequency. The RF modulation power is chosen to generate a sideband intracavity photon number much smaller than the carrier photon number (${\ncav}_{\text{,+}} \ll \ncav$) while maintaining a large count rate $\sim10^5$~c.p.s. at the SPDs to minimize data integration times. This corresponds to modulation indices in the range of $\beta \sim 10^{-3}$ for our system parameters (measurements were performed on device D). The modulation frequency $\Delta_\text{p}$ is swept over a range of about $1$~MHz to map out the transparency window. This range is large enough to include the optomechanically-broadened mechanical linewidth which sets the width of the transparency window, but much narrower than the bandwidth of the FFP filters ($\approx 50$~MHz), allowing for the filters to be stably locked at a single position in the center of the optical cavity line. Figure~\ref{fig:SI_EITfig} shows the normalized reflection level for various optical probe power levels $\ncav$, as well as fits to the data. The extracted total mechanical linewidth $\gamma = \gammai + \gammap + \gammaOM$ is plotted in Fig.~\ref{fig:SI_EITfig}. At low probe-power, $\gamma/2\pi$ saturates to a value $\approx 4$~kHz, which represents time-averaged broadening of the intrinsic mechanical linewidth due to mechanical frequency jitter as summarized in Fig.~\ref{fig:fig3} of the Main Text. With $\kappa$ and $\ncav$ calibrated, the linear portion of the curve which is dominated by back-action damping is fitted to extract the optomechanical coupling rate $\gzero/2\pi  = 1.15$~MHz. 

\subsection{Blue-Detuned Pumping.}

In the limit of high phonon amplitude, we perform ringdown using a pulse sequence consisting of a blue-detuned excitation pulse followed by a red-detuned readout (or \textit{probe}) pulse. Two separate laser sources are used, as shown in Figure~\ref{fig:fig2}, for generating the excitation and readout pulses in order to allow a fixed detuning of each laser to avoid instabilities associated with rapid stabilization of the laser frequency on the timescale of the pulse sequencing (tens of $\mu$s). Owing to the extremely narrow instantaneous mechanical mode linewidth, a very small back-action amplification rate $\gammaOM/2\pi \lesssim 8$~Hz is required to drive the mechanics into the self-oscillation regime. This enables operation at low driving pulse photon number $n_\text{c,blue} = 0.15 \gg n_\text{c,thresh} = 2 \times 10^{-3}$, in order to minimize the effective temperature and coupling rate of the absorption-induced phonon bath. The steady-state phonon amplitude in the presence of the driving pulse is saturated to $\nbar \approx 5 \times 10^4$. The driving pulse is turned off and after a variable delay time $\Toff$ a red-detuned pulse from the readout laser is used to probe the mode occupancy.

\begin{figure}[bpt]
	\begin{center}
	\centering
	\includegraphics[width=0.75\textwidth]{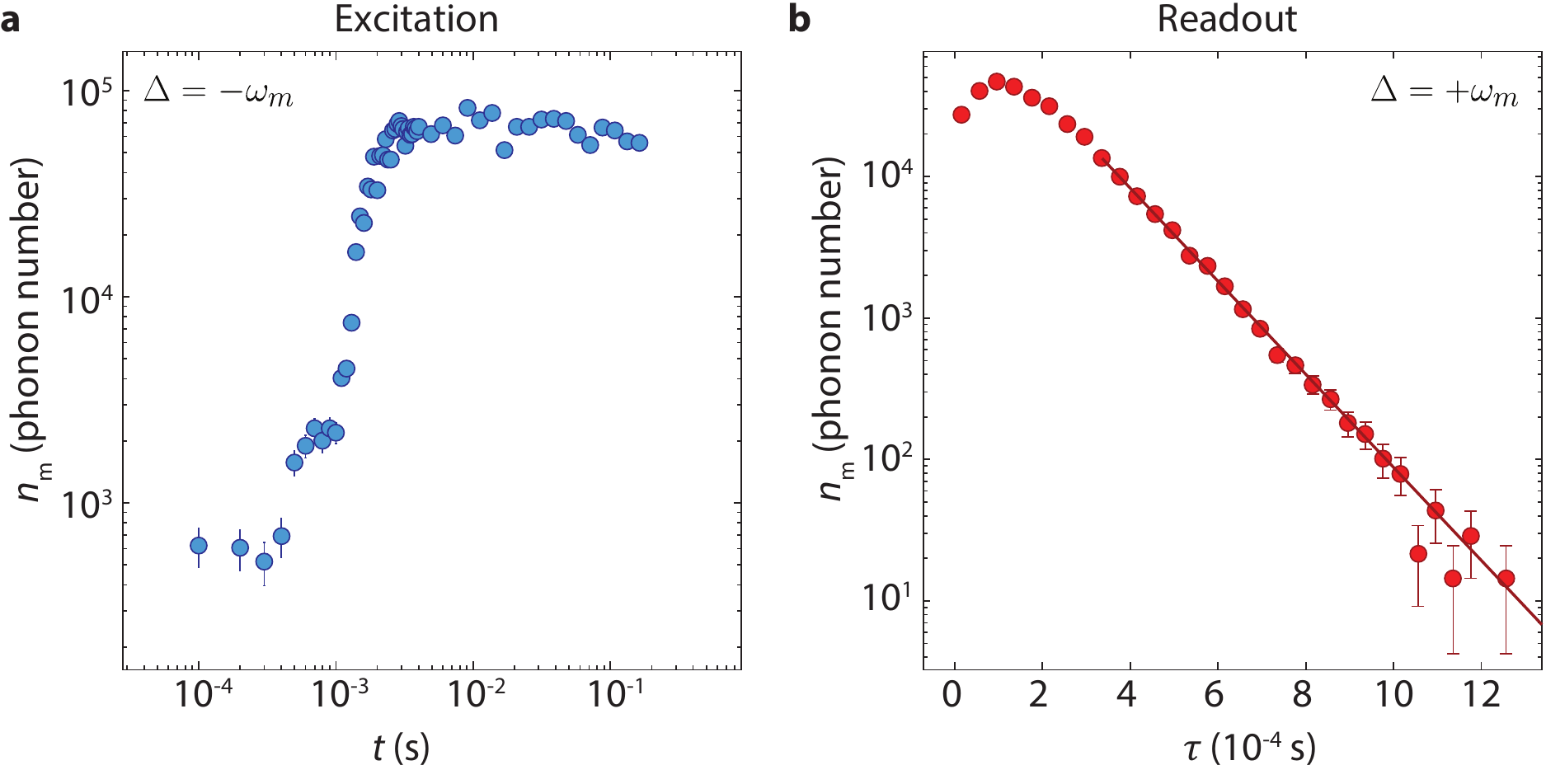}
	\caption[Mode occupancy during excitation and readout pulses for high-amplitude ringdown]{\textbf{Measurement of the mode occupancy during excitation and readout pulses for high-amplitude ringdown.} \textbf{a}, A blue-detuned ($\Delta = -\omega_m$) laser with pulse-on photon number $n_\text{c,blue} = 0.15$ drives the mechanics into self-oscillation in a timescale of $2$~ms, with a saturated phonon occupancy $\nbar \approx 5 \times 10^4$. \textbf{b}, A red-detuned ($\Delta = + \omega_m$) probe laser with pulse-on photon number $n_\text{c,red} = 0.30$ serves to reads out the mode occupancy after a variable delay time $\Toff$. Measurements are performed on Device D.}
	\label{fig:blue_red_pulseOccupancyMeas}
	\index{figures}
	\end{center}
\end{figure}

The readout photon number $n_\text{c,red} = 0.30$ is again chosen small to minimize absorption bath effects, as well as to give a total count rate $\Gamma \propto \Gamma_\text{SB,0} \nbar \propto \ncav \nbar$ within the dynamic range of the single-photon detector, which in our amplifier setup has sensitivity to a maximum count rate of $\sim 2 \times 10^6$~c.p.s. With the present setup efficiencies and device parameters (see caption of Figure~\ref{fig:blue_red_pulseOccupancyMeas}), the detected count rate is approximately $\Gamma = 31$~c.p.s. per phonon per photon at $\Delta = \pm \omega_m$, and the resulting effective upper bound on $n_\text{c,red}$ at which the SPD can efficiently detect is $2.2$~photons. Now, after the readout pulse is used to probe the mode occupancy, the mode occupancy is cooled via dynamical back-action to near its local bath temperature in preparation for the subsequent series of driving and readout pulses (effectively `re-setting' the measurement). In practice, a single red-detuned pulse is used for both readout and cooling (re-setting). In Figure~\ref{fig:blue_red_pulseOccupancyMeas} we show the phonon occupancy during both the driving pulse and the readout pulse. Note that the excited occupancy saturates to $\nbar \approx 5 \times 10^4$ from an initial occupancy $\nbar \lesssim 1 \times 10^3$, corresponding to our estimated decay ratio of $50$ from one pulse period to the next. 

\subsection{Modulated Pump-Probe Excitation and Ringdown.}

For finer control of the mechanical mode amplitude during excitation, microwave-frequency modulation of the excitation pulse was used to amplify the mechanics to a fixed phonon amplitude which is tunable by the depth of optical modulation placed on the pump laser tone. This technique allows us to probe the intrinsic energy decay constant $\gammanotO$ in the regime lying intermediate between the level of single-phonons and the saturated high-phonon-amplitude limit of self-oscillation. In this measurement scheme, a radio-frequency (RF) signal generator is used to drive an electro-optic intensity modulator (EOM) at the mechanical resonance frequency $\omega_m/2\pi$ to generate the probe sideband. The excitation pulse consists of a red-detuned pump carrier tone which is weakly modulated (RF driving power $-4$~dBm applied to an EOM with $V_{\pi} = 4.1$~V, giving a modulation index $\beta = 0.11$) to generate a probe sideband at the cavity resonance frequency. Interference between the pump carrier and probe sideband generates a time-dependent radiation pressure force at the difference frequency $\omega_m/2\pi$, which resonantly excites the acoustic mode. A second pulsed laser source is then used to generate the readout optical pulse, which is a red-detuned pulse of fixed frequency and power. Using this method as shown in Fig.~\ref{fig:fig2}, we measure an initial excited mechanical occupation of $\nbar = 1.2 \times 10^3$ having a decay rate $\gammanotO/2\pi = 0.108$~Hz at the lowest temperatures $\Tf = 10$~mK, corresponding to a mechanical $Q$-factor of $4.92 \times 10^{10}$ ($\taucoh = 1.47$~s). The decay is observed to be exponential over at least 1.5 orders of magnitude, consistent with measurements in both the high- and low-phonon number limit.

In both the case of blue-detuned driving and RF-modulated driving ring-up techniques, the total repetition rate of the pulse sequence $1/\Tper$ is fixed while only the variable delay $\Toff$ between the driving pulse and readout pulse is varied. This fixing of the overall duty-cycle of the pulse sequence is performed to eliminate systematic variations in the local absorption-induced bath temperature $\Tbath$ which, in steady-state, is expected to depend on the average power circulating in the cavity (see Fig. \ref{fig:SI_bath_properties}). We find that the measured ringdown time constant is approximately consistent over more than three orders of magnitude in starting phonon population, from $\nbar \lesssim 10$ in the case of thermally-excited ringdown measurements to $\nbar > 2 \times 10^4$ in the case of coherently-excited phonon populations.

\section{3-phonon-scattering damping model.}
\label{App:F}

The anharmonicity of the atomic lattice in solid-state materials leads to frequency mixing of the approximate harmonic modes (the phonons) of the lattice.  This frequency mixing - of all different orders - within a continuum of modes leads to different forms of phonon damping depending on the damped phonon frequency ($\omega_{\mone}$), wavelength ($\lambda_{q_{\mone}}$), and the lattice temperature~\cite{Woodruff1961,Srivastava_book}. At low temperatures where phonon relaxation times ($\tau$) are long, and at relatively high phonon frequencies, the dominant source of phonon damping due to the anharmonic lattice potential results from 3-phonon scattering processes in the so-called Landau-Rumer limit ($\omega_{\mone}\tau \gg 1$)~\cite{Landau1937,terHaar_book2013}.  In this limit a single-mode relaxation time (SMRT) approximation~\cite{Srivastava_book} can be made in which only the damped phonon mode under consideration is disturbed from equilibrium and the other two phonon modes involved in the scattering are assumed to be frozen at their equilibrium occupancies.  Using the SMRT approximation one can calculate the 3-phonon-scattering damping rate from second-order perturbation theory of the quantum mechanical model of the anharmonic lattice.  At higher temperatures where the thermal phonon relaxation rate ($1/\tau$) is very fast, or for very low frequency phonons, this approximation breaks down and one enters the Akhiezer limit of phonon damping where $\omega_{\mone}\tau \ll 1$~\cite{Akhiezer1939}.  In this limit a phenomenological model is employed in which the strain wave of a phonon mode induces a redistribution of thermal phonons via the lattice anharmoniticity, and damping occurs due to relaxation of the thermal phonons back towards thermal equilibrium.  If in addition the phonon wavelength is long relative to the mean free path of thermal phonons ($l_{th}$), then a local temperature can be defined and damping can also occur via diffusion of thermal phonons.  In this limit, $\omega_{\mone}\tau \ll 1$ and $l_{th}/\lambda_{q_{\mone}} \ll 1$, energy in the acoustic wave is carried away in heat flow due to temperature gradients on the scale of $\lambda_{q_{\mone}}$, and the resulting relaxation process is called thermoelastic damping~\cite{Lifshitz2000}.

As we are concerned with microwave frequency acoustic waves and sub-Kelvin temperatures, the dominant phonon-phonon scattering damping is expected to arise from 3-phonon scattering under the SMRT approximation.  In what follows we present a model of such Landau-Rumer damping utilizing leaky quasi-modes~\cite{Moiseyev2011}.  This quasi-mode picture arises naturally in the context of the OMC cavity structure, in which localized acoustic modes are weakly coupled to the continuum of phonon modes in the surrounding substrate via the peripheral clamping of the Si device layer to the underlying Si dioxide BOX layer (c.f., Fig.~\ref{fig:FEM_sim_layout_mesh}).  We follow closely the derivation of 3-phonon scattering in Ref.~\cite{Srivastava_book}, although with slight adjustments to the notation to accommodate the use of quasi-normal modes.  The notation developed here will also be used in the analysis of two-level system damping described in the next sub-section.      

For a displacement vector field $u_{\alpha}(\vecb{r})$, with $u$ the local amplitude of atomic displacement in direction $\alpha$ from equilibrium, the stored potential energy to second and third order in the displacement field can be written as,

\begin{equation}
\mathcal{V}_{2} \equiv \text{2nd-order elastic (potential) energy} = \frac{1}{2} \int \text{d}^3r J\indices{^{\alpha\gamma}_{\beta\delta}} \frac{\partial u_{\alpha}}{\partial r_{\beta}} \frac{\partial u_{\gamma}}{\partial r_{\delta}},
\label{eq:V2}    
\end{equation}

\noindent and

\begin{equation}
\mathcal{V}_{3} \equiv \text{third-order elastic (potential) energy} = \frac{1}{3!} \int \text{d}^3r A\indices{^{lmn}_{ijk}} \frac{\partial u_{l}}{\partial r_{i}} \frac{\partial u_{m}}{\partial r_{j}} \frac{\partial u_{n}}{\partial r_{k}}.
\label{eq:V3}    
\end{equation}

\noindent Here, $J\indices{^{\alpha\gamma}_{\beta\delta}}$ is in general a rank 4 tensor whose coefficients are the 2nd-order elastic coefficients of the material which relate strain to stress and have units of energy density.  $A\indices{^{lmn}_{ijk}}$ is rank 6 tensor with coefficients arising from the lowest order anharmoniticity of the lattice.  $\partial u_{\alpha/\partial r_{\beta}}$ is a rank 2 tensor representing the local strain created by the displacement vector field $u_{\alpha}(\vecb{r})$.

From these expression we can define the total elastic energy density for a classical acoustic wave oscillating harmonically in mode $s$ as,

\begin{equation}
h(\rttensortwo{e}_{\mone}(\vecb{r})) \equiv \text{(classical) strain field elastic energy density} = \frac{1}{2} J\indices{^{\alpha\gamma}_{\beta\delta}} \left(e_{\mone}(\vecb{r})\right)\indices{_\alpha^\beta} \left(\left(e_{\mone}(\vecb{r})\right)\indices{_\gamma^\delta}\right)^{*},  
\label{eq:strainEdensity}    
\end{equation}

\noindent where $\rttensortwo{e}_{\mone}(\vecb{r})$ is a complex strain tensor field related to the real (physical) strain tensor field by,

\begin{equation}
\frac{\partial \left(u_{\mone}(\vecb{r})\right)_{i}}{\partial r_{j}}  \equiv \text{(classical) strain tensor field of mode $s$} \equiv \Re\left(\left(e_{\mone}(\vecb{r})\right)\indices{_i^j}\right)  = \Re\left(\rttensortwo{e}_{\mone}(\vecb{r})\right). 
\label{eq:straintensor}    
\end{equation}

\noindent Note that we have used the fact that for a harmonic wave the cycle averaged potential and kinetic energies are equal (and thus the total wave energy is twice the potential energy), and $h(\rttensortwo{e}_{\mone}(\vecb{r}))$ should therefore be strictly considered as the energy density averaged over a single cycle in time and a single wavelength in space.  We also define a normalized complex strain field for mode $s$ having a peak strain value of approximately unity (exactly unit for tensor-averaged fields) and a peak energy density of $\bar{J}$,

\begin{equation}
\overline{\rttensortwo{e}_{\mone}(\vecb{r})} \equiv \text{normalized classical strain field for mode $s$} = \frac{(\bar{J})^{1/2}\rttensortwo{e}_{\mone}(\vecb{r})}{\left( \max[h(\rttensortwo{e}_{\mone}(\vecb{r}))] \right)^{1/2} },
%\left(\int h(\rttensortwo{e}_{\mone}(\vecb{r})) \text{d}^3 r \right)^{1/2}
\label{eq:normstrainfield}    
\end{equation}

\noindent where $\bar{J}$ is the tensor-average of the harmonic elastic coefficients,

\begin{equation}
\bar{J} \equiv \text{tensor-averaged 4th-order elastic tensor} = \tensoravg{\frac{1}{2}\rttensortwo{J}} = \frac{1}{2 \cdot 3^4} \left(\sum_{\alpha,\gamma,\beta,\delta} \left( J \indices{^{\alpha\gamma}_{\beta\delta}}\right)^2 \right)^{1/2}.
\label{eq:Javg}    
\end{equation}

\noindent The effective mode volume over which the strain energy of mode $s$ is localized can also be defined as,

\begin{equation}
V_{\mone} \equiv \text{effective mode volume of mode $s$} = \frac{\int h(\rttensortwo{e}_{\mone}(\vecb{r})) \text{d}^3r}{\max[h(\rttensortwo{e}_{\mone}(\vecb{r}))]}.  
\label{eq:modevolume}
\end{equation}

The peak strain amplitude of mode $s$ containing half a quanta of energy, i.e., the `vacuum' strain level, is given by,

\begin{equation}
e_{\text{vac,}s} \equiv \text{vacuum strain field amplitude for mode $s$} = \sqrt{\frac{\hbar \omega_{\mone}}{2 \bar{J} V_{\mone}}}, 
\label{eq:vacstrain}    
\end{equation}

\noindent where $\hbar\omega_{\mone}$ is the mode $s$ energy quantum.  From the peak strain amplitude of vacuum and the normalized strain field we can define a quantum strain field operator,

\begin{equation}
\hat{\rttensortwo{e}}_{\mone}(\vecb{r}) \equiv \text{quantum strain tensor field operator for mode $s$} = (e_{\text{vac,}s})\left[ \hat{b}_{\mone} \overline{\rttensortwo{e}_{\mone}(\vecb{r})} + \hat{b}^{\dagger}_{\mone} \left(\overline{\rttensortwo{e}_{\mone}(\vecb{r})}\right)^{*}  \right]  
\label{eq:qstrainfield}    
\end{equation}

\noindent where $\hat{b}_{\mone}$ and $\hat{b}^{\dagger}_{\mone}$ annihilate and create individual phonon quanta in mode $s$.  The corresponding quantum interaction Hamiltonian for 3-phonon scattering can then be written in terms of triplets of quantum strain field operators directly from the third-order elastic potential energy relation in Eq.~(\ref{eq:V3}),

\begin{equation}
\hat{\mathcal{H}}_{\text{3-ph}} \equiv \text{3-phonon interaction Hamiltonian} = \frac{1}{3!} \sum_{s\, \mtwo \mthr} \int \text{d}^3r \left( A\indices{^{lmn}_{ijk}}\right) \left(\hat{\rttensortwo{e}}_{\mone}(\vecb{r}) \right)\indices{_{l}^{i}} \left(\hat{\rttensortwo{e}}_{\mtwo}(\vecb{r}) \right)\indices{_{m}^{j}} \left( \hat{\rttensortwo{e}}_{\mthr}(\vecb{r}) \right)\indices{_{n}^{k}}. 
\label{eq:H3ph}    
\end{equation}

\subsection{Type-I scattering processes}

3-phonon scattering, as it pertains to damping of a particular mode $s$, can be categorized into two classes of processes~\cite{Srivastava_book}.  Type-I scattering involves the mode of interest, mode $\mone$, as a `daughter' phonon which combines with another `sibling' phonon (mode $\mtwo$) to create a higher frequency `parent' phonon (mode $\mthr$).  The reverse process is also of type-I.  Type-II scattering has the mode $\mone$ of interest as the high frequency parent phonon.  The 3-phonon interaction Hamiltonian for type-I scattering is given in terms of phonon creation and annihilation operators as,

\begin{equation}
\hat{\mathcal{H}}^{\text{3-ph}}_{s+\mtwo \rightleftharpoons \mthr} \equiv A^{\mthr}_{s  \mtwo } \hat{b}_{\mone} \hat{b}_{\mtwo} \hat{b}^{\dagger}_{\mthr} +  A_{\mthr}^{s  \mtwo } \hat{b}^{\dagger}_{\mone} \hat{b}^{\dagger}_{\mtwo} \hat{b}_{\mthr},
\label{eq:H3phI}    
\end{equation}

\noindent where

\begin{equation}
A^{\mthr}_{s  \mtwo } = \left( A_{\mthr}^{s  \mtwo } \right)^{*} \equiv \left[(e_{\text{vac,}{\mone}}) (e_{\text{vac,}{\mtwo}}) (e_{\text{vac,}{\mthr}})\right]  \int  \left( A\indices{^{lmn}_{ijk}}\right) \left(\overline{\rttensortwo{e}_{\mone}(\vecb{r})}\right)\indices{_{l}^{i}} \left(\overline{\rttensortwo{e}_{\mtwo}(\vecb{r})}\right)\indices{_{m}^{j}} \left(\left(\overline{\rttensortwo{e}_{\mthr}(\vecb{r})}\right)^{*}\right)\indices{_{n}^{k}}  \text{d}^3r. 
\label{eq:A3phI}    
\end{equation}

Calculating to 2nd-order in perturbation theory, the energy shift in the phonon Fock state $| n_{\mone}, n_{\mtwo}, n_{\mthr}\rangle$ is given by,

\begin{multline}
\left\langle\left(\delta E_{n_{\mone},n_{\mtwo},n_{\mthr}}\right)^{\text{I}}_{\text{3-ph}}\right\rangle = \sum_{\mtwo \mthr} \left[ \frac{\left|\left\langle n_{\mone}-1, n_{\mtwo}-1, n_{\mthr}+1 | A^{\mthr}_{s  \mtwo } \hat{b}_{\mone} \hat{b}_{\mtwo} \hat{b}^{\dagger}_{\mthr} | n_{\mone}, n_{\mtwo}, n_{\mthr} \right\rangle\right|^2}{ \hbar\left((\omega_{\mone} + \omega_{\mtwo} - \omega_{\mthr}) - i(\Gamma_{\mone} + \Gamma_{\mtwo} - \Gamma_{\mthr})\right)} \right.\\
+ \left. \frac{\left|\left\langle n_{\mone}+1, n_{\mtwo}+1, n_{\mthr}-1 | A_{\mthr}^{s  \mtwo } \hat{b}^{\dagger}_{\mone} \hat{b}^{\dagger}_{\mtwo} \hat{b}_{\mthr} | n_{\mone}, n_{\mtwo}, n_{\mthr} \right\rangle\right|^2}{ \hbar\left(( \omega_{\mthr} - \omega_{\mone} - \omega_{\mtwo}) - i(\Gamma_{\mthr} - \Gamma_{\mone} - \Gamma_{\mtwo})\right)}\right].
\label{eq:deltaE3phIgen}    
\end{multline}

\noindent Note that we have used a modified form of non-Hermitian perturbation theory~\cite{Moiseyev2011,Field1993}, suitable for leaky quasi-normal modes, in which the finite linewidth of the phonon quasi-modes are included in the denominator of Eq.~(\ref{eq:deltaE3phIgen}).  Also implicit in our use of 2nd-order perturbation theory is that the phonon-phonon coupling is weak.  Collecting terms and specifically identifying mode $\mone$ as the breathing acoustic mode of the OMC cavity (labeled by $m$), we have for the complex level shift,

\begin{equation}
\left\langle\left(\delta E_{n_{m},n_{\mtwo},n_{\mthr}}\right)^{\text{I}}_{\text{3-ph}}\right\rangle =  \sum_{\mtwo \mthr} \left(\frac{\left|A^{\mthr}_{m\, \mtwo}\right|^2 \left[ n_{m} n_{\mtwo} (n_{\mthr} + 1) - (n_{m}+1)(n_{\mtwo}+1)n_{\mthr} \right]}{\hbar\left((\omegam + \omega_{\mtwo} - \omega_{\mthr}) - i(\Gamma_{m} + \Gamma_{\mtwo} - \Gamma_{\mthr})\right)}\right). 
\label{eq:deltaE3phI}    
\end{equation}

We now invoke the single mode relaxation time approximation, and assume that only mode $m$ is perturbed from equilibrium,

\begin{equation}
\left\langle\left(\delta E_{m,\mtwo,\mthr}\right)^{\text{I}}_{\text{3-ph}}\right\rangle_{\text{smrt}} =  \sum_{\mtwo \mthr} \left(\frac{\left|A^{\mthr}_{m\, \mtwo}\right|^2 \left[ (\bar{n}_{m} + \delta n_{m}) \bar{n}_{\mtwo} (\bar{n}_{\mthr} + 1) - (\bar{n}_{m} + \delta n_{m}+1)(\bar{n}_{\mtwo}+1)\bar{n}_{\mthr} \right]}{\hbar\left((\omegam + \omega_{\mtwo} - \omega_{\mthr}) - i(\Gamma_{m} + \Gamma_{\mtwo} - \Gamma_{\mthr})\right)}\right), 
\label{eq:deltaE3phIsmrt}    
\end{equation}

\noindent where $\bar{n}$ are the thermal equilibrium mode occupancies and $\delta n_{m}$ is the perturbation in phonon number of the breathing mode from equilibrium.  Taking the difference between the complex level shift for $\delta n_{m}+1$ and $\delta n_{m}$, we find for the single photon energy shift in mode $m$,   

\begin{equation}
\left\langle\hbar\left(\delta \tildeomegam \right)^{\text{I}}_{\text{3-ph}}\right\rangle_{\text{smrt}} =  \sum_{\mtwo \mthr} \left(\frac{\left|A^{\mthr}_{m\, \mtwo}\right|^2 \left[ \bar{n}_{\mtwo}  - \bar{n}_{\mthr} \right]}{\hbar\left((\omegam + \omega_{\mtwo} - \omega_{\mthr}) - i(\Gamma_{m} + \Gamma_{\mtwo} - \Gamma_{\mthr})\right)}\right).
\label{eq:deltaE3phIsmrt2}    
\end{equation}

\noindent Assuming the phonon mode linewidths are energy-damping limited ($\Gamma_{s} = \gamma_{s}/2$), we find for the change in the energy damping rate of mode $m$ due to type-I 3-phonon scattering, 

\begin{align}
\left\langle\left(\delta\gamma_{m}\right)^{\text{I}}_{\text{3-ph}}\right\rangle_{\text{smrt}} \equiv \frac{-2\Im \left\langle\hbar\left(\delta \tildeomegam \right)^{\text{I}}_{\text{3-ph}}\right\rangle_{\text{smrt}}}{\hbar \delta n_{m}}  
& = \frac{1}{\hbar^2} \sum_{\mtwo \mthr} \left(\frac{\left| A^{\mthr}_{m\, \mtwo} \right|^2(\gamma_{\mthr} - \gamma_{\mtwo} - \gamma_{m}) \left[ \bar{n}_{\mtwo} - \bar{n}_{\mthr} \right] }{(\omegam + \omega_{\mtwo} - \omega_{\mthr})^2 + (\frac{\gamma_{m} + \gamma_{\mtwo} - \gamma_{\mthr}}{2})^2}\right) \nonumber \\
& \approx \frac{1}{\hbar^2} \sum_{\mtwo \mthr} \left(\frac{\left| A^{\mthr}_{m\, \mtwo} \right|^2(\gamma_{\mthr}) \left[ \bar{n}_{\mtwo} - \bar{n}_{\mthr} \right] }{(\omegam + \omega_{\mtwo} - \omega_{\mthr})^2 + (\frac{\gamma_{m} + \gamma_{\mtwo} - \gamma_{\mthr}}{2})^2}\right),
\label{eq:gammam3phI_sum}    
\end{align}

\noindent where in the last approximate equality we have neglected the unperturbed damping of the two `child' phonons ($m$ and $\mtwo$) and only included the quasi-mode damping of the `parent' phonon ($\mthr$) into which the mode $m$ decays in the type-I process.

%\begin{equation}
%\left\langle\left(\delta \gamma_{m}\right)^{\text{I}}_{\text{3-ph}}\right\rangle_{\text{smrt}} = \frac{1}{\hbar^2} \sum_{\mtwo %\mthr} \left(\frac{\left| A^{\mthr}_{m\, \mtwo} \right|^2(\gamma_{\mthr}/2 - \gamma_{m}) \left[ \bar{n}_{\mtwo} - \bar{n}_{\mthr} %\right] }{(\omegam + \omega_{\mtwo} - \omega_{\mthr})^2 + (\frac{\gamma_{m} + \gamma_{\mtwo} - \gamma_{\mthr}}{2})^2}\right) 
%\label{eq:gammam3phI_sum}    
%\end{equation}

% \begin{equation}
% (\delta \omegam)^{\text{I}}_{\text{3-ph}} \equiv \frac{\Re \left\langle\left(\delta E_{m}\right)^{\text{I}}_{\text{3-ph}}\right\rangle_{\text{smrt}}}{\hbar (\bar{n}_{m}+\delta n_{m})}  = \frac{(1+\bar{n}_{m}/\delta n_{m})^{-1}}{\hbar^2} \sum_{\mtwo \mthr} \left( \frac{\left|A^{\mthr}_{m\, \mtwo}\right|^2 (\omegam + \omega_{\mtwo} - \omega_{\mthr}) \left[ n_{\mtwo} - n_{\mthr} \right] }{(\omegam + \omega_{\mtwo} - \omega_{\mthr})^2 + (\frac{\Gamma_{m} + \Gamma_{\mtwo} - \Gamma_{\mthr}}{2})^2}\right)
% \label{eq:df3phI}    
% \end{equation}

Equation~(\ref{eq:gammam3phI_sum}]) can be approximately evaluated using the relation between the `mode-averaged' ($\langle \cdot \rangle_{\text{m}}$) third-order elastic constants and the Gr\"{u}neisen parameter~\cite{Srivastava_book},

\begin{equation}
\left\langle \left|A\indices{^{lmn}_{ijk}}\right|^2 \right\rangle_{\text{m}} \equiv \text{mode-averaged three-phonon scattering strength} \approx 4\rhoSi^2\vSibar^4\grun^2
\label{eq:avg3phscatt}    
\end{equation}

\noindent where mode averaging is taken over different bulk phonon mode directions and polarizations.  This allows us to write for the 3-phonon scattering amplitude,

\begin{equation}
A^{\mthr}_{m  \mtwo } \approx \left(2\rhoSi\vSibar^2\grun \right) \left[(e_{\text{vac,}{m}}) (e_{\text{vac,}{\mtwo}}) (e_{\text{vac,}{\mthr}})\right]\left(\GammathreephI \Vm \right),
\label{eq:A3approx}    
\end{equation}

\noindent in which $\GammathreephI$ is a mode overlap factor, or equivalently, for plane wave modes, a phase-matching term.  The mode overlap factor is less than or equal to unity and depends approximately upon the tensor-averaged normalized strain fields of each of the three modes participating in the scattering process,

\begin{equation}
\GammathreephI = \frac{1}{\Vm}\int  \left(\left\langle \overline{\rttensortwo{e}_{m}(\vecb{r})} \right\rangle_{\text{t}} \right) \left(\tensoravg{ \overline{\rttensortwo{e}_{\mtwo}(\vecb{r})} } \right) \left(\left\langle \overline{\rttensortwo{e}_{\mthr}(\vecb{r})} \right\rangle_{\text{t}} \right)^{*} \text{d}^3r \lesssim 1. 
\label{eq:Gamma3phI}    
\end{equation}

\noindent As with the elastic constants, we define a tensor-averaged strain field as,

\begin{equation}
\langle \rttensortwo{e}_{\mone}(\vecb{r}) \rangle_{\text{t}} \equiv \text{tensor-averaged local strain field amplitude of mode $s$}= \frac{1}{3^2} \left(\sum_{i,j} \left(e_{\mone}(\vecb{r})\indices{_i^j}\right)^2 \right)^{1/2}.
\label{eq:avstsq}    
\end{equation}

\subsection{Type-II scattering processes}

Following a similar procedure for type-II scattering ($m \rightleftharpoons \mtwo + \mthr$) yields a complex energy level shift in Fock state $|n_{m},n_{\mtwo},n_{\mthr} \rangle$,

\begin{equation}
\left\langle\left(\delta E_{m,\mtwo,\mthr}\right)^{\text{II}}_{\text{3-ph}}\right\rangle =  \frac{1}{2} \sum_{\mtwo \mthr} \left(\frac{\left|A^{m}_{\mtwo \mthr}\right|^2 \left[ n_{\mtwo} n_{\mthr} (n_{m} + 1) - (n_{\mtwo}+1)(n_{\mthr}+1)n_{m} \right]}{\hbar\left((\omega_{\mtwo} + \omega_{\mthr} - \omegam) - i(\Gamma_{\mtwo} + \Gamma_{\mthr} - \Gamma_{m})\right)}\right). 
\label{eq:deltaE3phII}    
\end{equation}

\noindent Assuming the SMRT approximation and taking the difference between the energy level shifts for displaced phonon numbers of $\delta n_{m} + 1$ and $\delta n_{m}$ yields the single photon complex energy level shift,

\begin{equation}
\left\langle \hbar \left(\delta \tildeomegam \right)^{\text{II}}_{\text{3-ph}}\right\rangle_{\text{smrt}} =  \frac{1}{2} \sum_{\mtwo \mthr} \left(\frac{\left|A^{m}_{\mtwo \mthr}\right|^2 \left[ - ( 1 + \bar{n}_{\mtwo} + \bar{n}_{\mthr}) \right]}{\hbar\left((\omega_{\mtwo} + \omega_{\mthr} - \omegam) - i(\Gamma_{\mtwo} + \Gamma_{\mthr} - \Gamma_{m})\right)}\right), 
\label{eq:deltaE3phIIsmrt}    
\end{equation}

\noindent and the perturbation in the energy damping rate of mode $m$ due to type-II 3-phonon scattering,

\begin{equation}
\left\langle \left(\delta\gamma_{m}\right)^{\text{II}}_{\text{3-ph}} \right\rangle_{\text{smrt}} = \frac{1}{2\hbar^2} \sum_{\mtwo \mthr} \left(\frac{\left| A^{m}_{\mtwo \mthr}  \right|^2(\gamma_{\mtwo} + \gamma_{\mthr} -\gamma_{m})  \left[ 1 + \bar{n}_{\mtwo} + \bar{n}_{\mthr}\right]  }{( \omega_{\mtwo} + \omega_{\mthr} - \omegam )^2 + (\frac{\gamma_{\mtwo} + \gamma_{\mthr} - \gamma_{m}}{2})^2}\right).
\label{eq:gammam3phII_sum}    
\end{equation}

\noindent In Section~\ref{sec:numerical_modeling_damping} we use Eqs.~(\ref{eq:gammam3phI_sum}) and (\ref{eq:gammam3phII_sum}) to numerically evaluate the expected damping of the breathing mode due to 3-phonon scattering with numerically simulated quasi-normal modes of the OMC cavity.

% \begin{equation}
% (\delta \omegam)^{\text{II}}_{\text{3-ph}} = \frac{(1+\bar{n}_{m}/\delta n_{m})^{-1}}{2\hbar^2} \sum_{\mtwo \mthr} \left( \frac{ \left| A^{m}_{\mtwo \mthr}  \right|^2 ( \omegam - \omega_{\mtwo} - \omega_{\mthr} )  \left[ 1 + \bar{n}_{\mtwo} + \bar{n}_{\mthr}\right]   }{( \omega_{\mtwo} + \omega_{\mthr} - \omegam )^2 + (\frac{\Gamma_{\mtwo} + \Gamma_{\mthr} - \Gamma_{m}}{2})^2}\right)
% \label{eq:df3phII}    
% \end{equation}

\subsection{3-phonon scattering in bulk Si}

In order to compare the estimated 3-phonon scattering in the restricted geometry of the OMC cavity to that of a bulk material, here we consider a simplified model of 3-phonon scattering in bulk Si in which we treat Si as an isotropic acoustic material.  We are primarily interested in Normal ($\mathcal{N}$), type-I scattering processes.  $\mathcal{N}$ processes due to the low temperature, and thus low frequency of the acoustic phonons involved in the scattering, and type-I scattering due to the suppression of type-II scattering processes in the effectively one-dimensional OMC cavity for phonon frequencies below that of the breathing mode.  As derived in Ref.~\cite{Srivastava_book}, for such a bulk material system the acoustic damping of mode $\mone$ under the single mode relaxation time approximation can be written as,

\begin{equation}
\left(\gamma_{\mone}\right)^{\mathcal{N}-\text{I}}_{\text{3-ph,bulk}}  = \frac{\hbar v_{p}  \qavg{\grun}^2 }{4\pi \rhoSi  \qavg{\vSibar^2}} \sum_{p^{'} p^{''}} \left(v_{\ptwo}^2 v_{\pthr}^2\right)^{-1} \int\limits_{\mathcal{R}[\ptwo,\pthr]} \mathrm{d}\omega_{\mtwo} \left(\omega_{\mtwo}\right)^2\left(\omega_{\mtwo} + \omega_{\mone}\right)^2\left( \bar{n}[\hbar\omega_{\mtwo}/k_B T] - \bar{n}[\hbar(\omega_{\mtwo}+\omega_{\mone})/k_B T] \right),
\label{eq:gamma3phI_bulkSi}    
\end{equation}

\noindent where $\pone$, $\ptwo$, and $\pthr$ label the acoustic polarization of the modes $\mone$, $\mtwo$, and $\mthr$, respectively.  The region of integration, $\mathcal{R}[\pone,\ptwo]$, depends upon the acoustic velocity dispersion version frequency and polarization (having assumed an isotropic bulk model there is no directional dispersion).  Rewriting in terms of normalized frequencies ($y=\omega(\hbar/k_{B}T)$), yields the following simplified form of the phonon damping, 

\begin{equation}
\left(\gamma_{\mone}\right)^{\mathcal{N}-\text{I}}_{\text{3-ph,bulk}}  = \frac{\hbar v_{p}  \qavg{\grun}^2 (k_{B}T/\hbar)^5}{4\pi \rhoSi  \qavg{\vSibar^2}} \sum_{p^{'} p^{''}} \left(v_{\ptwo}^2 v_{\pthr}^2\right)^{-1} \left\{ \int\limits_{\mathcal{R}[\ptwo,\pthr,T]} \mathrm{d}y_{\mtwo} \left(y_{\mtwo}\right)^2\left(y_{\mtwo} + y_{\mone}\right)^2\left( \bar{n}[y_{\mtwo}] - \bar{n}[y_{\mtwo}+y_{\mone}] \right) \right\}.
\label{eq:gamma3phI_bulkSi_simple}    
\end{equation}

\noindent The integral in curly brackets is unitless and depends on temperature through both the integration range and the constant $y_{\mone} = \hbar\omega_{\mone}/k_{B}T$.

Neglecting frequency dispersion, $\mathcal{R}[\ptwo,\pthr]$ takes on a relatively simple form for the various acoustic polarization scenarios.  Due to polarization dispersion, the only allowed Normal, type-I scattering processes are: $L_{\mone} + L_{\mtwo} \rightleftharpoons L_{\mthr}$, $L_{\mone} + T_{\mtwo} \rightleftharpoons L_{\mthr}$, $T_{\mone} + L_{\mtwo} \rightleftharpoons L_{\mthr}$, and $T_{\mone} + T_{\mtwo} \rightleftharpoons L_{\mthr}$, where $L$ ($T$) corresponds to longitudinal (transverse/shear) polarization acoustic waves.  The breathing mode is of mixed polarization character, so all four combinations are potentially relevant for comparison to the numerical calculations performed using the quasi-modes of the OMC structure.  Defining normalized wavevector magnitudes for the three acoustic waves ($\xone=|\vecb{q}_{\mone}|/q_D$, $\xtwo=|\vecb{q}_{\mtwo}|/q_D$, $\xthr=|\vecb{q}_{\mthr}|/q_D$), the integration range $\mathcal{R}[\pone,\ptwo]$ for $\xtwo$ is given by:

\begin{align}
    L_{\mone} + L_{\mtwo} \rightleftharpoons L_{\mthr}\text{: } & \xtwo = \{0,1-x \} \rightarrow \omega_{\mtwo} \simeq \{0,\omega_{D} \}, \label{eq:LLLrange} \\
    T_{\mone} + L_{\mtwo} \rightleftharpoons L_{\mthr}\text{: } & \xtwo = \{x_{1}x,1-rx \} \rightarrow \omega_{\mtwo} \simeq \{(x_{1}/r)\omega_{\mone},\omega_{D} \}, \label{eq:TLLrange}\\
    T_{\mone} + T_{\mtwo} \rightleftharpoons L_{\mthr}\text{: } & \xtwo = \{ x_{1}x,x/x_{1} \} \rightarrow \omega_{\mtwo} \simeq \{(x_{1}/r)\omega_{\mone},(1/x_{1}r)\omega_{\mone} \}, \label{eq:TTLrange}\\
    L_{\mone} + T_{\mtwo} \rightleftharpoons L_{\mthr}\text{: } & \xtwo = \{ 0,x/x_{3} \} \rightarrow \omega_{\mtwo} \simeq \{0,(1/x_{3}r)\omega_{\mone} \}, \label{eq:LTLrange}
\end{align}

\noindent where $q_{D}=\pi/a$ is the Debye wavevector for an atomic lattice constant $a$, $\omega_{D} = q_{D}\qavg{\vSibar}$, $r = \vSit/\vSil$ ($\approx 0.69$ in the [100] direction), $x_{1} = (1-r)/(1+r) \approx 0.18$, and $x_{3} = (1-r)/2 \approx 0.15$.  For a breathing mode at $5$~GHz, the corresponding lower frequency cut-off for the integration range of both $TLL$ and $TTL$ scattering would approximately be $1.3$~GHz.  The upper frequency cut-off of $TTL$ and $LTL$ scattering would be $40$~GHz and $47$~GHz, respectively. At temperatures below approximately $50$~mK these processes would turn off, and they would saturate for temperatures above approximately $2$~K.  In what follows we consider the $LLL$ scattering combination for comparison as it has effectively unlimited integration range and thus will contribute at both low and high temperatures relative to $T = \hbar\omega_{\mone}/k_{B}$ ($\approx 200$~mK for the breathing mode).   

\section{TLS damping model.}
\label{App:G}

\begin{figure*}[btp]
\begin{center}
\includegraphics[width=0.75\columnwidth]{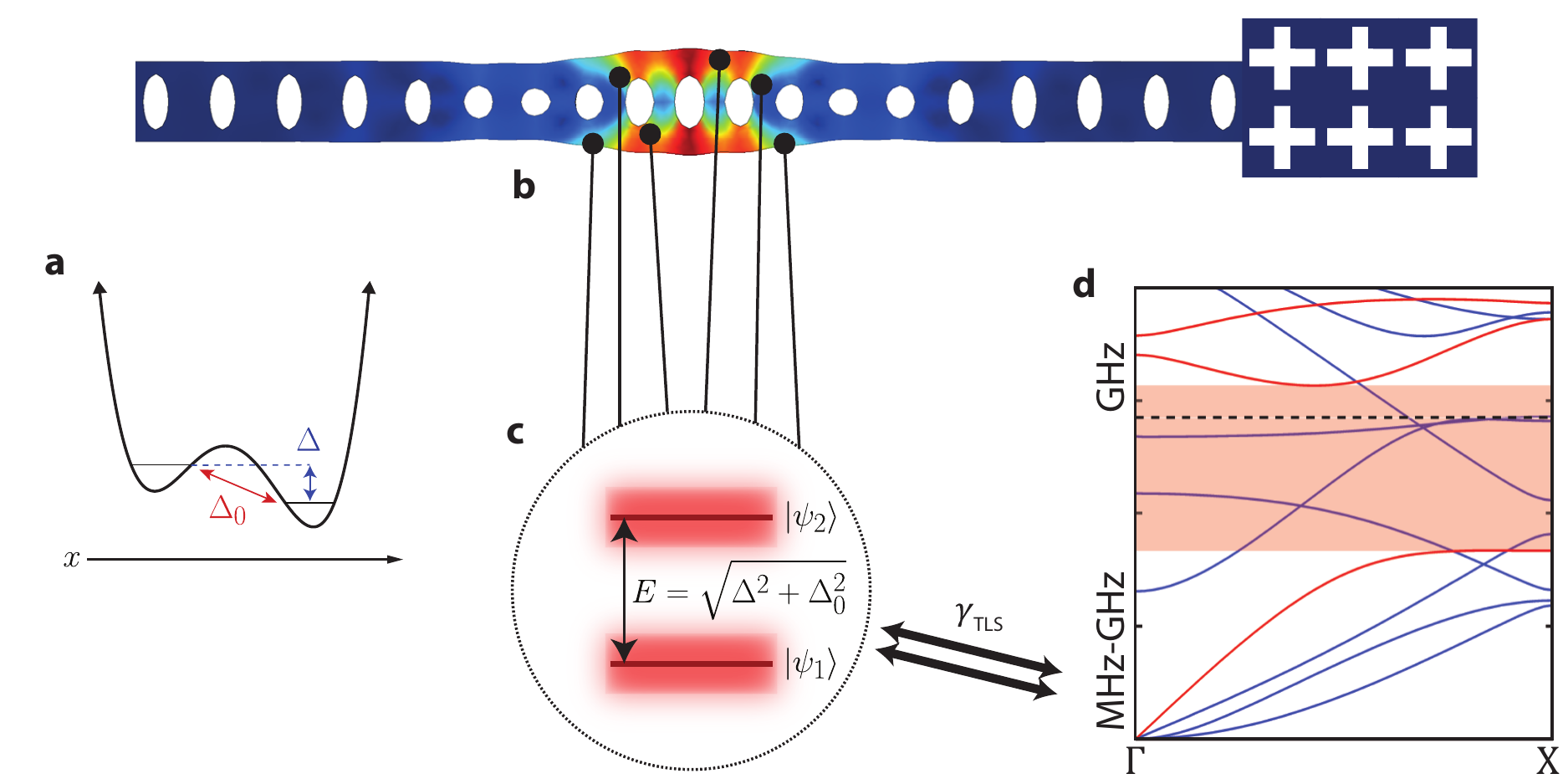}
\caption{\textbf{Simulation of TLS strain coupling to OMC cavity.} \textbf{a}, Double potential well energy profile of a tunneling-state (TS), showing the asymmetry energy $\Delta$ and the tunneling energy $\Delta_{0}$ between left and right localized potentials. \textbf{b} FEM-simulated mode profile of the breathing mode of the OMC cavity, indicating surface-localized TS states which are strain coupled through deformation potential $\bar{\gamma}_{\text{TLS}}$.  \textbf{c}, TS energy diagram, showing the transition energy, $E=(\Delta^2 + \Delta_{0}^2)^{1/2}$, between hybridized modes $|\psi_{1}\rangle$ and $|\psi_{2}\rangle$ which are mixtures of left and right localized states of the TS double potential well.  \textbf{d},  Acoustic bandstructure of the OMC nanobeam, with blue (red) bands correspond to even (odd) vector parity acoustic modes with respect to in-plane mirror symmetry.  The dashed black curve corresponds to the localized breathing mode frequency.  The shaded orange region corresponds to the bandgap of the surrounding acoustic shield.   
}  
\label{fig:SI_TLScartoon}
\end{center}
\end{figure*}

In addition to phonon-phonon scattering, another possible form of damping for the acoustic breathing mode is due to coupling to tunneling states (TS) or two-level systems (TLS).  TS (or similar TLS) states correspond to a generic defect state in a solid-state material, typically an amorphous material, in which two local arrangements of atoms are nearly degenerate in energy.  The two different atomic arrangements can have both a permanent electric and acoustic dipole associated with them, and atoms can tunnel between the two different arrangements.  The TS and TLS models are two different phenomelogical models that are used to describe a wide variety of microscopic situations.  Generically, in the TS model one has an asymmetry energy $\Delta$ which corresponds to the energy difference between the lowest energy level in each of the local potential energy profiles defining the two independent atomic arrangements, and a tunneling energy $\Delta_{0}$ related to the energy barrier between the two local atomic arrangements (see Fig.~\ref{fig:SI_TLScartoon}(a)).  One diagonalizes the two lowest energy states of the two atomic arrangements into hybridized modes $|\psi_{1}\rangle$ and $|\psi_{2}\rangle$, whose energy difference is dependent upon a longitudinal dipole matrix element and which can be coupled via a transition dipole matrix element.  In the TS model the ratio of the longitudinal dipole coupling to transition dipole coupling strength depends on the ratio of asymmetry energy to tunneling energy.  The TLS model treats the longitudinal and transition dipole couplings as independent.   

In the diagonal basis of the TS with asymmetry energy $\Delta$ and tunneling energy $\Delta_{0}$, the interaction between the TS and a stress wave of phonon mode $s$ is,

\begin{equation}
\hat{\mathcal{H}}_{\text{int, TS}-s} \approx \left( \frac{\Delta_{0}}{E}\hat{\sigma}_{x} + \frac{\Delta}{E}\hat{\sigma}_{z}\right) \bar{\gamma}_{\text{TS}} \hat{\bar{e}}_{\mone}(\vecb{r}_{0}),   
\label{eq:HTSint}    
\end{equation}

%\tensoravg{\hat{\rttensortwo{e}}_{\mone}(\vecb{r}_{0})}

\noindent where $\vecb{r}_{0}$ is the point-like location of the TS and $E = (\Delta^2 + \Delta_{0}^2)^{1/2}$ is the TS transition energy.  Here we treat the stress interaction as approximately scalar, hence, $\bar{\gamma}_{\text{TS}}$ and $\hat{\bar{e}}_{\mone}$ are the tensor-averaged deformation potential and stress operator, respectively. The corresponding TLS interaction Hamiltonian is given by,

\begin{equation}
\hat{\mathcal{H}}_{\text{int, TLS}-s} \approx \left( M\hat{\sigma}_{x} + D\hat{\sigma}_{z}\right) \hat{\bar{e}}_{\mone}(\vecb{r}_{0}),   
\label{eq:HTLSint}    
\end{equation}

\noindent where $M$ is a transverse coupling potential and $D$ is a longitudinal coupling potential. We will follow a TLS model in what follows as it simplifies some of the analysis; however, it is important to note that the TLS model is more constrained than the TS model in that the ratio of transverse and longitudinal coupling is fixed for a given TLS energy.  This has the effect of eliminating the wide range of possible excited state decay rates for TS of a fixed energy.  In fitting our data with a TS model we found that a model with rather narrow $\Delta_{0}$ distribution whose mean scales approximately with $E$ fit best, which is effectively a TLS model.  

We define corresponding (tensor-averaged) transverse and longitudinal vacuum coupling rates as,

\begin{equation}
\gts(\vecb{r}_{0}) = \frac{M}{\hbar} (e_{\text{vac},s})\tensoravg{\overline{\rttensortwo{e}_{\mone}(\vecb{r}_{0})}}
\label{eq:gts}
\end{equation} 

\noindent and

\begin{equation}
\gls(\vecb{r}_{0}) = \frac{D}{\hbar} (e_{\text{vac},s})\tensoravg{\overline{\rttensortwo{e}_{\mone}(\vecb{r}_{0})}}
\label{eq:gls}
\end{equation} 

\noindent respectively, allowing us to write for the interaction Hamiltonian,

\begin{equation}
\hat{\mathcal{H}}_{\text{int, TLS}-s} \approx \hbar\left[ \gts(\vecb{r}_{0})\hat{\sigma}_{x} + \gls(\vecb{r}_{0})\hat{\sigma}_{z}\right] \left( \hat{b}_{\mone} + \hat{b}^{\dagger}_{\mone}\right).   
\label{eq:HTLSint2}    
\end{equation}

\noindent Including the bare TLS and phonon energy terms, we have for the total Hamiltonian,

\begin{equation}
\hat{\mathcal{H}}_{\text{TLS}-s} \approx \frac{\hbar\omegaTLS}{2}\hat{\sigma}_{z} + \hbar\omega_{\mone}\left(\hat{b}_{\mone}^{\dagger}\hat{b}_{\mone}+1/2\right) + \hbar\left[ \gts(\vecb{r}_{0})\hat{\sigma}_{x} + \gls(\vecb{r}_{0})\hat{\sigma}_{z}\right] \left( \hat{b}_{\mone} + \hat{b}^{\dagger}_{\mone}\right),   
\label{eq:HTLS_s}    
\end{equation}

\noindent where $\omegaTLS$ is the bare transition frequency of the TLS ($E=\hbar\omegaTLS$).  The $\hat{\sigma}_{x}$ interaction term leads to `resonant' decay into the phonon bath and a bath-dependent level shift of the TLS which can be treated using 2nd-order perturbation theory.  The $\hat{\sigma}_{z}$ interaction term gives rise to `relaxation'-type processes of higher-order in perturbation theory. 

\subsection{TLS decay into the phonon bath}
\label{subsubsec:TLSdecay}
We first consider a single TLS interacting with the phonons as a dissipative bath (the roles will be reversed when we consider the damping of a given phonon mode).  We assume that the TLS decay primarily through resonant $\hat{\sigma}_{x}$ interactions with the phonon bath, neglecting the $\hat{\sigma}_{z}$ interaction term.  Also, owing to the finite-size of the acoustic cavity structure studied here, we work in a discrete basis of phonon quasi-normal modes.  The bare phonon modes of the acoustic cavity have a complex frequency due to coupling to the phonons of the substrate, $\tilde{\omega}_{\mone} = \omega_{\mone} - i\Gammas$, where $\omega_{\mone}$ is the real angular frequency and $\Gammas=\gammas/2 + \Gammasphi$ is the phonon \emph{amplitude} decoherence rate given by the sum of half the energy decay rate ($\gammas$) and the pure dephasing rate ($\Gammasphi$) of the phonon mode.  From 2nd-order perturbation theory~\cite{Lefebvre-Brion2004,Moiseyev2011} we find a complex frequency shift of the TLS transition given by,

\begin{equation}
\left(\delta \tildeomegaTLS\right)_{\mone} \approx \left( 2 \gts^2 (n_{\mone} + 1/2)\right)\left[ \frac{1}{\tildeDeltaTLSs} + \frac{1}{\tildeDeltaTLSs + 2\tildeomegas} \right], 
\label{eq:dfTLS_s_complex}
\end{equation}

\noindent where $n_{\mone}$ is the phonon occupancy of mode $s$ and $\tildeDeltaTLSs \equiv \tildeomegaTLS - \tildeomegas$ is the near-resonant complex detuning. Here we have included the non-resonant term as it contributes non-negligibly to the TLS frequency shift (the real part of Eq.~(\ref{eq:dfTLS_s_complex})) when summing over contributions from phonon modes of large detuning. Implicit in our use of 2nd-order perturbation theory is that the TLS-$s$ coupling remain in a small coupling limit ($|\gts|/|\tildeDeltaTLSs| \ll 1$) where non-degenerate perturbation theory is accurate.  One can also utilize quasi-degenerate perturbation theory~\cite{Lefebvre-Brion2004} to determine the complex frequency shift without restriction on the strength of the coupling; however, the formulae are more complex and require careful elimination of non-physical solutions.  For simplicity of presentation, here we limit ourselves to the small coupling limit.  Below, in performing numerical calculations with specific TLS ensembles we found that the small coupling limit is adequate due to the low spectral density of phonon quasi-modes and TLS, and consequently the very unlikely situation where $|\gts|/|\tildeDeltaTLSs| \gtrsim 1$. 

One can arrive at a similar result in Hamiltonian form by rotating to a dressed TLS basis that diagonalizes Eq.~(\ref{eq:HTLS_s}) in the $N$-excitation  manifold to 2nd-order in the small parameter $\gts/|\tildeomegaTLS - \tildeomegas|$,

\begin{equation}
%\hat{\mathcal{H}}^{\text{eff,res.}}_{\text{TLS}-s} \approx \hbar\left(\tildeomegaTLS + \frac{\gts^2}{\tildeDeltaTLSs}\right)\frac{\hat{\sigma}_{z}}{2} + \hbar\left(\tildeomegas + \frac{\gts^2}{\tildeDeltaTLSs}\hat{\sigma}_{z}\right) \left(\hat{b}_{\mone}^{\dagger}\hat{b}_{\mone}+1/2\right),  
\hat{\mathcal{H}}^{\text{eff,res.}}_{\text{TLS}-s} \approx \hbar\left(\tildeomegaTLS \right)\frac{\hat{\sigma}_{z}}{2} + \hbar\left(\tildeomegas + \frac{\gts^2}{\tildeDeltaTLSs}\hat{\sigma}_{z}\right) \left(\hat{b}_{\mone}^{\dagger}\hat{b}_{\mone}+1/2\right),   
\label{eq:HTLS_s_diag}    
\end{equation}

\noindent where we have included only the `resonant' $\sigma_{x}$ interaction for now.  Due to our use of quasi-normal modes for both TLS and mode $s$, $\hat{\mathcal{H}}^{\text{eff,res.}}_{\text{TLS}-s}$ is an effective Hamiltonian with complex energy eigenvalues.  We see from this effective Hamiltonian that the resonant $\sigma_{x}$ interaction leads both to a dressing of the TLS and the phonon mode: (i) viewed as a Stark-like shift of the TLS, the dressed complex frequency of the TLS is $\tildeomegaTLSdressed = \tildeomegaTLS + (2\gts^2/\tildeDeltaTLSs) \langle\hat{b}^{\dagger}_{\mone}\hat{b}_{\mone} + 1/2 \rangle$, (ii) viewed as a TLS state-dependent shift of the phonon frequency, the dressed complex frequency of the mode $s$ is $\tildeomegasdressed = \tildeomegas + (\gts^2/\tildeDeltaTLSs )\langle\hat\sigma_{z}\rangle$.  The absence of the non-resonant term [$(\tildeDeltaTLSs + 2\tildeomegas)^{-1}$] in Eq.~(\ref{eq:HTLS_s_diag}) is due to our restriction to the $N$-excitation manifold.

Returning to Eq.~\ref{eq:dfTLS_s_complex} and focusing on the damping effect of the phonon mode $s$ on the TLS, we extract the imaginary component of $\delta \tildeomegaTLS$ corresponding to the phonon-induced decoherence rate of the TLS,

\begin{equation}
\left(\delta \GammaTLStwo\right)_{\mone} = -\Im \left[(\delta \tildeomegaTLS)_{\mone}\right] \approx \frac{2 \gts^2\left(\Gammas-\GammaTLStwo\right) (n_{\mone} + 1/2)}{(\omegaTLS-\omegas)^2 + (\GammaTLStwo - \Gammas)^2}. 
\label{eq:GammaTLSstwo}
\end{equation}

\noindent Assuming the TLS primarily decohere through the phonon bath such that the bare $\GammaTLStwo \approx 0$ (neglecting other bath contributions and TLS-TLS dephasing, for instance), and neglecting pure dephasing of the phonon mode ($\Gammasphi = 0$), we can write for the phonon-induced energy decay rate of the TLS due to mode $s$,

\begin{equation}
\left(\delta \GammaTLSone\right)_{\mone} \approx 2\left(\delta \GammaTLStwo\right)_{\mone} \approx \left[\frac{\gts^2}{(\omegaTLS-\omegas)^2 + (\gammas/2)^2}\right] \left( \gammas (2n_{\mone} + 1)\right). 
\label{eq:GammaTLSsone}
\end{equation}

\noindent For a phonon bath in thermal equilibrium at temperature $T$ we have that $2n_{\mone} + 1 = \coth(\hbar\omegas/2k_{B}T)$. Summing over the discrete set of quasi-normal phonon modes allows us to write for the total phonon-induced $\GammaTLSone$ as a function of phonon bath temperature,

\begin{equation}
\left(\delta\GammaTLSone\right)_{\text{ph}} \approx \sum_{\mone} \left[\frac{\gts^2\gammas}{(\omegaTLS-\omegas)^2 + (\gammas/2)^2}\right] \left( \coth(\hbar\omegas/2k_{B}T) \right). 
\label{eq:GammaTLSoneT}
\end{equation}

One recovers the standard result for a TLS interacting with a continuum phonon bath~\cite{Phillips1987} by integrating Eq.~(\ref{eq:GammaTLSoneT}) weighted by the appropriate phonon density of states per unit frequency, $\rho_{\text{ph}}[\omegas]$,

\begin{equation}
\left(\delta \GammaTLSone\right)_{\text{ph,cont.}} \approx 2\pi \rho_{\text{ph}}[\omegaTLS](\gts[\omegaTLS])^2 \left( \coth[\hbar\omegaTLS/2k_{B}T] \right), 
\label{eq:GammaTLSonecontinuum}
\end{equation}

%\stackrel{\text{cont.}}{\approx}

\noindent For a three-dimensional (3D) bulk material the (polarization-averaged) phonon bath density of states is $\rho_{\text{ph}} = (V/(2\pi^2\bar{v}^3))\omegas^2$, where $\bar{v}$ is an average acoustic velocity in the material.  The phonon modes of a homogeneous bulk are plane waves with vacuum strain amplitude $e_{\text{vac},s} = (\hbar\omegas/2\bar{J}V)^{1/2}$, where $\bar{J}$ is a (2nd-order in strain) elastic energy density coefficient or bulk modulus of the material.  The acoustic velocity and bulk modulus can be related to the bulk material mass density, $\rhobarm = \bar{J}/\vbar^2$.  Substituting these values into Eq.~(\ref{eq:GammaTLSonecontinuum}) yields for an average TLS coupled to a phonon bath in a 3D bulk,

\begin{equation}
\left(\delta \GammaTLSone\right)_{\text{ph,3D}} \approx \left(\frac{\bar{M}^2 \omegaTLS^3}{2\pi\hbar\bar{J}\bar{v}^3}\right) \left( \coth[\hbar\omegaTLS/2k_{B}T] \right) = \left(\frac{ \bar{M}^2 \omegaTLS^3}{2\pi\hbar\bar{\rho}_{\text{m}}\bar{v}^5}\right) \left( \coth[\hbar\omegaTLS/2k_{B}T] \right), 
\label{eq:GammaTLSone3D}
\end{equation}

\noindent where $\bar{M}$ is an averaged (over TLS orientation and acoustic polarization) transverse coupling potential.  For future reference we also quote here the corresponding result for a quasi two-dimensional (2D) material, corresponding to a plate of large area and thickness $t$ smaller than the acoustic wavelength, 

\begin{equation}
\left(\delta\GammaTLSone\right)_{\text{ph,2D}} \approx \left(\frac{ \bar{M}^2 \omegaTLS^2}{2\hbar\bar{\rho}_{\text{m}}\bar{v}^4 t }\right) \left( \coth[\hbar\omegaTLS/2k_{B}T] \right),
\label{eq:GammaTLSone2D}
\end{equation}

\noindent and a quasi one-dimensional (1D) material, corresponding to a beam of long length and small cross-sectional dimension $\bar{w}$ relative to the acoustic wavelength,

\begin{equation}
\left(\delta\GammaTLSone\right)_{\text{ph,1D}} \approx \left(\frac{ \bar{M}^2 \omegaTLS}{2\hbar\bar{\rho}_{\text{m}}\bar{v}^3 \bar{w}^2}\right) \left( \coth[\hbar\omegaTLS/2k_{B}T] \right).
\label{eq:GammaTLSone1D}
\end{equation}

\noindent In a prelude to what follows, we note that the frequency scaling with bath dimension will set the temperature scaling of the non-resonant TLS `relaxation' damping of acoustical modes; hence, a 3D phonon bath will yield a $T^3$ dependence, a 2D bath a quadratic $T^2$ dependence, and a 1D bath will result in a linear $T$ dependence.

\subsection{`Resonant' TLS damping and frequency shift of acoustic cavity quasi-modes}
\label{subsubsec:resTLSdampingshift}

In contrast to the analysis of the prior sub-section, we now reverse roles and consider the TLS to be a bath for the acoustic phonon modes of the structure.  In particular, we are interested in the localized high-$Q$ phonon mode which lies within the phononic bandgap of the acoustic radiation shield.  As such, this phonon mode should have much smaller intrinsic (radiation) damping to the substrate than the average phonon bath mode considered in the previous sub-section.  We first consider the effects of the $\hat{\sigma}_{x}$ TLS-phonon interaction.  Referring to Eq.~(\ref{eq:HTLS_s_diag}), the `resonant' ($\hat{\sigma}_{x}$) contribution to the TLS state-dependent shift in the complex phonon frequency is given by,

\begin{equation}
\left(\delta \tildeomegas\right)_{\text{TLS,res}} \approx \left( \gts^2[\vecb{r}_{\text{TLS}}] \langle \hat{\sigma}_{z} \rangle \right)\left[ \frac{1}{\tildeDeltaTLSs} + \frac{1}{\tildeDeltaTLSs + 2\tildeomegas} \right], 
\label{eq:df_s_complex}
\end{equation}

\noindent where $\vecb{r}_{\text{TLS}}$ is the spatial location of the TLS in the acoustic cavity and we have added in by hand the non-resonant [$(\tildeDeltaTLSs + 2\tildeomegas)^{-1}$] term as found in the perturbation analysis of Eq.~(\ref{eq:dfTLS_s_complex}).  For a TLS in thermal equilibrium at temperature $T$ one has $\langle \hat{\sigma_{z}} \rangle = -\tanh[\hbar\omegaTLS/2k_{B}T]$.  Considering interaction with an ensemble of TLS and summing over this ensemble yields for the TLS state-dependent shift in the real part of the frequency of the phonon mode $s$,

\begin{multline}
(\delta \omegas)_{\text{res}} \approx \sum_{\text{TLS}} \Re \left[( \delta \tildeomegas )_{\text{TLS}} \right] =  -\sum_{\text{TLS}} \left( \gts^2[\vecb{r}_{\text{TLS}}]  \tanh[\hbar\omegaTLS/2k_{B}T] \right) \\
\times \left[ \frac{\omegaTLS-\omegas}{(\omegaTLS-\omegas)^2 + (\GammaTLStwo - \Gammas)^2} + \frac{\omegaTLS+\omegas}{(\omegaTLS+\omegas)^2 + (\GammaTLStwo + \Gammas)^2} \right].
\label{eq:df_s_res}
\end{multline}

\noindent Substituting for $\GammaTLStwo$ the estimated energy decay rate due to coupling to the rest of the phonon bath found in Eq.~(\ref{eq:GammaTLSsone}) and a pure dephasing rate ($\GammaTLSphi$), and assuming the phonon mode $m$ of interest has a much smaller decoherence rate than the \emph{dressed} TLS, we have that,

\begin{multline}
(\delta \omegam)_{\text{res}} \approx -\sum_{\text{TLS}} \left( \gtm^2[\vecb{r}_{\text{TLS}}]  \tanh[\hbar\omegaTLS/2k_{B}T] \right) \\
\times \left[ \frac{\omegaTLS-\omegam}{(\omegaTLS-\omegam)^2 + ((\delta \GammaTLSone)_{\text{ph}}/2 + \GammaTLSphi)^2} + \frac{\omegaTLS+\omegam}{(\omegaTLS+\omegam)^2 + ((\delta \GammaTLSone)_{\text{ph}}/2 + \GammaTLSphi)^2} \right].
\label{eq:df_m_res}
\end{multline}

\noindent Similarly, the energy damping rate of mode $m$ due to resonant interaction processes with the TLS bath is given by,

\begin{equation}
(\delta \gammam )_{\text{res}} \approx \sum_{\text{TLS}} -2\Im \left[( \delta \tildeomegam )_{\text{TLS}} \right] \approx \sum_{\text{TLS}}  \left[  \frac{\left( \gtm^2[\vecb{r}_{\text{TLS}}]  \tanh[\hbar\omegaTLS/2k_{B}T] \right) (\delta \GammaTLSone)_{\text{ph}}}{(\omegaTLS-\omegam)^2 + ((\delta \GammaTLSone)_{\text{ph}}/2 + \GammaTLSphi)^2} \right],
\label{eq:dgamma_m_res}
\end{equation}

\noindent where we have neglected the non-resonant term [$(\tildeDeltaTLSs + 2\tildeomegas)^{-1}$] due to its much weaker contribution to the Lorentzian damping function.  Note that we have not included the $\GammaTLSphi$ contribution in the numerator of Eq.~(\ref{eq:dgamma_m_res}) as it adds pure dephasing to the phonon mode $m$.

One recovers the standard result for damping of a phonon interacting with a continuum TLS bath~\cite{Phillips1987} by integrating Eq.~(\ref{eq:dgamma_m_res}) weighted by the TLS density of states per unit \emph{angular} frequency in the acoustic mode volume, $\ndTLSm/2\pi \equiv \hbar\ndTLS\eta_{\text{surf}}\Vm$,

\begin{equation}
(\delta \gammam )_{\text{res, cont.}} \approx  (2\pi)(\hbar\eta_{\text{surf}}\ndTLS\Vm) \overline{\gtm^2[\vecb{r}_{m}]}  \tanh[\hbar\omegam/2k_{B}T] \approx \left(\frac{\pi \bar{M}^2 \omegam}{\rhobarm\vbar^2}\right) (\eta_{\text{surf}}\ndTLS) \tanh[\hbar\omegam/2k_{B}T],
\label{eq:dgamma_m_res_cont}
\end{equation}

\noindent where $\eta_{\text{surf}}\ndTLS$ is the effective bulk TLS density per unit volume per unit energy, and the average transverse coupling rate for TLS in the acoustic mode volume is approximately, $\overline{\gtm^2[\vecb{r}_{m}]} \approx (\bar{M}/\hbar)^2(\hbar\omegam/2\rhobarm\vbar^2\Vm)$.  Following a similar averaging over the cavity mode volume and integration over a TLS density in Eq.~(\ref{eq:df_m_res}), one obtains the corresponding frequency shift of the breathing mode due to resonant interaction with a continuum of TLS,

\begin{multline}
(\delta \omegam )_{\text{res, cont.}} \approx  -(\hbar\eta_{\text{surf}}\ndTLS\Vm) \overline{\gtm^2[\vecb{r}_{m}]} \\
\times \Re{ \left\{\int_0^{\omega_{\text{max}}} \mathrm{d}\omegaTLS \tanh[\hbar\omegaTLS/2k_{B}T] \left(\frac{1}{(\omegaTLS-\omegam) + i\GammaTLStwo} + \frac{1}{(\omegaTLS+\omegam) - i\GammaTLStwo} \right) \right\}},
\label{eq:df_m_res_cont}
\end{multline}

\noindent where $\omega_{\text{max}}$ is the maximum transition frequency of the TLS ensemble.  The integral can be evaluated using the digamma function~\cite{GaoPhD}, yielding the following simplified result,

\begin{equation}
(\delta \omegam )_{\text{res, cont.}} \approx \left(\frac{\bar{M}^2 \omegam}{\rhobarm\vbar^2}\right) (\eta_{\text{surf}}\ndTLS) \left( \Re{\left\{ \Psi \left[ \frac{1}{2} + i\frac{\hbar\omegam}{2\pi k_{B} T}\right] \right\}} - \ln{\left[\frac{\hbar\omega_{\text{max}}}{2\pi k_{B} T}\right]} \right).
\label{eq:df_m_res_cont_digamma}
\end{equation}

\subsection{`Relaxation' TLS damping and frequency shift of acoustic cavity quasi-modes}
\label{subsubsec:relTLSdampingshift}

Relaxation damping of acoustic cavity modes results not from direct energy exchange with nearly-resonant TLS, but rather from the shift in the TLS transition frequencies due to the $\sigma_{z}$ interaction of Eq.~(\ref{eq:HTLSint}).  This shift, which is linear in the stress amplitude of the acoustic modes and oscillates in time with the frequency of the acoustic mode, displaces the TLS from equilibrium.  During this oscillatory displacement out of equilibrium the TLS will relax back towards equilibrium at a rate given by the TLS energy decay rate, stealing away energy from the acoustic mode in the process.  Microscopically this is a process involving higher-order perturbation interactions between the TLS and the phonon bath.  As such, here we follow the standard semi-classical analysis of relaxation damping by considering the acoustic dipole response of the TLS to a (classical) strain field~\cite{Phillips1987,GaoPhD}. 

The magnitude of the longitudinal acoustic dipole of a TLS is given by $\langle \rttensortwo{p}_{a} \rangle_{\text{t}} \equiv D \langle \hat{\sigma}_{z} \rangle$.  For a small amplitude, harmonic strain field oscillating in mode $s$, $\langle \rttensortwo{e}_{\mone}[\vecb{r};\omega_{\mone}]\rangle_{\text{t}}$, the harmonically oscillating component of the longitudinal acoustic dipole is linearly related to the applied strain at the site of TLS through a (tensor-averaged) susceptibility, $\langle \chirel[\omega_{\mone}] \rangle_{\text{t}} \equiv \langle \delta\rttensortwo{p}_{a}[\omega_{\mone}] \rangle_{\text{t}}/\langle \rttensortwo{e}_{\mone}[\vecb{r}_{\text{TLS}};\omega_{\mone}]\rangle_{\text{t}}$, where  $\langle \delta\rttensortwo{p}_{a}[\omega_{\mone}] \rangle_{\text{t}} = D (\delta\langle \hat{\sigma}_{z}[\omega_{\mone}] \rangle)$, $\delta \langle \hat{\sigma}_{z} \rangle = (\langle \hat{\sigma}_{z} \rangle - \langle \hat{\sigma}_{z} \rangle_{\text{eq.}})$, and $\langle \hat{\sigma}_{z} \rangle_{\text{eq.}} = -\tanh[\hbar\omegaTLS/2k_{B}T]$ is the TLS inversion in thermal equilibrium.  Solving the Bloch equations assuming a finite relaxation rate to equilibrium of $\GammaTLSone$, the displacement of the inversion from equilibrium follows the applied harmonic strain with a phase lag,

\begin{equation}
\delta \langle \hat{\sigma}_{z} \rangle \approx \frac{\partial \langle \hat{\sigma}_{z} \rangle_{\text{eq.}}}{\partial \omegaTLS} \frac{\partial \omegaTLS}{\partial \langle \rttensortwo{e}_{\mone}\rangle_{\text{t}}} \left( \frac{1-i\omega_{\mone}(\GammaTLSone)^{-1}}{1+(\omega_{\mone}(\GammaTLSone)^{-1})^2} \right) \langle  \rttensortwo{e}_{\mone}\rangle_{\text{t}}.
\label{eq:inversion_rel}
\end{equation}

\noindent From Eq.~(\ref{eq:HTLS_s}) one can show that $\partial \omegaTLS/\partial \langle \rttensortwo{e}_{\mone}\rangle_{\text{t}} =  (2D/\hbar)$, and $\partial \langle \hat{\sigma}_{z} \rangle_{\text{eq.}}/\partial \omegaTLS = (\hbar/2k_{B}T)\sech^2[\hbar\omegaTLS/2k_{B}T]$.  This yields for the relaxation susceptibility,

%\langle \hat{\sigma}_{z} \rangle (2\bar{D})

%change in the TLS inversion level away from its thermal equilibrium value, $\delta\langle p_{a} \rangle \approx \delta \langle \sigma_{z} \rangle (2D)$    

%(tensor-averaged) susceptibility, $\langle \chirel(\omega) \rangle_{\text{t}}\equiv \langle \hat{\sigma}_{z} \rangle (2\bar{D})/\langle \rttensortwo{e}(\omega)\rangle_{\text{t}}$, can be defined to describe the relaxation response of an average TLS back to its thermal equilibrium state in the presence of an oscillating (tensor-averaged) strain field $\langle \rttensortwo{e}(\omega) \rangle_{\text{t}}$~\cite{Phillips1987,GaoPhD},

\begin{equation}
\langle \chirel[\omega_{\mone};\omegaTLS] \rangle_{\text{t}} = \left(\frac{D^2}{k_{B}T}\right)\left(\frac{1-i\omega_{\mone}(\GammaTLSone)^{-1}}{1+(\omega_{\mone}(\GammaTLSone)^{-1})^2} \right)\sech^2[\hbar\omegaTLS/2k_{B}T].
\label{eq:TLSrelsusc}
\end{equation}

The complex energy shift of the acoustic mode due to its interaction with the TLS is given by $(\delta \tilde{E}_{\mone})_{\text{TLS}} \approx  \langle \rttensortwo{p}_{a} \rangle_{\text{t}} (\langle \rttensortwo{e}_{\mone}[\vecb{r}_{\text{TLS}}]\rangle_{\text{t}})^{*}$.  Noting the complex energy shift can be related to a complex frequency shift in the acoustic resonance through the stored phonon number ($n_{\mone}$), $(\delta \tilde{E}_{\mone})_{\text{TLS}} = \hbar(\delta \tilde{\omega}_{\mone}n_{\mone})$, and writing the local applied strain amplitude in terms of phonon number, $\langle \rttensortwo{e}_{\mone}[\vecb{r}_{\text{TLS}}]\rangle_{\text{t}} = (e_{\text{vac,}s})(\langle \overline{\rttensortwo{e}_{\mone}[\vecb{r}_{\text{TLS}}]} \rangle_{\text{t}})\sqrt{n_{\mone}}$, yields for the complex frequency shift in quasi-mode $s$ due to relaxation interactions with a single TLS,

\begin{equation}
(\delta \tilde{\omega}_{\mone})_{\text{TLS,rel}} \approx \frac{(\langle \overline{\rttensortwo{e}_{\mone}[\vecb{r}_{\text{TLS}}]} \rangle_{\text{t}})^2  (e_{\text{vac,}s})^2 }{\hbar} \langle \chirel[\omega_{\mone};\omegaTLS] \rangle_{\text{t}}.
\label{eq:omegarel}
\end{equation}

\noindent For a given quasi-mode $m$ of interest, interacting with an ensemble of TLS, the corresponding frequency shift due to relaxation processes is given by, 

\begin{align}
(\delta \omegam)_{\text{rel}}  = \sum_{\text{TLS}}\Re \left[(\delta \tilde{\omega}_{m})_{\text{TLS,rel}}\right] & \approx \sum_{\text{TLS}} \frac{\omegam (\langle \overline{\rttensortwo{e}_{m}[\vecb{r}_{\text{TLS}}]} \rangle_{\text{t}})^2 \Re \left[\langle \chirel[\omegam;\omegaTLS] \rangle_{\text{t}}\right]}{2\rhobarm\vbar^2\Vm}  \label{eq:df_m_rel} \\ \nonumber
& \approx \sum_{\text{TLS}} \frac{ (\langle \overline{\rttensortwo{e}_{m}[\vecb{r}_{\text{TLS}}]} \rangle_{\text{t}})^2 D^2 (\GammaTLSone)^{2} }{2\omegam\rhobarm\vbar^2\Vm k_{B}T} \sech^2[\hbar\omegaTLS/2k_{B}T] \\ \nonumber
&= \sum_{\text{TLS}} \left(\frac{\glm[\vecb{r}_{\text{TLS}}]}{\omegam} \right)^2 \left(\frac{\hbar(\GammaTLSone)^2}{k_{B}T} \right) \sech^2[\hbar\omegaTLS/2k_{B}T].  
\end{align}

\noindent Similarly, the relaxation energy damping rate of mode $m$ is given by,

\begin{align}
(\delta \gammam )_{\text{rel}} = -2\sum_{\text{TLS}} \Im \left[(\delta \tilde{\omega}_{m})_{\text{TLS,rel}}\right] & \approx -\sum_{\text{TLS}} \frac{\omegam (\langle \overline{\rttensortwo{e}_{m}[\vecb{r}_{\text{TLS}}]} \rangle_{\text{t}})^2 \Im \left[\langle \chirel[\omegam;\omegaTLS] \rangle_{\text{t}}\right] }{\rhobarm\vbar^2\Vm}  \label{eq:dgamma_m_rel} \\ \nonumber
& \approx \sum_{\text{TLS}} \frac{(\langle \overline{\rttensortwo{e}_{m}[\vecb{r}_{\text{TLS}}]} \rangle_{\text{t}})^2 D^2 \GammaTLSone}{\rhobarm\vbar^2\Vm k_{B}T} \sech^2[\hbar\omegaTLS/2k_{B}T] \\ \nonumber
& = \sum_{\text{TLS}} \left(\frac{2\glm^2[\vecb{r}_{\text{TLS}}]}{\omegam} \right) \left(\frac{\hbar\GammaTLSone}{k_{B}T} \right) \sech^2[\hbar\omegaTLS/2k_{B}T].
\end{align}

\noindent Since we are concerned with high frequency, microwave phonon modes and TLS at cryogenic temperatures, we have safely assumed that we are in the non-adiabatic limit, $\omegam\GammaTLSone^{-1} \gg 1$.

Assuming resonant phonon damping is the dominant decay mechanism for TLS, and substituting $(\delta\GammaTLSone)_{\text{ph,cont.}}$ for $\GammaTLSone$ in Eqs.~(\ref{eq:df_m_rel}-\ref{eq:dgamma_m_rel}), yields the standard relations for the frequency shift and energy damping for an acoustic mode interacting with a TLS bath in a bulk material which supports a phonon continuum. For a uniform spectral density of states for the TLS, as assumed here, integrating over $\omegaTLS$ yields a phonon relaxation damping of $(\delta \gammam )_{\text{rel}} \sim T^{d}$, where $d$ is the dimension of the phonon bath.  This is a result of the fact that $\sech^2[\hbar\omegaTLS/2k_{B}T]$ effectively limits the TLS frequency integral to frequencies below $\approx k_{B}T$ (i.e., relaxation damping is limited to thermally occupied TLS), and within this range of frequencies $\coth[\hbar\omegaTLS/2k_{B}T] \approx 2k_{B}T/\hbar\omegaTLS$.  This observation is especially important for the nanoscale optomechanical structures under study, where at the temperatures and corresponding frequencies considered, the relevant phonon bath density of states varies between something approximating a 1D bath at temperatures below $100$~mK to something approximating a 2.5D bath at temperatures between $100$~mK and $1$~K. The influence of the geometric patterning of the optomechanical structure also plays a role, modifying the phonon density of states in extreme ways by introducing phononic bandgaps and flat band regions.  As we will show below using numerical methods to calculate the full spectrum of quasi-acoustic-modes of the optomechanical structure, the temperature dependence of the relaxation damping rate due to TLS can indeed be significantly modified from that in bulk.           

%Considering a localized phonon mode $s$, such as the mode of interest in this work, only those TLS that lie roughly within the effective mode volume of the phonon mode will interact significantly.  The stress those TLS feel is approximately given by $\hat{\bar{e}}_{\mone}(\vecb{r}_{0}) \approx e_{\text{vac},s}(\hat{b}_{\mone} + \hat{b}^{\dagger}_{\mone})$.    

\section{Numerical modeling of TLS interactions and acoustic damping in the OMC cavity}
\label{sec:numerical_modeling_damping}

In order to more accurately account for the complex geometry of the optomechanical crystal structure studied and its impact on the phonon mode spectrum and TLS dynamics, we have performed numerical simulations of the acoustic resonances (quasi-modes) of the structure for frequencies below $100$~GHz.  One half of the simulated structure is shown schematically in Fig.~\ref{fig:SI_TLScartoon}, and includes one of the pair of OMC cavities, the optical coupling waveguide, and the phononic shield which make up a single `device'.  In practice, such a device is clamped from below at its periphery through the connection of the top Si device layer to the underlying $3$~$\mu$m thick SiO$_{x}$ buried oxide (BOX) layer, which itself is grown on top of a thick ($500$~$\mu$m) Si handle layer.  Simulations are performed using the COMSOL finite-element-method solver, using the Amazon AWS cloud computing resources to simulate various parts of the phonon spectrum in parallel.  Acoustic perfectly matched layers (PML) at the periphery of the structure are used as a radiation boundary condition.  

\begin{figure*}[btp]
\begin{center}
\includegraphics[width=1.0\columnwidth]{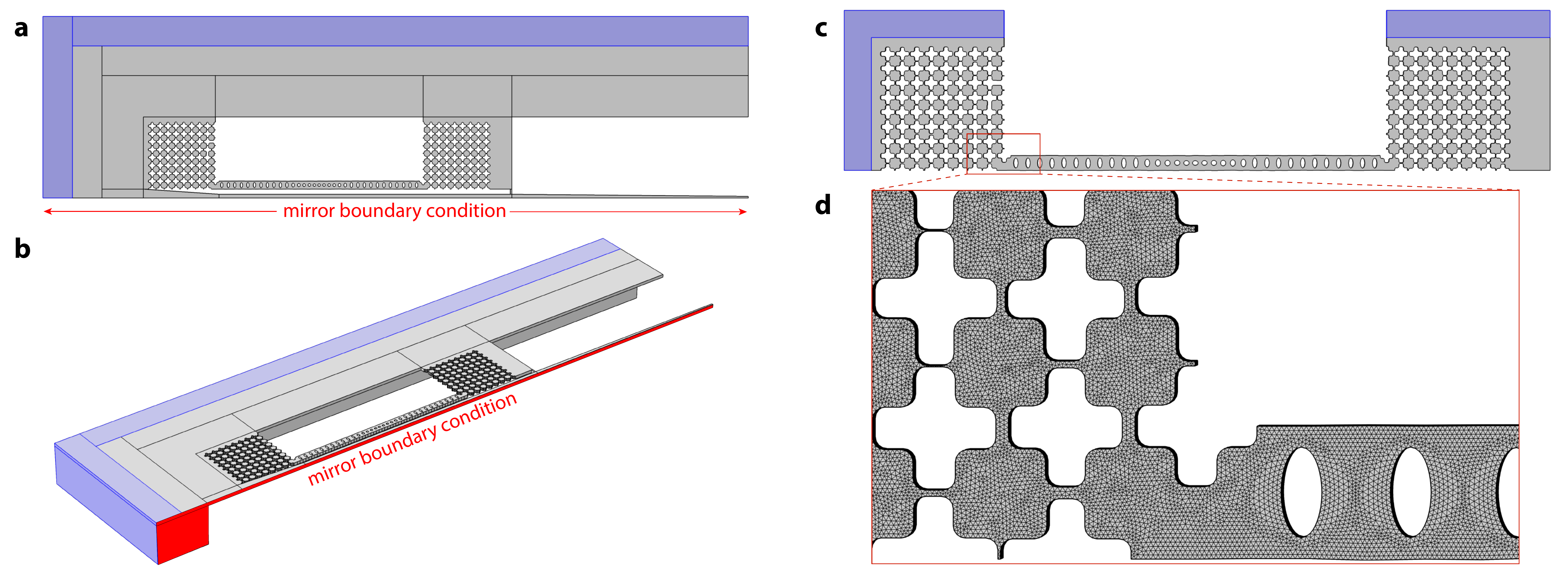}
\caption{\textbf{FEM simulation layout and mesh.} \textbf{a}, Top view of the FEM-simulated structure for phonon frequencies below $10$~GHz.  Blue regions correspond to PML radiation boundaries.  The structure is mirrored about the center axis of the coupling waveguide (red arrows and label).  A maximum mesh size of $20$~nm is utilized in the OMC cavity and phononic shield regions.  A maximum mesh size of $250$~nm is set in the surrounding periphery and PML regions.  The mesh resolution is smoothly varied between the two regions.  \textbf{b}, Isometric view of the same structure in (\textbf{a}), showing the underlying BOX clamping layer and the corresponding PML layers.  The bottom of the BOX layer has a fixed boundary condition applied to it.  \textbf{c}  Top view of the reduced FEM-simulation volume for phonon frequencies between $10$-$100$~GHz.  Again, the blue region is an acoustic PML.  All other boundaries other than PML are set to free boundaries.  The maximum mesh size is now $20$~nm throughout the entire structure, including the PML.  \textbf{d} Zoom in of the red box region in (\textbf{c}), showing the dense meshing of the nanobeam and shield.  
} 
\label{fig:FEM_sim_layout_mesh}
\end{center}
\end{figure*}

A series of different meshing schemes are used to cover the acoustic frequency spectrum up to $100$~GHz.  For all frequencies we utilize a fine mesh with maximum element size of $20$~nm in the OMC cavity and phononic shield regions.  This yields a meshing resolution of roughly $3$-$4$ points per wavelength even at the highest ($100$~GHz) frequencies considered.  As a result of memory and computing time limitations, we adjust the meshing and structure layout as a function of frequency in the rest of the structure outside of the OMC cavity and shield.  For acoustic frequencies below $10$~GHz the full structure shown in Fig.~\ref{fig:SI_TLScartoon}(b) is simulated, which includes at its periphery a micron thick (in depth and height) BOX layer followed by the PML radiation boundary.  For these frequencies a lower mesh density (maximum mesh size of $250$~nm) is utilized in the BOX and PML regions, providing a minimum of roughly $4$ points per wavelength resolution for modes up to $10$~GHz.  The simulated quasi-modes of the structure are therefore damped through their acoustic radiation into the BOX layer, with no further acoustic reflections considered (such as at the Si handle layer).  For frequencies above $10$~GHz we remove the BOX clamping region and apply the PML layers right at the boundary of the phononic shield, as shown in Fig.~\ref{fig:SI_TLScartoon}(c).  At these higher frequencies we utilize the same fine mesh in the PML as in the OMC cavity region (maximum mesh size of $20$~nm).  The thought in doing this is that the BOX clamping region plays less of a role for these short wavelength phonons that can effectively propagate within the thin Si device layer without major reflection at this boundary, hence the removal of the BOX layer and application of the PML region in the Si device layer right at the exit of the acoustic shield.  More importantly is that we provide a fine mesh in the PML to avoid unintended reflections.  Beyond $100$~GHz the resulting memory requirements and computation time are prohibitive, and given the temperature range of interest ($\lesssim 1$~K), $100$~GHz frequency is a natural cut-off point.  In addition to the above meshing strategy, in order to reduce the memory and computation time we apply a mirror boundary condition along the center of the structure, running down the middle of the coupling waveguide, and double the number of modes recorded in simulation.    

\begin{figure*}[btp]
\begin{center}
\includegraphics[width=0.75\columnwidth]{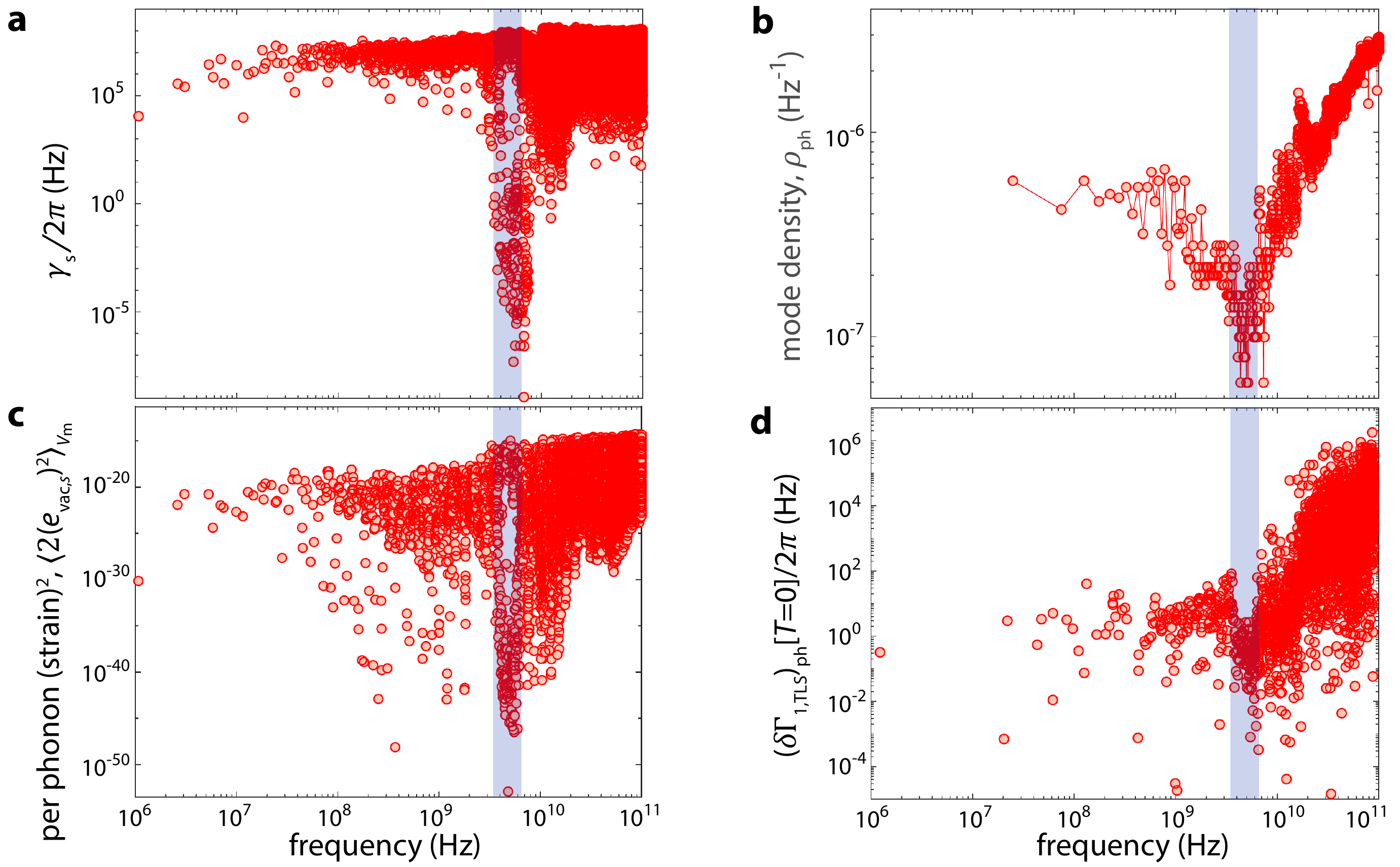}
\caption{\textbf{Model phonon and TLS properties.}  In this model, phonons of the Si optomechanical slab structure are simulated using a FEM numerical solver.  The simulation volume consists of the two optomechanical cavities, fiber coupling waveguide, and 10 period phononic shield.  As in the fabricated devices, the Si slab is clamped at the periphery of the optomechanical structure to the underlying SiO$_{\text{x}}$ BOX layer of the SOI.  An acoustic radiation boundary condition consisting of a perfectly-matched layer allows for radiation to escape into the external Si slab and underlying substrate.  All phonon resonances up to $100$~GHz frequency are calculated.  The lowest phononic bandgap of the shield is shown as a transparent blue band.
\textbf{a},  Acoustic radiation damping rate for phonon quasi-modes of the optomechanical structure.
\textbf{b},  Spectral phonon mode density, spectrally averaged over $20$~MHz bin size. 
\textbf{c},  Per-phonon squared strain value within mode volume $\Vm$ for each of the phonon-quasi modes.  Here, the sum of the square of the strain components are averaged over $201$ positions within the acoustic mode volume of the breathing mechanical mode. 
\textbf{d}, For an exemplary ensemble instance, zero temperature TLS decay rate of each of the $\NTLS = 3920$ randomly oriented and distributed within $\Vm$ due to resonant ($\hat{\sigma}_{x}$) coupling to phonon quasi-modes of the optomechanical structure.  
} 
\label{fig:sim_ph_TLS_prop}
\end{center}
\end{figure*}

For each acoustic resonance found in the simulation we not only record the frequency ($\omega$) and energy damping rate ($\gamma$), but also calculate the per phonon strain tensor at $101$ locations in the acoustic mode volume $\Vm$ of the high-$Q$ breathing mode.  These locations were chosen to be in random locations in the Si device layer, but within $\delta w=15$~nm of the Si-air interface, as this is where we expect TLS to be located due to etch damage. A fixed set of $101$ positions are evaluated for all acoustic modes.  Plots showing the resulting energy damping rate, effective phonon mode density, and average squared strain in the mode volume $\Vm$ as a function of phonon frequency are shown in Figs.~\ref{fig:sim_ph_TLS_prop}(a-c).  Noteworthy in these plots is the position of the fundamental phononic bandgap of the acoustic shield, which is shown as a semi-transparent blue band from approximately $3.5$ to $6.5$~GHz.  As can be seen, a significant change in the local strain amplitude and mode density occurs around this phononic bandgap.  Below the bandgap, the mode density is roughly constant at 1 mode every $5$~MHz (a bin size of $50$~MHz was used when estimating this spectral quasi-mode density), consistent with that of an effectively 1D system.  Within the bandgap of the acoustic shield, the mode density drops, and then above the bandgap the mode density rises a little faster than linearly with frequency, corresponding to that of a 2.4-D system.  Below the bandgap the per-phonon strain in mode volume $\Vm$ is quite small, and then rises within and above the phononic bandgap frequency due to the localization of modes within the shield.  The energy damping rate plot shows that a portion of the modes become substantially less damped in and around the bandgap as expected.

\subsection{Numerical simulation of 3-phonon scattering}
\label{subsec:3phononnumerical}

\begin{table}[btp]
\begin{center}
\caption{\textbf{3-phonon scattering model parameters.} \label{tab:3phonon} }
%\begin{ruledtabular}
\begin{tabular}{|l|l|l|l|}
\hline\hline
%\toprule
 Parameter & Description & Value & Refs./Notes \\
\hline
  $\modeavg{\GammathreephI}$ & average 3-phonon mode overlap factor & 0.01 & $\cong$ phase matching term \\ 
  $\modeavg{\grun}$ & mode averaged Gr\"{u}neisen parameter at cryogenic temperature ($T < 4K$) & 0.24 & \cite{Gauster1971} \\ 
  $\vSit$ & transverse acoustic phonon velocity in bulk Si ([100] dir., [011] pol.) & $8.4 \times 10^{3}$~m/s & \cite{Hall1967,McSkimin1953} \\
  $\vSil$ & longitudinal acoustic phonon velocity in bulk Si ([100] dir. and pol.) & $5.8 \times 10^{3}$~m/s & \cite{Hall1967,McSkimin1953} \\
  $\qavg{\vSibar^2}$ & average square of acoustic phonon velocity in bulk Si ([100] dir.) & ($7.6 \times 10^{3}$~m/s)$^{2}$ &  \\
  $\rhoSi$ & Si mass density & $2.33 \times 10^{3}$~kg/m$^{3}$ & \cite{Hall1967} \\
  $\qavg{\esqm}$ &  average strain squared of vacuum for mode $m$ within $\Vm$ & $2.25 \times 10^{-8}$ & COMSOL sim. \\
  
  \hline\hline
%\bottomrule
\end{tabular}
%\end{ruledtabular}
\end{center}
\end{table}

\begin{figure*}[btp]
\begin{center}
\includegraphics[width=0.75\columnwidth]{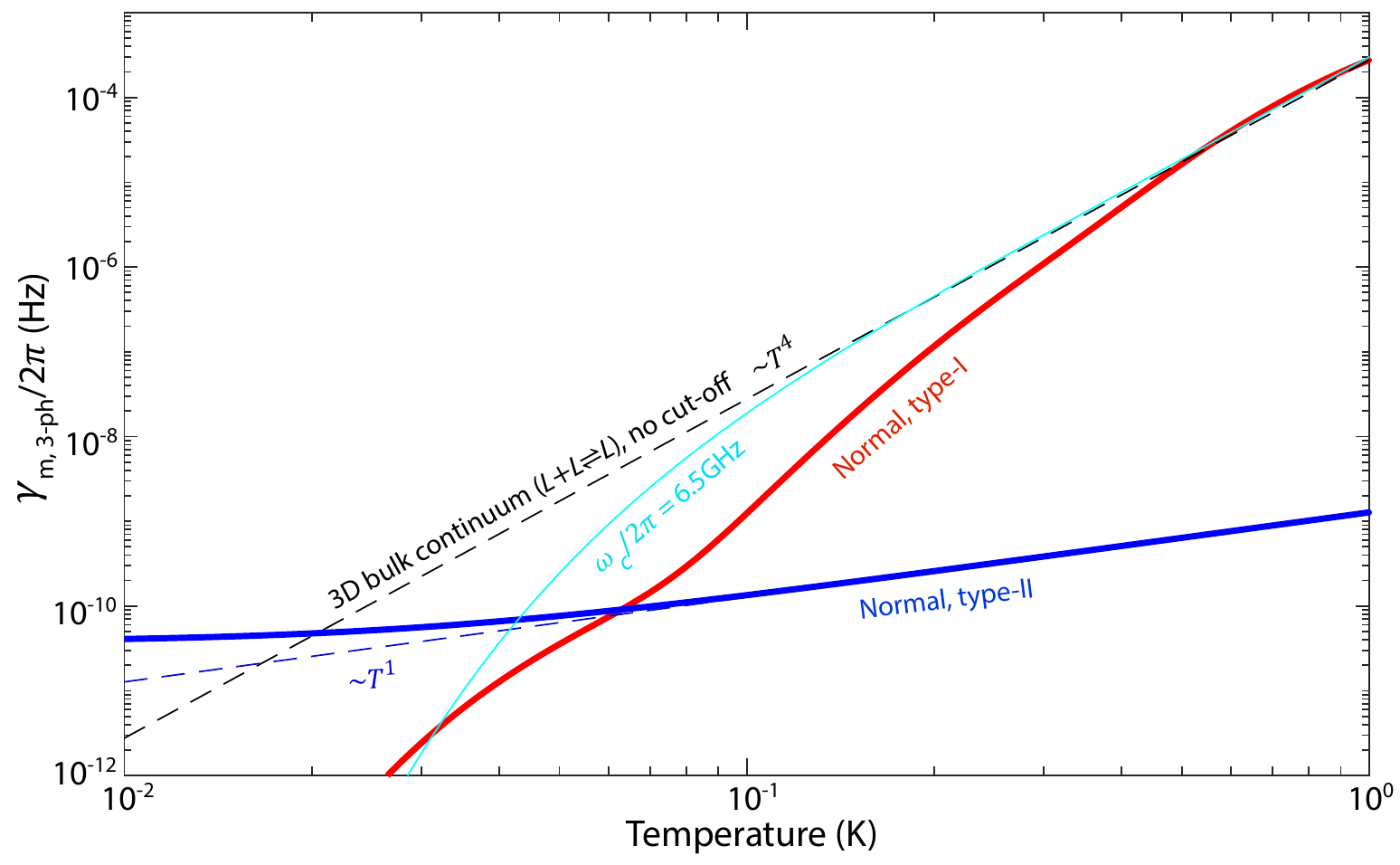}
\caption{\textbf{3-phonon scattering model.}  Simulation of the acoustic damping due to 3-phonon scattering of the localized breathing mechanical mode with other quasi-modes of the OMC cavity structure of Fig.~\ref{fig:sim_ph_TLS_prop}.  Parameters used in the modeling are listed in Tab.~\ref{tab:3phonon}.  Both type I (solid red curve) and type II (solid blue curve) scattering processes involving the breathing mode are modeled.  For comparison, we also plot the estimated 3-phonon-scattering damping rate due to type-I processes for longitudinally polarized phonons ($L + L \rightleftharpoons L$) in bulk Si.  For the bulk simulation we plot the estimated damping rate without a low-frequency cut-off (dashed black curve), and with a low-frequency cut-off (solid cyan curve) corresponding to the top of the first phononic bandgap ($\omega_c/2\pi = 6.5$~GHz).  In all cases we only include Normal scattering processes, and neglect Umklapp scattering, due to the low temperature range considered ($T\lesssim 1$~K).    
} 
\label{fig:3phonon}
\end{center}
\end{figure*}

Utilizing Eqs.~(\ref{eq:gammam3phI_sum}), (\ref{eq:gammam3phII_sum}), and (\ref{eq:gamma3phI_bulkSi_simple}), along with the parameters listed in Tab.~\ref{tab:3phonon} and the COMSOL-simulated acoustic modes of the OMC cavity structure (c.f., Fig.~\ref{fig:FEM_sim_layout_mesh}), in Fig.~\ref{fig:3phonon} we plot the calculated acoustic damping of the localized $5$~GHz breathing mode due to 3-phonon scattering at temperatures below $1$~K.  The resulting temperature-dependent damping rate using the numerically computed acoustic quasi-modes of the OMC cavity are shown as a solid red curve for $\mathcal{N}$-I processes and a solid blue curve for $\mathcal{N}$-II processes.  Damping arising from $\mathcal{N}$-II scattering processes is largely suppressed in the OMC cavity structure due to the reduced density of phonon states lying below the breathing mode frequency, a consequence of the effectively reduced dimensionality of the OMC nanobeam at these acoustic wavelengths.  At temperatures below $T = \hbar\omegam/k_{B}\approx 200$~mK the $\mathcal{N}$-II  damping rate approaches the spontaneous decay rate of the breathing mode, whereas at higher temperatures the damping rate increases linearly with temperature.  The $\mathcal{N}$-I damping rate, on the other hand, increases rapidly towards a high temperature scaling of $\sim T^4$.  Again, at temperatures below $T = \hbar\omegam/k_{B}\approx 200$~mK the $\mathcal{N}$-I damping rate drops rapidly due to the reduced density of available phonon states in the OMC structure.  This is another manifestation of the `phonon bottleneck' effect discussed above in regards to the optically-induced hot phonon bath.  Energy deposited into high frequency phonons decays to lower frequency phonons through processes like the 3-phonon mixing studied here, however, when the phonon wavelengths approach the dimension of the structure the reduced density of phonon states results in a precipitous drop in the nonlinear phonon scattering, effectively trapping the energy in phonons above a certain cut-off frequency.    

For comparison purposes we have plotted the estimated damping for an isotropic Si bulk material involving $L + L \rightleftharpoons L$ acoustic scattering.  The black dashed curve (solid cyan curve) is the bulk damping rate without (with) a low-frequency cut-off for the range of integration in the $\mathcal{N}$-I process.  The cut-off frequency for the solid cyan curve is $\omega_{c}/2\pi=6.5$~GHz, chosen to match the top of the first phononic bandgap of the acoustic shield in the OMC cavity structure (c.f., Fig.~\ref{fig:sim_ph_TLS_prop}).  The temperature scaling for the bulk scattering (above cut-off) is approximately $T^4$.  An average overlap factor $\qavg{\GammathreephI} = 0.01$ was assumed in the quasi-mode modeling, resulting in a reasonable correspondence with the $\mathcal{N}$-I bulk damping at temperatures above $400$~mK.

Although the 3-phonon scattering rapidly rises with temperature ($\sim T^4$), for temperatures below $1$~K where the size-scale of the OMC cavity structure comes into play, the magnitude of 3-phonon scattering is estimated to be substantially smaller than that measured in our experiment.  In the following subsection we consider a more likely source of the observed damping, two-level systems, which act as an intermediate bath between phonons, greatly increasing the predicted damping rate.

\subsection{Numerical modeling of TLS-phonon interactions in the OMC cavity}
\label{subsubsec:nummodel_TLSdamping}

The coupling of the high-$Q$ acoustic breathing mode of the OMC cavity to TLS defect states depends on a number factors.  First and foremost there is the spatial and spectral density of the TLS, which determine the number of interacting defect states with the breathing mode.  In order to constrain the TLS density to a realistic value we consider here that the majority of the TLS are associated with defects in a near-surface layer of the etched Si structure making up the OMC cavity.  We assume no TLS defects in the bulk of the crystalline Si layer.  The thickness of the defective surface layer of Si depends greatly on its preparation.  

In our case, we have used an inductively-coupled reactive ion etch (ICP-RIE) to pattern the $220$~nm thick Si device layer.  The ICP-RIE etch utilizes an SF$_6$:C$_4$F$_8$ gas chemistry, with low RF power ($\approx 30$~W) and low DC-bias voltage ($\approx 70$~V), to reduce optimize the shape of the etched sidewall and to attempt to reduce etch-induced damage on the sidewalls of the etched Si.  Nonetheless, it is well known that these etch processes still produce a variety of damage to the exposed near-surface layers of Si in the process.  Typically in RIE etching~\cite{Yabumoto1981,Oehrlein1989,Oehrlein1992,Lee1989}, a surface consisting of a super-surface top layer of fluoro-carbons ($\sim 5$~nm) and Si-oxygen ($\sim 1.5$~nm) is followed by a sub-surface heavily damaged layer containing Si-carbon (among other impurities) that can penetrate into the bulk to depths of tens of nanometers. Here we assume an etched sidewall damage layer thickness of $\delta w = 15$~nm.  In order to reduce Si oxide growth on the top and bottom surface layers of the released Si device (i.e., those layers that do not see the ICP-RIE etch) we `flash' the sample with an anhydrous vapor HF etch prior to inserting into the vacuum of the dilution refrigerator (time between flash and vacuum pump down in the cryostat is $\lesssim 45$~minutes).  Ideally, this removes surface oxide layers and leaves a hydrogen-terminated Si surface nominally free of oxides.  To be conservative, however, we also assume a $\delta t=0.25$~nm surface oxide layer~\cite{Morita1989,Morita1994} on the top and bottom surfaces of the released Si device layer.  This yields a volume fraction of damaged Si in our devices which is $\eta_{\text{surf}} = 0.29$.  

Assuming a bulk TLS density commensurate with that in vitreous Si dioxide, $\ndTLS = 1.04$ states/J/m$^3$~\cite{Phillips1987}, this yields a spectral density of TLS that lie within the acoustic cavity volume of the breathing mode of only $\ndTLSm\sim 20$~states/GHz.  For a TLS population that is uniformly distributed spectrally~\cite{Kleiman1987}, this yields only $\sim 2000$ TLS with transition frequencies lying below $100$~GHz that are in the cavity mode volume.  Note that we simulate an ensemble TLS size of $\NTLS = 3920$, taking into account TLS that are within a spatial region of twice that of the cavity mode volume.              

The resonant interaction of those TLS that lie spatially in the breathing mode cavity volume $\Vm$ and have their transition frequency spectrally nearby the $\omegam/2\pi \approx 5$~GHz acoustic resonance frequency is determined by the magnitude of the transverse coupling deformation potential, $M$.  These TLS not only act as a bath to damp the breathing mode, but their temporal fluctuations from ground to excited state and back, lead to fluctuations in the acoustic environment of the breathing mode, producing both a frequency jitter of the mechanical mode and an overall shift in the resonance frequency that depends on the TLS temperature through its average excited state population~\cite{Phillips1987,GaoPhD}.  The acoustic frequency shift due to non-resonant TLS interacting through longitudinal $\hat{\sigma_{z}}$-coupling is negligible compared to the resonant $\hat{\sigma_{x}}$-coupling term.  As such, the magnitude of the transverse coupling parameter is chosen in our model to be $M=0.07$~eV, yielding an average transverse vacuum coupling rate to the breathing mode for TLS in $\Vm$ of $\langle \gtm/2\pi \rangle \sim 100$~kHz.  The corresponding dispersive shift due to the nearest resonant TLS (on average) is then approximately $\langle \dfmmax \rangle \cong \ndTLSm (\gtm/2\pi)^{2} \sim 1.5$~kHz.  This level of dispersive shift is in line with the both the measured frequency jitter ($\Delta_{1/2} \approx 3.5$~kHz) and the frequency shift around $T \approx \hbar\omegam/\kB$ for device D shown in Fig.~\ref{fig:fig3}(d).

The non-resonant relaxation interactions of TLS with the breathing mode is via a longitudinal coupling deformation potential, $D$.  With the transverse coupling rate set nominally by the measured frequency jitter and temperature-dependent frequency shift of the breathing mode, the longitudinal deformation potential is adjusted to approximately match the measured acoustic damping rate of the breathing mode at the lowest fridge measurement temperatures ($\Tf \approx 7$~mK).  A value of $D=5.6$~eV (angle-averaged $\bar{D} = 3.23$~eV) gives a reasonable fit to the measured data of Fig.~\ref{fig:fig3}(a).  Typical values in the literature for averaged $D$ and $M$ parameters are on the order of $1$~eV~\cite{Kleiman1987,Phillips1987}, although these values are hard to distinguish separately from the TLS density~\cite{Phillips1988}.  The large value of $D$ and small value of $M$ indicate a set of TLS (or TS) states which have a large asymmetry energy and small tunneling energy.  Other parameters and their assumed values in our model are listed, along with references and comments, in Table~\ref{tab:TLSmodel}.

\begin{table}[btp]
\begin{center}
\caption{\textbf{TLS damping model parameters.}  \label{tab:TLSmodel}}
%\begin{ruledtabular}
\begin{tabular}{|l|l|l|l|}
\hline\hline
%\toprule
 Parameter & Description & Value & Refs./Notes \\
\hline
  $D$ & longitudinal coupling deformation potential & $5.6$~eV & fit value (see also \cite{Phillips1987,Kleiman1987,Phillips1988})\\ 
  $\bar{D}$ & angle-averaged longitudinal coupling & $3.23$~eV & $\equiv D/\sqrt{3}$ \\
  $M$ & transverse coupling deformation potential & $0.07$~eV & fit value (see also \cite{Phillips1987,Kleiman1987,Phillips1988})\\ 
  $\bar{M}$ & angle-averaged transverse coupling & $0.04$~eV & $\equiv M/\sqrt{3}$ \\
  $\delta w$ & Si device layer etched sidewall damage layer thickness & $15$~nm & \cite{Yabumoto1981,Oehrlein1989,Oehrlein1992,Lee1989} \\
  $\delta t$ & Si device layer top and bottom oxide thickness & $0.25$~nm &  \cite{Morita1989,Morita1994}\\
  $\eta_{\text{surf}}$ & damaged material/surface oxide volume fraction & $0.29$ & estimated from $\delta t$, $\delta w$\\
  $m$ & high-$Q$ breathing mode label & N/A & \\
  $\omegam$ & mode $m$ frequency & $5.3$~GHz & COMSOL sim. \\
  $\Vm$ & mode $m$ acoustic mode volume & $0.11$~($\mu$m)$^3$ & COMSOL sim. \\
  $\langle \esqm \rangle $ &  average strain squared of vacuum for mode $m$ within $\Vm$ & $2.25 \times 10^{-8}$ & COMSOL sim. \\
  $\ndTLS$ & TLS density per unit volume per unit energy & $1.04$~states/J/m$^3$ & \cite{Phillips1987} \\
  $\ndTLSm$ & TLS density per unit frequency in $\Vm$ ($\equiv (\hbar 2 \pi)\ndTLS\eta_{\text{surf}}\Vm$) & $22.1$~states/GHz & calculated \\
  $\NTLS$ & estimated number of TLS in $2\times \Vm$ with $f \leq 100$~GHz & $3920$ & calculated \\
  $\GammaTLSphi/2\pi$ & TLS pure dephasing rate & $10$~kHz & \cite{Phillips1987,Black1977} \\
  $\LambdaTLS$ & TLS excitation rate (due to optical pumping) & $2$~Hz & $\sim (2\times 10^{-3})(\gammap \nbathp)$ at $\ncav=0.1$ \\
  $\langle \gtm/2\pi \rangle $ & average transverse vacuum coupling rate to mode $m$ for TLS in $\Vm$ & $\sim 100$~kHz & calculated \\
  $\langle \glm/2\pi \rangle $ & average longitudinal vacuum coupling rate to mode $m$ for TLS in $\Vm$ & $\sim 8.6$~MHz & calculated \\
  $\langle \dfmmax \rangle $ & average dispersive shift of mode $m$ for nearest resonant TLS & $\sim 1.5$~kHz & calculated ($\cong \ndTLSm (\gtm/2\pi)^{2}$) \\  
  
  \hline\hline
%\bottomrule
\end{tabular}
%\end{ruledtabular}
\end{center}
\end{table}

The calculated zero-temperature energy decay rate of an ensemble of $~3920$ TLS, randomly chosen from $101$ fixed positions within the breathing mode cavity mode volume, with randomly oriented acoustic dipoles, and with randomly chosen frequency below $100$~GHz is displayed Fig.~\ref{fig:sim_ph_TLS_prop}(d).  This calculation, following Eq.~(\ref{eq:GammaTLSoneT}), uses the simulated acoustic strain and radiative decay rate of the localized quasi-mode phonons of the suspended and peripherally-clamped OMC structure whose properties are also displayed in Fig.~\ref{fig:sim_ph_TLS_prop}.  Several points are worth noting here.  The first is that the small number of TLS in the small acoustic mode volume $\Vm$ and the small number of localized quasi-normal phonon modes at low frequency means a significant spectral fluctuation in the decay rate of TLS with transition frequency below $\sim 1$~GHz.  Within the phononic bandgap of the OMC cavity ($3.5-6.5$~GHz), there is a dramatic reduction in the decay rate of TLS, down to levels on the order of $1$~Hz.  Above the phononic bandgap, the TLS zero-temperature decay rate rises rapidly, roughly as the cube of the TLS transition frequency, consistent with the approximate 2D phonon quasi-mode density to which the TLS are coupled (c.f., Fig.~\ref{fig:sim_ph_TLS_prop}(c).   

\begin{figure*}[btp]
\begin{center}
\includegraphics[width=1.0\columnwidth]{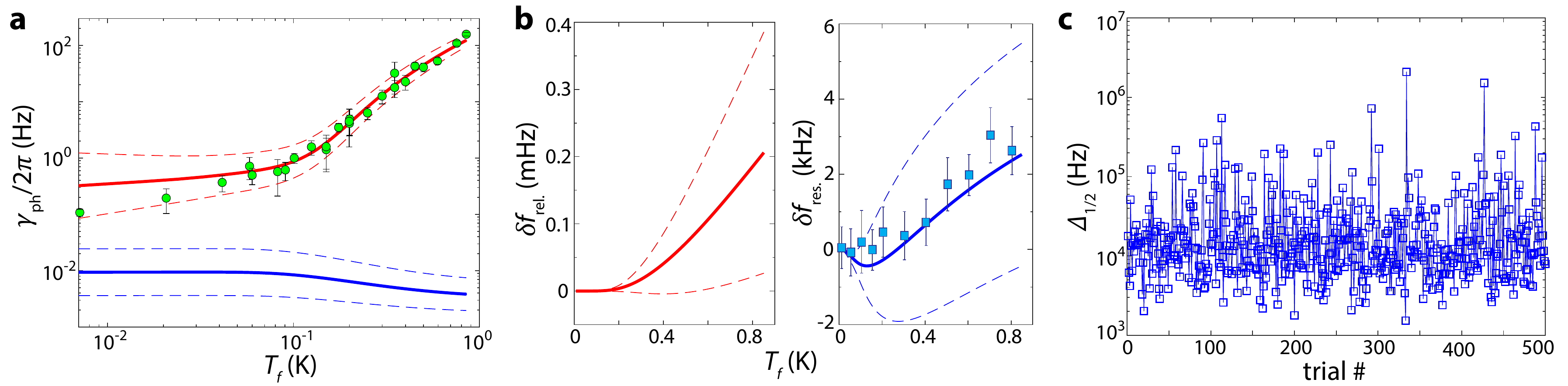}
\caption{\textbf{Modeled breathing mode interactions with a TLS bath.} Modeling was performed using the acoustic mode and TLS properties found in Tab.~\ref{tab:TLSmodel}.  Simulations were performed using 100 ensemble instances of $\NTLS = 3920$ randomly positioned (sampled from a set of 101 fixed positions) and oriented TLS within the breathing mode acoustic volume, $\Vm$.  The transition energy of each TLS are also sampled from a random distribution. 
\textbf{a},  Relaxation (red curves) and resonant (blue curves) damping of the acoustic breathing mode versus TLS bath temperature.  Solid (dashed) curves represent the average ($1$-$\sigma$ standard deviation in log-space) of the 500 simulation trials run. Measured damping of device D is shown as filled green circles. 
\textbf{b},  Breathing mode frequency shift versus temperature due to relaxation (left plot) and resonant (right plot) TLS interactions.  The solid cyan curve is a curve resulting from single TLS ensemble trial. Measured frequency shift of mode of device D is shown as filled blue squares. 
\textbf{c}, Full-width half-maximum of the time-averaged frequency jitter of the breathing acoustic mode resulting from resonant interactions with all of the TLS for each of the $500$ trial ensembles. Here we assume a $\LambdaTLS = 2$~Hz excitation rate of each TLS due to weak optical absorption.
} 
\label{fig:sim_ph_damp_df_jitter}
\end{center}
\end{figure*}

Figure~\ref{fig:sim_ph_damp_df_jitter} displays the resulting damping (c.f., Eqs.(\ref{eq:dgamma_m_res},\ref{eq:dgamma_m_rel})) and frequency shift (c.f., Eqs.(\ref{eq:df_m_res},\ref{eq:df_m_rel})) of the high-$Q$ breathing mode versus temperature due to coupling with the TLS bath in the acoustic mode volume $\Vm$.  In these simulations we performed 500 trial runs of random TLS ensembles.  Both the average and standard deviation of the damping and frequency shift are shown.  Also shown are the effects of both `resonant' $\hat{\sigma}_{x}$-interactions and `relaxation' $\hat{\sigma}_{z}$-interactions with the TLS.  As can be clearly seen, the resonant TLS damping of the breathing mode at the lowest temperatures $\Tf \lesssim 100$~mK is predicted to be roughly an order of magnitude smaller than the relaxation damping.  In addition, as the temperature is increased above that of $\hbar\omegam/\kB \approx 200$~mK, the resonant damping term begins to saturate due to the thermal excitation of TLS.  The overall suppression of the resonant TLS damping is due to the presence of the acoustic bandgap, which dramatically reduces the decay rate of TLS nearly-resonant with the breathing mode due to a lack of localized quasi-normal phonon modes in the gap.  Instead, the typically weaker relaxation damping from non-resonant TLS outside the acoustic bandgap dominates the simulated breathing mode damping. 

The correspondence of the simulated TLS relaxation damping of the breathing mode with the measured breathing mode damping is striking not only in the overall magnitude of the predicted damping but also in its temperature dependence.  At temperatures below $\Tf \approx 100$~mK, where the thermally excited TLS that contribute to relaxation damping have transition frequencies below the acoustic bandgap and interact with a quasi-1D phonon bath, the damping is seen to have a reduced, linear to sub-linear dependence with temperature.  Above $\Tf \approx 100$~mK, thermal excitation of TLS with transition frequencies above that of the acoustic bandgap begin to contribute to the damping.  In this spectral range the phonon density of states in the OMC cavity structure is approximately linear with frequency, resulting in TLS decay times which scale quadratically with transition frequency.  This sets the temperature scaling of the breathing mode relaxation damping, which is also seen to scale approximately quadratically with temperature above $\Tf \approx 100$~mK.

\begin{figure*}[btp]
\begin{center}
\includegraphics[width=0.70\columnwidth]{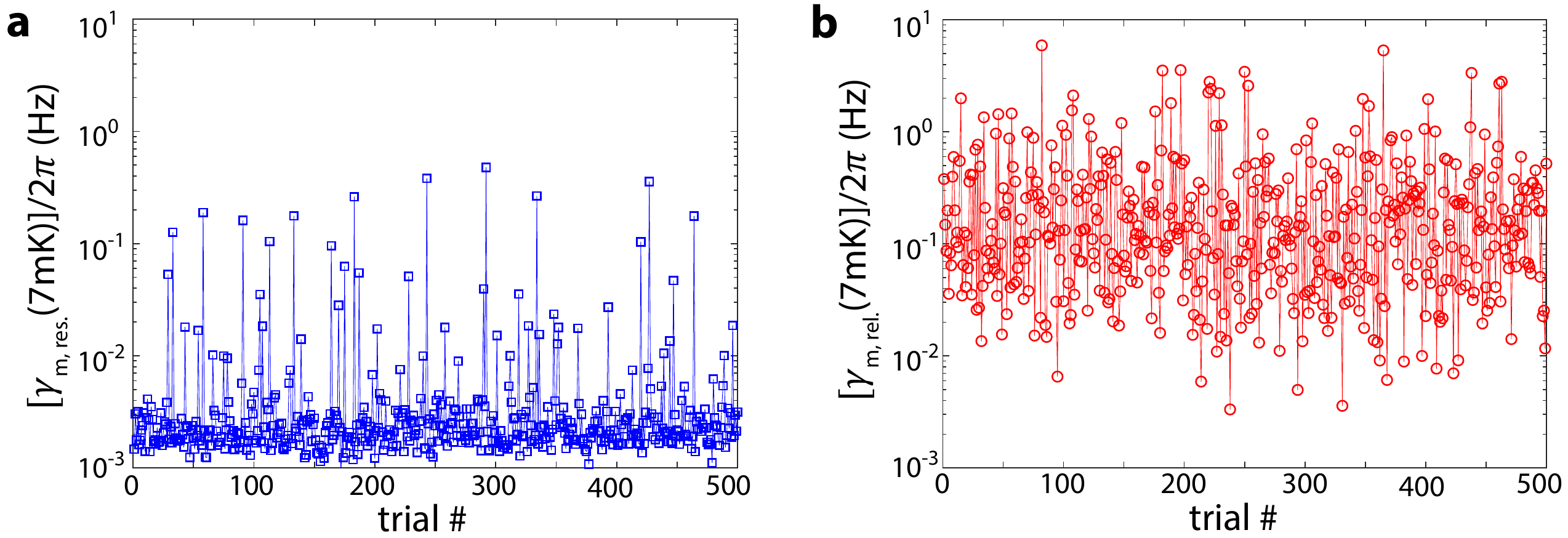}
\caption{\textbf{Fluctuation from trial-to-trial in the simulated low temperature damping at 7mK.} \textbf{a}, Simulated TLS-resonant-interaction damping at temperature $T=7$~mK of breathing mode as a function of different TLS ensemble trials.  \textbf{b}, Simulated TLS-relaxation damping at temperature $T=7$~mK of breathing mode as a function of different TLS ensemble trials.
} 
\label{fig:sim_ph_damp_versus_trial}
\end{center}
\end{figure*}

Another feature of the simulated TLS damping of the breathing mode is the large fluctuation from trial-to-trial in the low temperature ($\Tf \lesssim 100$~mK) portion of the damping versus temperature curve.  This shows up as a large variance in magnitude and temperature trend of the relaxation damping at these low temperatures in simulation.  The dashed curves in Fig.\ref{fig:sim_ph_damp_df_jitter}(a) represent the range of temperature curves within one standard deviation of the mean curve \emph{on a log scale}.  In order to better appreciate what the level of fluctuations in the predicted low temperature damping are for both resonant and relaxation TLS damping of the breathing mode, we plot in Fig.~\ref{fig:sim_ph_damp_versus_trial} the trial-to-trial variations of the simulated damping factors at $\Tf=7$~mK.  The variations are substantial, with a standard deviation in log-space of a little over an order of magnitude.  This is consistent with our measured observations, as can be gleaned from the plot of measured breathing mode energy damping rates versus phononic bandgap shield number in Fig.~\ref{fig:fig2}(d).  For acoustic shield periods greater than about $6$, where radiation damping is estimated to be minimal, the measured damping rates also range over a little over an order of magnitude.  This variance is explained in the TLS model by the very small number (handful) of TLS with transition frequency below $100$~MHz within the breathing mode volume, which owing to their random positioning and orientation, results in a large variation in breathing mode damping.    

The simulated frequency shift versus temperature, for both resonant and relaxation interactions with the TLS ensemble, is shown in Fig.~\ref{fig:sim_ph_damp_df_jitter}(b).  In these plots we have referenced the frequency shift to that at the lowest measured temperature ($\Tf\approx7$~mK).  The relaxation component of the frequency shift (left panel) is estimated to be in the milli-Hz range, far below the kHz-scale frequency shift due to resonant interactions with the TLS (right panel).  The resonant TLS frequency shift of the breathing mode versus temperature has a simulated mean curve averaged over the $500$ ensemble trials that roughly follows the digamma function response of Eq.~(\ref{eq:df_m_res}), with the breathing mode frequency initially shifting lower and then begin to increase around $\hbar\omegam/\kB \approx 200$~mK, followed by a monotonic (logarithmic) increase in frequency for higher temperatures.  The simulated frequency shift versus temperature curve has a large variance however (range of curves within standard deviation of mean curve are bounded by dashed curves), depending sensitively on the magnitude and sign of the detuning of the TLS closest to resonance with the breathing mode.  The measured frequency shift with temperature of device D is plotted alongside the simulated resonant component of the TLS-induced frequency shift in Fig.~\ref{fig:sim_ph_damp_df_jitter}(b), showing good correspondence with the mean curve.

\begin{figure*}[btp]
\begin{center}
\includegraphics[width=\textwidth]{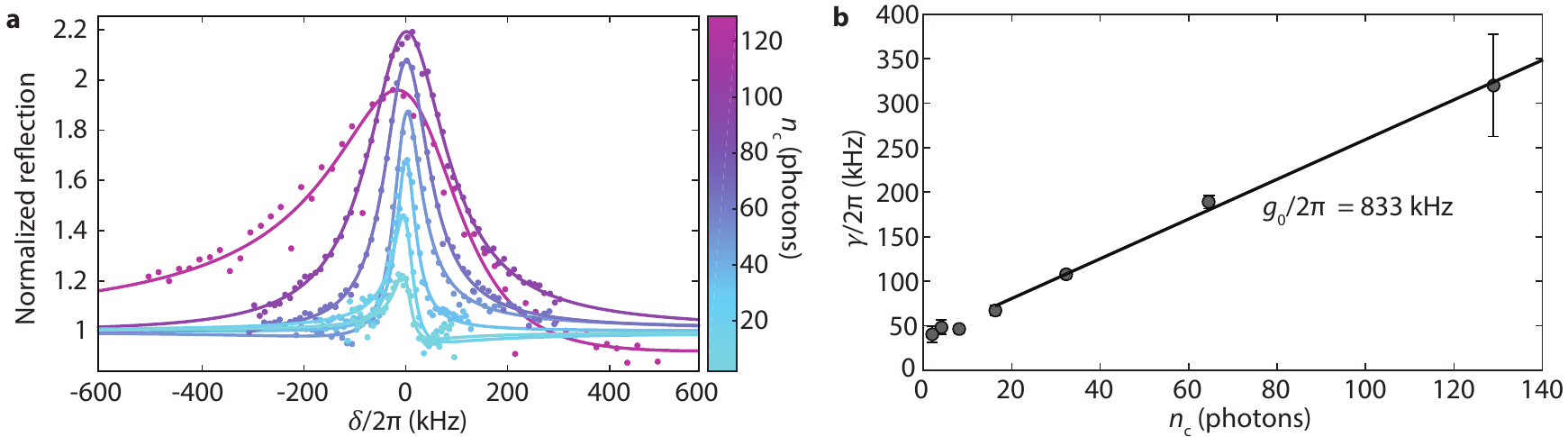}
\caption{\textbf{EIT linewidth measurements of Device E at $\Tf = 10$~mK.}  \textbf{a},  Normalized reflection amplitude for probe photons as a function of pump photon number. The reflection peak represents an EIT-like transparency window approximately centered within the bare optical cavity line ($\sim 1$~GHz wide). Asymmetry in the trace at $\ncav = 129$ can be attributed to an effective detuning shift, likely caused by thermal shifting of the cavity at high input power. These EIT measurements were performed on a device E having seven acoustic shield periods and mechanical $Q = 1.5\times 10^{10}$ measured via ringdown. \textbf{b}, Plot of the fit mechanical linewidth versus pump photon number from the EIT curves of (\textbf{a}). At $\ncav \lesssim 10$, the time-averaged mechanical linewidth saturates to $40$~kHz due to mechanical frequency jitter. At higher $\ncav$, the mechanical mode is broadened by optomechanical back-action, the slope of which yields $\gzero/2\pi = 833$~kHz.}
\label{fig:deviceE_EIT_linewidth}
\end{center}
\end{figure*}

As discussed in the main text, and shown in Fig.~\ref{fig:fig3}(d), the measured frequency jitter when averaged over minutes has a spectral full-width at half-maximum (FWHM) of $\Delta_{1/2} \approx 3.5$~kHz, approximately independent of temperature ($\Tf=7-850$~mK) and optical probing power ($\ncav=0.3-0.02$).  Over $\lesssim 0.1$~s timescales one can resolve this frequency jitter in the time domain (c.f., main text Fig.~\ref{fig:fig3}(c)).  In order to estimate the amount of frequency jitter that the acoustic breathing mode might incur due to interactions with TLS we have assumed in our simulations that all TLS are being excited at a rate faster than the measurement averaging time of a few minutes.  Assuming all TLS are fluctuating independently, we can then write using Eq.~(\ref{eq:df_m_res}) without the temperature dependence and assuming that $\langle \hat{\sigma_{z}}^2 \rangle = 1$, 

\begin{multline}
\Delta_{1/2} = 2(2\log{[2]}) \Bigg( \sum_{\text{TLS}} \left( \gtm^2[\vecb{r}_{\text{TLS}}]  \right)^2 \\
\times \left\{ \frac{\omegaTLS-\omegam}{(\omegaTLS-\omegam)^2 + ((\delta \GammaTLSone)_{\text{ph}}/2 + \GammaTLSphi)^2} + \frac{\omegaTLS+\omegam}{(\omegaTLS+\omegam)^2 + ((\delta \GammaTLSone)_{\text{ph}}/2 + \GammaTLSphi)^2} \right\}^2 \Bigg)^{1/2},
\label{eq:dfjitterTLS}
\end{multline}

\noindent where the prefactor $2(2\log{[2]})$ accounts for the conversion between the standard deviation and the FWHM of a normal distribution.  

TLS excitation occurs naturally through thermal excitation, although the rate of excitation in that case depends strongly on temperature, which we do not observe in our measurements. However, we also know that the optical probing of the mechanics can lead to optical-absorption-induced excitation of a hot bath that damps the mechanics.  It seems reasonable then to assume that the same optical absorption would also excite a broad spectrum of TLS, thus leading to their contribution to the breathing mode frequency jitter. In the model we have taken the optical-absorption-induced pumping rate of the TLS to be $\LambdaTLS=2$~Hz, which for context represents $0.2\%$ of the hot bath heating rate $\gammap\nbathp$ of the breathing mode at $\ncav=0.1$.  So even if the TLS are driven much more weakly than the breathing mode due to optical absorption effects, one would need to probe at lower optical probe powers than currently accessible in our experiments ($\ncav=0.02$) to see a reduction in the time-averaged frequency jitter of the mechanics, where time averaging is performed over a second or longer.  Note that pumping-induced saturation effects of the TLS have also been included in the simulation although their effects on both the damping and the overall frequency shift is minor.    

The resulting simulated frequency jitter FWHM of the breathing mode, $\Delta_{1/2}$, is plotted in Fig.~\ref{fig:sim_ph_damp_df_jitter}(d) for each of the 500 trial TLS ensembles.  The variation from trial-to-trial of the frequency jitter is substantial, with the jitter ranging from kHz to (in rare cases) MHz.  Although the measured frequency jitter of device D, comprehensively studied in the main text, lies on the low end of this spectrum at $\Delta_{1/2} \approx 3.5$~kHz, this device is also on the low end of the range of measured values in our experience.  As an example, the measured linewidth of another high-$Q$ device (device E in Table~\ref{tab:devices}) is shown in Fig.~\ref{fig:deviceE_EIT_linewidth}.  In this case, the linewidth at low optical probe power is found to saturate to a FWHM of $\sim 40$~kHz, closer to the mean simulated value.  The large fluctuation in the measured frequency jitter from device-to-device, consistent with the model, is again an indication of the sensitivity of the breathing mode to a select few TLS in the near-resonant regime.

%The most relevant TLS in this regard are the TLS nearest in resonance with the breathing mode, which also lie in the acoustic bandgap of the phononic shield of the OMC structure ($\omegaTLS/2\pi \sim 3.5$-$6.5$~GHz).  

\end{document}